\documentclass[nonacm,sigplan]{acmart}

\usepackage{algorithm}
\usepackage{algorithmic}
\usepackage{color}
\usepackage{pifont}
\usepackage{multirow}
\usepackage{graphicx}

\begin{document}

\title{MLDSE: Scaling Design Space Exploration Infrastructure for Multi-Level Hardware}

\author{Huanyu Qu}
\affiliation{
    \institution{University of Macau}
    \city{Macau}
    \country{China}
}
\affiliation{
    \institution{Guangdong Institute of Intelligence Science and Technology}
    \city{Hengqin, Guangdong}
    \country{China}
}
\email{yc37960@um.edu.mo}
\authornote{These authors contributed equally to this work.}

\author{Weihao Zhang}
\affiliation{
    \institution{Tsinghua University}
    \city{Beijing}
    \country{China}
}
\email{zwh18@tsinghua.org.cn}
\authornotemark[1]

\author{Junfeng Lin}
\affiliation{
    \institution{Tsinghua University}
    \city{Beijing}
    \country{China}
}
\email{linjf21@mails.tsinghua.edu.cn}

\author{Songchen Ma}
\affiliation{
    \institution{ACCESS -- Al Chip Center for Emerging Smart Systems, Hong Kong University of Science and Technology}
    \city{Hongkong}
    \country{China}
}
\email{songchenma@ust.hk}

\author{Hongyi Li}
\affiliation{
    \institution{Tsinghua University}
    \city{Beijing}
    \country{China}
}
\email{hy-li21@mails.tsinghua.edu.cn}

\author{Luping Shi}
\affiliation{
    \institution{Tsinghua University}
    \city{Beijing}
    \country{China}
}
\email{lpshi@tsinghua.edu.cn}
\authornote{Corresponding authors.}

\author{Chengzhong Xu}
\affiliation{
    \institution{University of Macau}
    \city{Macay}
    \country{China}
}
\email{czxu@um.edu.mo}
\authornotemark[2]

\begin{abstract}
To efficiently support large-scale NNs, multi-level hardware, leveraging advanced integration and interconnection technologies, has emerged as a promising solution to counter the slowdown of Moore's law. However, the vast design space of such hardware, coupled with the complexity of their spatial hierarchies and organizations, introduces significant challenges for design space exploration (DSE). Existing DSE tools, which rely on predefined hardware templates to explore parameters for specific architectures, fall short in exploring diverse organizations, spatial hierarchies, and architectural polymorphisms inherent in multi-level hardware. To address these limitations, we present \textbf{M}ulti-\textbf{L}evel \textbf{D}esign \textbf{S}pace \textbf{E}xploror (MLDSE), a novel infrastructure for domain-specific DSE of multi-level hardware. MLDSE introduces three key innovations from three basic perspectives of DSE: 1) \textbf{Modeling}: MLDSE introduces a hardware intermediate representation (IR) that can recursively model diverse multi-level hardware with composable elements at various granularities. 2) \textbf{Mapping}: MLDSE provides a comprehensive spatiotemporal mapping IR and mapping primitives, facilitating the mapping strategy exploration on multi-level hardware, especially synchronization and cross-level communication; 3) \textbf{Simulation}: MLDSE supports universal simulator generation based on task-level event-driven simulation mechanism. It features a hardware-consistent scheduling algorithm that can handle general task-level resource contention. Through experiments on LLM workloads, we demonstrate MLDSE's unique capability to perform three-tier DSE spanning architecture, hardware parameter, and mapping.
\end{abstract}

\maketitle 
\pagestyle{plain} 

\section{Introduction}
\label{sec:intro}
Large language models (LLMs) have demonstrated exceptional capabilities across a wide range of applications\cite{achiam2023gpt, openai2022chatgpt}. To support these large-scale neural networks (NNs), various LLM acceleration hardware\cite{heo2024neupims, choquette2021nvidia, choquette2023nvidia, qin2024mecla, peng2023chiplet, lee2024cost, lee2024tender, huang2024hecaton} and software toolchains\cite{kwon2023efficient, zhang2024llmcompass, zhong2024distserve, agrawal2023sarathi, agrawal2024vidur, agrawal2024etalon, kundu2024performance, lee2024infinigen} have been developed, where the design space exploration (DSE) tools\cite{zhang2024llmcompass, agrawal2023deap, zhu2024theseus, kundu2024performance, lin2024hex} play a pivotal role in designing efficient hardware accelerators and deploying strategies. However, the slowdown of Moore's law and the die size constraints have threatened the performance gains and the cost of single monolithic chips\cite{choquette2023nvidia, choquette2021nvidia, peng2023chiplet, arunkumar2017mcm}. As LLMs continue to grow in size and complexity, multi-level hardware\cite{talpes2023microarchitecture, peng2023chiplet, abts2022software, lie2021multi, tirumala2024nvidia, arunkumar2017mcm, jia2019dissecting, dennis2020think, cong2024attentionlego, huang2023hierarch} has emerged as a scalable solution, which arranges multiple processing or memory elements in hierarchical levels, such as many-core accelerators\cite{knowles2021graphcore, jia2019dissecting, modha2023neural, vasiljevic2021compute, li2022spacx, dennis2020think} and multi-chiplet accelerators\cite{shao2019simba, arunkumar2017mcm, smith202411, liao2021ascend, sharma2023heterogeneous}, to enhance scale-out capabilities. The complexity and diversity of multi-level hardware introduce additional design considerations beyond the scope of single LLM acceleration kernels, including: \textbf{Horizontal organization complexity.} Each hardware level consists of multiple computation, memory, communication elements or elements formed by the lower-level hardware, e.g., a multi-chip package is formed of multiple chiplets. The selection, arrangement, and size of these elements create complicated design trade-offs\cite{han2023big, rashidi2023unico}, e.g., computation-memory-communication co-optimization, communication with diverse interconnections, memory hierarchies with different capacities and access protocols. \textbf{Vertical level complexity.} Multi-level hardware involves multiple spatial levels with various granularities, where each level can have complicated within-level organization and different elements within the same level can have varied sub-spatial structures, leading to an exponentially expanding design space\cite{peng2023chiplet, abts2022software, tirumala2024nvidia, tan2021nn, cai2024gemini, arunkumar2017mcm, shao2019simba}. These levels range from multi-server clusters down to multi-core chips and individual cores\cite{orenes2024muchisim, zhang2024llmcompass, won2023astra, rashidi2020astra}, each influencing performance through complex cross-level interactions.

\begin{figure*}[!t]
  \centering
  \includegraphics[scale=0.56, keepaspectratio]{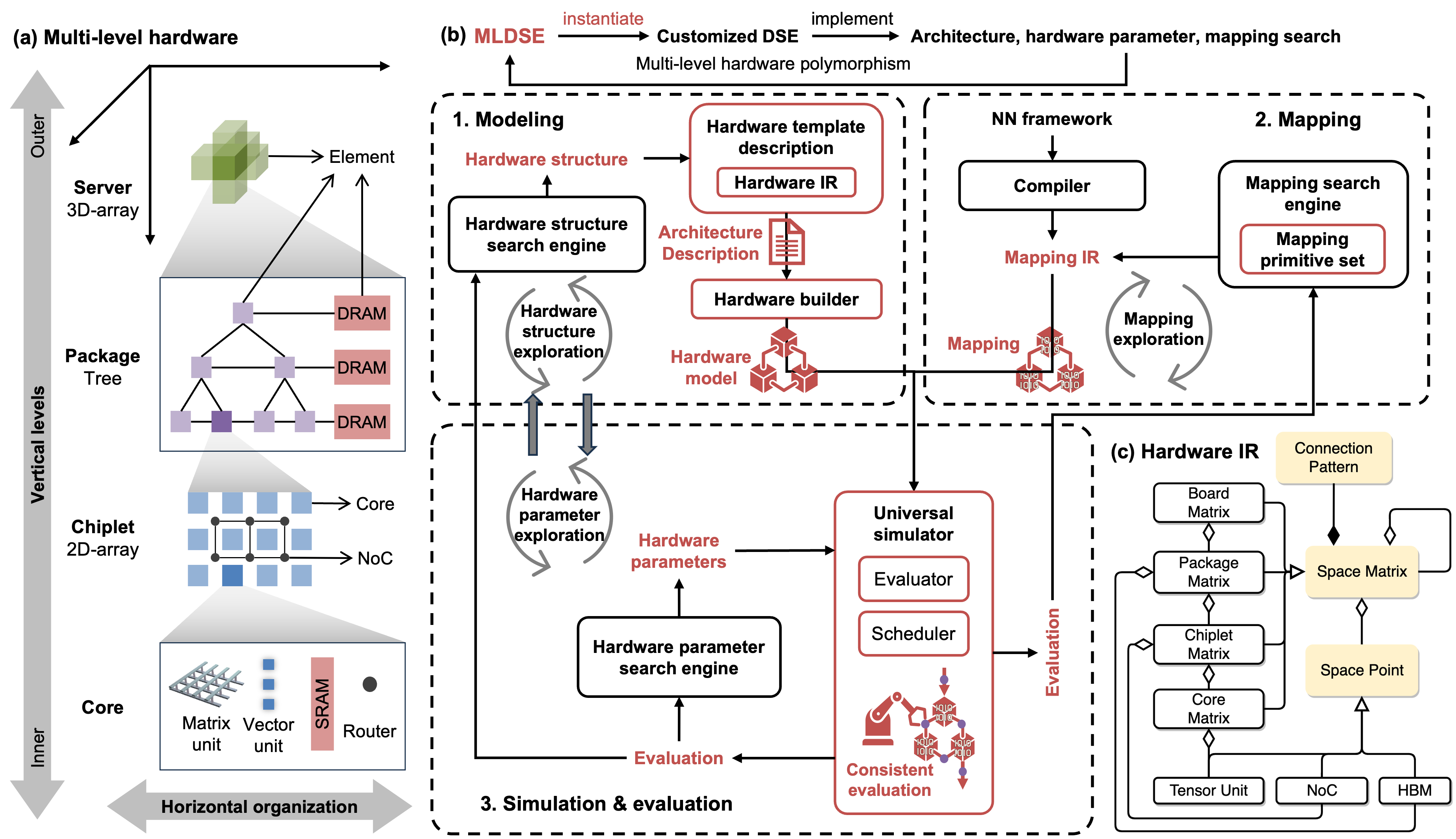}
  \caption{Overview of MLDSE infrastructure.}
  \label{fig:MLS_flow}
\end{figure*}

Depth of spatial levels, heterogeneous spatial structures of different hardware elements, organization of each level, and inter-level interactions all significantly impact overall performance. The structural diversity of multi-level hardware calls for \textbf{more versatile DSE tools that can effectively handle architecture-level polymorphism}. However, existing DSE tools often focus on parameter searches (e.g., memory capacity, compute parallelism, communication bandwidth) within predefined multi-level hardware templates\cite{tan2021nn, hao2023monad, cai2024gemini, zhang2024llmcompass, samajdar2023airchitect, kwon2020maestro, samajdar2018scale}. These tools may lack the capability to address the complexities at each hardware level, let alone the flexible interactions across levels. We classify existing DSE tools into four categories: 1) Tools that model specific multi-level hardware architectures, e.g., from tile to cluster in Muchisim\cite{orenes2024muchisim}. However, their fixed hardware templates limit the exploration of flexible intra-level organization and spatial hierarchy variations. 2) Tools that employ specific data structures to model a class of hardware, e.g., tree structure in Timeloop\cite{parashar2019timeloop}. While Timeloop allows flexible modeling of memory hierarchies in one level, it is primarily designed for single-core processors, making it challenging to extend to many-core scenarios or general multi-level hardware. 3) Tools that support hierarchical networks, e.g., topology building blocks in ASTRA-Sim2.0\cite{won2023astra}. However, ASTRA-Sim2.0's focus on cluster-level simulation, with the finest granularity at the NPU level, limits its ability to model arbitrary multi-level hardware. 4) Modular simulators\cite{herbst2024switchboard, matthews2020mosaicsim} provide frameworks for aggregating different simulators. However, simply aggregating existing tools cannot adequately address the intricate interactions within multi-level hardware. To address these limitations, we draw inspiration from the meta-compiler methodology of MLIR\cite{lattner2021mlir} and introduce the \textbf{M}ulti-\textbf{L}evel \textbf{D}esign \textbf{S}pace \textbf{E}xplorer (MLDSE). MLDSE is a \textbf{meta-DSE tool} designed for generalizability in generating and facilitating the DSE process for customized multi-level hardware. MLDSE is \textbf{not} a wrapper for existing tools; it is a dynamic \textbf{composer} that tackles challenges unaddressed by current solutions:

\textbf{The complexity of modeling multi-level hardware with flexible within-level organization, varied spatial hierarchy, and mixed modeling granularity:} MLDSE must handle the modeling of diverse and flexible multi-level hardware architectures, where each level has its unique organization, and the overall spatial hierarchy is not predefined. This requires MLDSE to recursively establish relationships between spatial levels, while ensuring that each level remains composable and easily modifiable. Moreover, the exploration of multi-level hardware often faces ever-changing regions of interest, which necessitates the mixed granularity modeling, e.g., treating GPUs as atomic elements while requiring fine-grained modeling of multi-chiplet accelerators from core to package levels when exploring a heterogeneous system\cite{hwang2020centaur, heo2024neupims, kim2019architecture, sharma2023heterogeneous}. 

\textbf{The challenge of developing a unified mapping representation for diverse multi-level hardware:} Supporting the diverse multi-level hardware mapping requires a highly versatile spatiotemporal mapping IR, which must accurately capture hardware element locations, task execution ordering, and spatiotemporal relationships between tasks and hardware elements for enabling complex mapping strategies\cite{chen2014diannao, chen2014dadiannao, pei2019towards, cai2024gemini, knowles2021graphcore, ma2022neuromorphic}. Moreover, MLDSE cannot rely on an architecture-specific search algorithm for mapping optimizations, because different architectures and workloads prefer different ones. Instead, a comprehensive set of mapping primitives is needed for constructing search algorithms to explore the multi-level hardware mapping space. The complexity increases with hierarchical synchronization mechanisms and interconnections in multi-level hardware, which are often overlooked in existing DSE tools.

\textbf{The difficulty in fast, accurate, and hardware-consistent evaluation of diverse multi-level hardware:} Many DSE tools\cite{cai2024gemini, tan2021nn, parashar2019timeloop, wu2019accelergy} use analytical models for fast evaluation. However, a meta-DSE tool like MLDSE cannot predefine hardware templates, making static evaluation models impractical for this scenario. Instead, MLDSE need to dynamically generate simulators for different multi-level hardware just-in-time (JIT) that accurately reflect real-world hardware behavior while remaining efficient enough to facilitate rapid DSE. This cannot be achieved by simply aggregating existing simulators, as this approach cannot adequately capture the complex dependencies, resource contention, and dynamic synchronization overheads inherent in multi-level hardware.

To overcome these challenges, MLDSE proposes the following key contributions in this paper:

\textbf{1. Multi-level hardware modeling with Recursive and composable rules.} MLDSE introduces a hardware IR and a hardware builder that recursively instantiate each level of user-defined hardware into composable hardware models with varying granularity.
    
\textbf{2. Comprehensive spatiotemporal mapping IR and mapping primitives.} MLDSE provides a spatiotemporal mapping IR and a set of mapping primitives to define the mapping optimization space of multi-level hardware, focusing on cross-level communication mapping and hierarchical synchronization mechanisms.

\textbf{3. JIT simulation generation for consistent evaluation.} MLDSE supports JIT simulation generation for various multi-level hardware, leveraging a task-level discrete-event mechanism. It features a hardware-consistent task scheduling algorithm to address general task-level resource contention on dataflow graphs.

\textbf{4. LLM Experiments with hardware insights.} Leveraging MLDSE's unique cross-architecture DSE capabilities, three different architectures are compared: 1) GPU-like shared memory vs distributed many-core architecture to quantitatively demonstrate how distributed hardware resources alleviate contention. 2) Evaluate the impact of spatial hierarchy variations on performance and cost.

\section{Background}
\subsection{Multi-Level Hardware for LLMs}

Following the scaling law\cite{kaplan2020scaling}, LLMs' increasing size demands greater hardware performance. NVIDIA's A100\cite{choquette2021nvidia} represents a large, costly, monolithic hardware for LLM training and inference. Its successor, H100\cite{choquette2023nvidia}, enhances single-chip performance through advancements in technology node and tensor core design. H100 systems utilize HBM3 for expanded memory bandwidth and fourth-generation NVLink for improved multi-GPU scalability. Further scaling is achieved with multi-chip-modulo (MCM) GPUs\cite{arunkumar2017mcm} and NVIDIA's B200\cite{tirumala2024nvidia}, which integrates two reticle-limited dies via chiplet, an approach also used in AMD's MI300X\cite{smith202411} to enable more complex spatial hierarchies for improving computing power. Beyond GPUs, Tesla's Dojo\cite{talpes2022dojo}, Grop\cite{abts2022groq}, Graphcore's IPU\cite{knowles2021graphcore}, IBM's Northpole\cite{modha2023neural}, Cerebras\cite{lie2022cerebras} also demonstrate how complex spatial hierarchies improve performance. Other initiatives\cite{shao2019simba, peng2023chiplet} use chiplets to build multi-level hardware for balancing performance and cost in large-scale computing systems. However, despite its promise, designing multi-level hardware presents significant challenges\cite{han2023big, rashidi2023unico} due to the expanded design space and intricate design trade-offs. Efficient tools are crucial for diagnosing bottlenecks and optimizing new multi-level hardware.

\subsection{Design Space Exploration}
\label{sec:DSE}

Various DSE tools\cite{tan2021nn, hao2023monad, cai2024gemini, zhu2024theseus, orenes2024muchisim, tan2024cocco, zhang2024llmcompass, rashidi2020astra} have been developed for different architectures. The common workflow for these architecture-specific DSE tools can be viewed as a sub-process in Figure \ref{fig:MLS_flow}: First, a hardware template needs to be defined for the target architecture. The DSE engine then tunes hardware parameters (e.g., memory capacity, bandwidth, systolic array size) on this template based on one or multiple workloads\cite{das2024multi}. From an initial hardware configuration, the mapping engine generates mapping strategies using architecture-specific search algorithms \cite{tan2024cocco, zhang2024llmcompass}. Each intermediate mapping is evaluated for performance, power, area, and cost (PPAC) by the evaluation engine to guide the mapping search. If the resulting PPCA is insufficient, the hardware DSE engine generates a new design point, iterating the whole process until an optimal configuration is found.

\section{Overview of MLDSE}
\label{sec:overview}

Figure \ref{fig:MLS_flow}(b) provides an overview of MLDSE. Unlike architecture-specific DSE tools limited to parameter tuning within predefined templates, MLDSE is a \textbf{meta-DSE} tool that supports multi-level hardware polymorphism, enabling exploration of a broader design space that spans architecture, hardware parameters, and mapping strategies. MLDSE achieves this through three key components:

\textbf{Hardware Modeling:} Enables flexible modeling of arbitrary multi-level hardware through a hardware IR. This IR enables the hierarchical composition of hardware components across an arbitrary number of spatial levels. The hardware builder then converts IR-defined architecture descriptions into operable hardware models, providing the foundation for subsequent DSE.

\textbf{Spatiotemporal Mapping:} Provides a unified abstraction for exploring multi-level hardware mapping. It includes a spatiotemporal mapping IR and a set of mapping primitives, emphasizing hierarchical synchronization and cross-level communication, to facilitate mapping strategy exploration of diverse multi-level hardware.

\textbf{Universal Simulator:} Provides fast, accurate evaluation of various hardware designs and mapping strategies. It uses task-level event-driven simulation to simulate any valid mapping on arbitrary multi-level hardware, with a hardware-consistent scheduler handling general task-level resource contention on task graphs. Our runtime event-driven design also naturally supports dynamic workloads.

While this paper focus on NN workloads, MLDSE extend to \textbf{any} parallel workload that can be represented as a task graph\cite{lin2023songc}.

\section{Multi-Level Hardware Modeling}
\label{sec:modeling}

\textbf{Hardware Template Description Using Hardware IR.} MLDSE defines "multi-level hardware" as a class of hardware with multiple spatial levels and flexible component organization within each level MLDSE introduces a hardware IR to flexibly model arbitrary multi-level hardware, acting as a superset of specific hardware templates. The basic unit of this IR \textit{SpaceMatrix} and \textit{SpacePoint}. Through recursive and composable rules, various hardware structures can be constructed by these two units.

\begin{figure}[!h]
  \centering
  \includegraphics[scale=0.36, keepaspectratio]{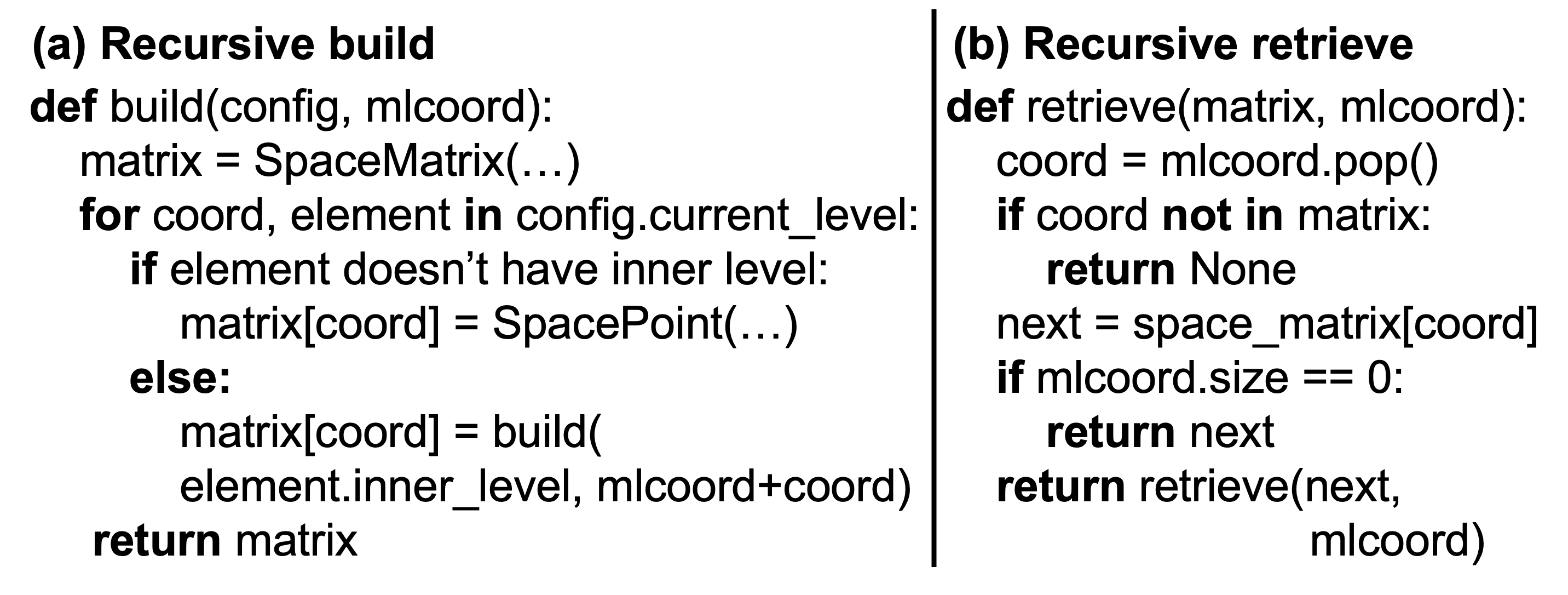}
  \caption{Recursive building and retrieving.}
  \label{fig:hardware_builder}
\end{figure}

\begin{figure*}[!t]
  \centering
  \includegraphics[scale=0.35, keepaspectratio]{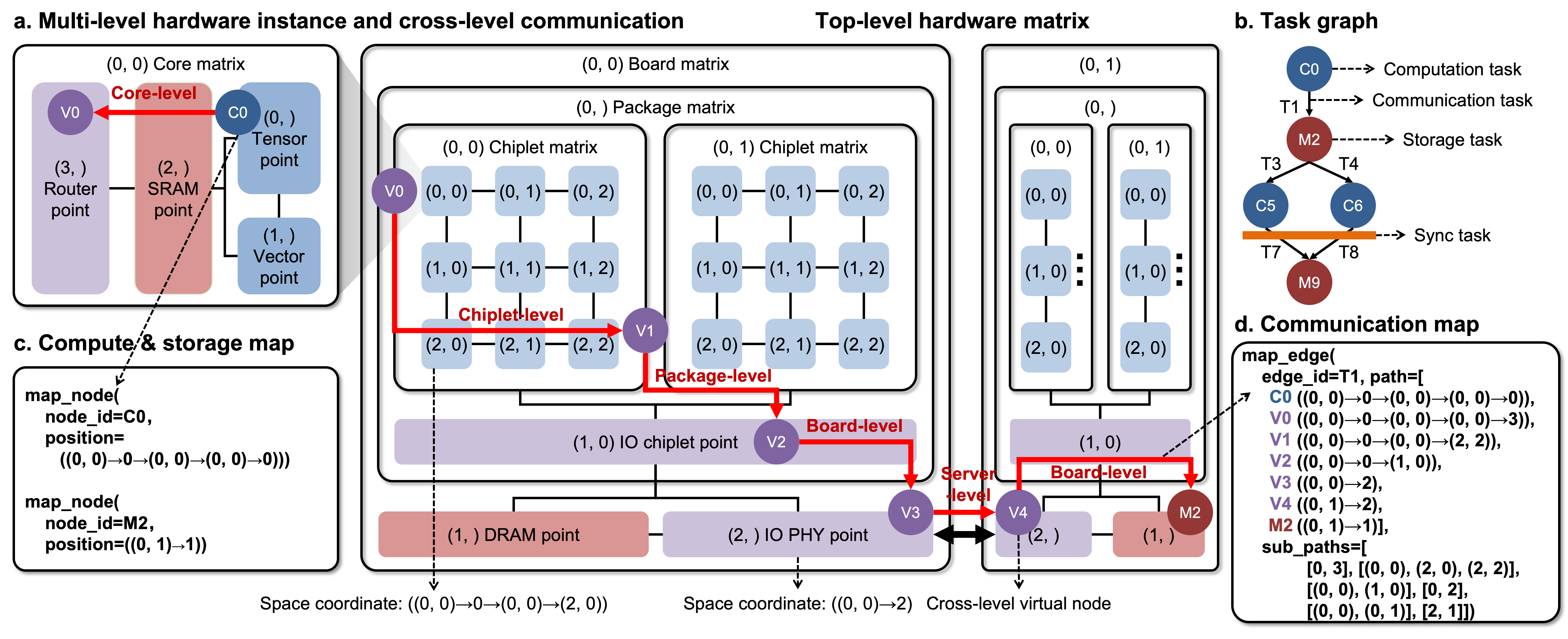}
  \caption{An example for recursive hardware modeling and cross-level communication task mapping.}
  \label{fig:edge_mapping}
\end{figure*}

\textbf{\textit{SpaceMatrix} and \textit{SpacePoint} Data Structures.} The hardware IR represents multi-level hardware as a nested structure, where each level consists of a collection of elements. An element can be either a specific hardware component or a collection of elements from the inner level (Figure \ref{fig:MLS_flow}(a)). We use \textit{SpaceMatrix} to represent a collection of elements and \textit{SpacePoint} for the finest-grained modeled elements, which do not contain other elements. A \textit{SpaceMatrix} is a multidimensional, recursive container; its dimensionality dictates the coordinate dimensionality of its elements. For example, a 2$\times$2 MCM hardware will be represented by a 2D \textit{SpaceMatrix} with dimensions [2, 2]. A 3D \textit{SpaceMatrix} indicates a hardware level with a 3D topology. As shown in Figure \ref{fig:MLS_flow}(c), a \textit{SpaceMatrix} can recursively contain other \textit{SpaceMatrices} or \textit{SpacePoints}. Each \textit{SpaceMatrix} specify its topological pattern (e.g., 2D-mesh, 3D-torus, bus, or tree) with a communication \textit{SpacePoint} (or several for multiple communication domains). The communication \textit{SpacePoint} defines the network's topology, bandwidth, latency and other routing strategies if exit. For instance, in MCM hardware, a dedicated \textit{SpacePoint} should represent the Network-on-Package (NoP), and each chiplet contains a \textit{SpacePoint} representing its Network-on-Chip (NoC). The recursive nature of \textit{SpaceMatrix} supports any depth of spatial levels, and different elements within a \textit{SpaceMatrix} can be flexibly interconnected and organized.

MLDSE's hardware modeling inherently support heterogeneous architectures. For example, as shown in Figure \ref{fig:edge_mapping}, the package at $((0, 0) \to 0)$ comprises three chiplets: two compute chiplets and one IO chiplet, where the two compute chiplets can also have distinct designs. In Section \ref{sec:space_level}, we distributed each Transformer layer across three chiplets (attention, FFN up-projection, FFN down-projection). MLDSE allows to model three distinct compute chiplets tailored for their specific roles, showcasing its capability to support and explore heterogeneous architectures.

\textbf{Hardware Builder.} Using the \textit{SpaceMatrix} and \textit{SpacePoint} data structures, the hardware builder automatically instantiates the hardware IR description into simulatable and operable models. Figure \ref{fig:hardware_builder}(a) illustrates the recursive build process. "Operable" means models include interfaces for accessing/manipulating hardware elements for exploration, mapping, and evaluation. For example, Figure \ref{fig:hardware_builder}(b) shows an hardware element retrieval interface. The builder also creates a multi-level spatial coordinate system to locate each element (\textit{mlcoord} in Figure \ref{fig:hardware_builder}). For instance, in two-level hardware with a 3D topology at the first level and a 2D mesh at the second, the multi-level space coordinate of an element in the second level is $((a, b, c) \to (c, d))$, where $(a, b, c)$ denotes the outer level position and $(c, d)$ the inner. Each \textit{SpacePoint} has a task container for workload mapping and links to an evaluator (e.g., analytical models, functional/cycle-accurate simulator, or RTL). Figure \ref{fig:edge_mapping}(a) provides an example of a specific multi-level hardware with four spatial levels (board $\to$ package $\to$ chiplet $\to$ core).

\section{Spatiotemporal Mapping}
\label{sec:mapping}

\subsection{Spatiotemporal Mapping IR}
\label{sec:mapping_ir}

MLDSE offers a versatile toolset for spatiotemporal mapping on multi-level hardware, covering three common mapping schemes on accelerators: temporal mapping\cite{chen2014diannao, chen2014dadiannao, liao2021ascend}, spatial mapping\cite{cai2024gemini, pei2019towards}, and spatiotemporal mapping\cite{knowles2021graphcore, ma2022neuromorphic}. To support diverse mapping strategies, MLDSE introduces a novel spatiotemporal mapping IR based on task graphs\cite{lin2023songc}, enabling fine-grained task decomposition that comprehensively captures the mapping space across multi-level hardware. Tasks in the task graph are represented at the tensor granularity. As shown in Figure \ref{fig:edge_mapping}(b), computation and storage tasks are depicted as nodes, and communication tasks are edges.

The mapping IR describes how task graphs are allocated to \textit{SpacePoints} within recursive \textit{SpaceMatrices}. Spatially, computation and storage tasks are assigned to \textit{SpacePoints} using multi-level space coordinates (Figure \ref{fig:edge_mapping}(c)). Communication tasks can span multiple spatial levels and space coordinates. The red lines in Figure \ref{fig:edge_mapping}(a) represent a cross-level communication task \textbf{T1}, transmitting data from compute task \textbf{C0} to storage task \textbf{M2}. To map a cross-level communication task, a series of critical coordinates along its path need to be specified. These critical coordinates are entry and exit points at each spatial level, dividing the original task into a sequence of sub-tasks. Each sub-task represents a communication segment that resides entirely within a single spatial level. Notably, for our simulator, each task is mapped to one and only one \textit{SpacePoint}. For storage replicated across memories, the corresponding storage task is also duplicated. For communication tasks, sub-paths are represented as isolated tasks derived from the original task and placed into corresponding communication \textit{SpacePoints}. This setup is crucial for the scheduling algorithm in Section \ref{sec:scheduler}.

\begin{figure}[!b]
  \centering
  \includegraphics[scale=0.36, keepaspectratio]{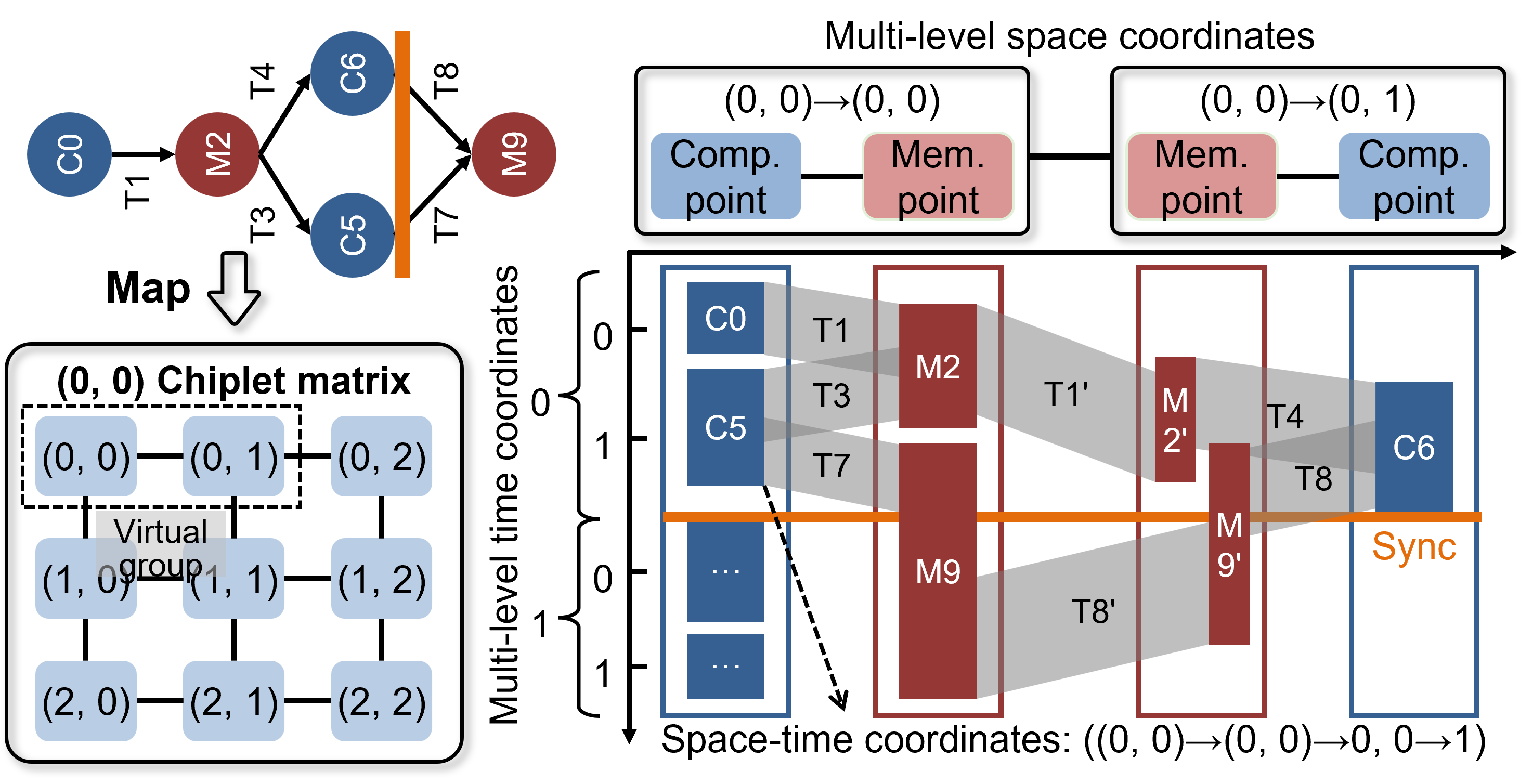}
  \caption{Spatiotemporal mapping IR in MLDSE.}
  \label{fig:st_mapping}
\end{figure}

For temporal mapping, the key is determining the execution orders of tasks (or life cycle for storage tasks), which relies on: 1) Data dependencies from the task graph, which introduces basic temporal relationships among tasks. 2) Resource exclusivity. Some hardware components can only be occupied by one task at a time, requiring scheduling for tasks sharing the same resources. 3) Synchronization mechanisms, typically determined by hardware execution models\cite{ma2022neuromorphic, knowles2021graphcore} and mapping strategies\cite{lin2024nnscaler, lin2024tessel}. In our hardware modeling, each \textit{SpacePoint} contains a task queue for computation and communication tasks and a pool for storage tasks. The first two types of ordering are automatically managed by our simulator and not explicit in the mapping IR. Synchronization tasks can be injected into the mapping IR among \textit{SpacePoints} to enable flexible sync/async mechanisms. Figure \ref{fig:st_mapping} illustrates a mapping where a synchronization task spans multiple \textit{SpacePoints}, ensuring that all tasks after this synchronization barrier cannot proceed until the synchronization is accomplished. However, for hardware with complex sync/async mechanisms, using synchronization tasks can become inefficient. We thus propose multi-level space-time coordinates. As shown in Figure \ref{fig:st_mapping}, virtual synchronization groups should be defined in hardware models through \textit{SpaceMatrix} modeling. These groups can be virtual hardware element collections or actual hardware hierarchies (e.g., chiplet, package). MLDSE's \textit{SpaceMatrix} enables modeling virtual groups not corresponding to physical hierarchies, e.g., in TianjicX chip\cite{ma2022neuromorphic}, users can partition many-core arrays into virtual groups for multi-NN resource isolation, which is unattainable for existing tools\cite{orenes2024muchisim, rashidi2020astra, won2023astra} that have rigid spatial hierarchies. Beyond space coordinates, tasks are assigned multi-level time coordinates. For an $n$-level time coordinate $(t_n, t_{n-1}, ..., t_1)$, a change in level $i$ ($i > 1$) triggers synchronization within the corresponding virtual group. In Figure \ref{fig:st_mapping}, the time coordinate of task \textbf{C5}'s successor shifts from $(0, 1)$ to $(1, 0)$. Since the second level changes, a synchronization occurs within the virtual group containing \textbf{C5}'s \textit{SpacePoint}. This is more efficient than explicit synchronization tasks for complex hardware mechanisms.

\subsection{Mapping Action Primitives}
\label{sec:mapping_primitive}

\begin{figure}[!t]
  \centering
  \captionsetup{name=Table}
  \renewcommand{\thefigure}{\thetable}
  \addtocounter{table}{1}
  \addtocounter{figure}{-1}
  \caption{Four types of mapping primitives.}
  \includegraphics[scale=0.31, keepaspectratio]{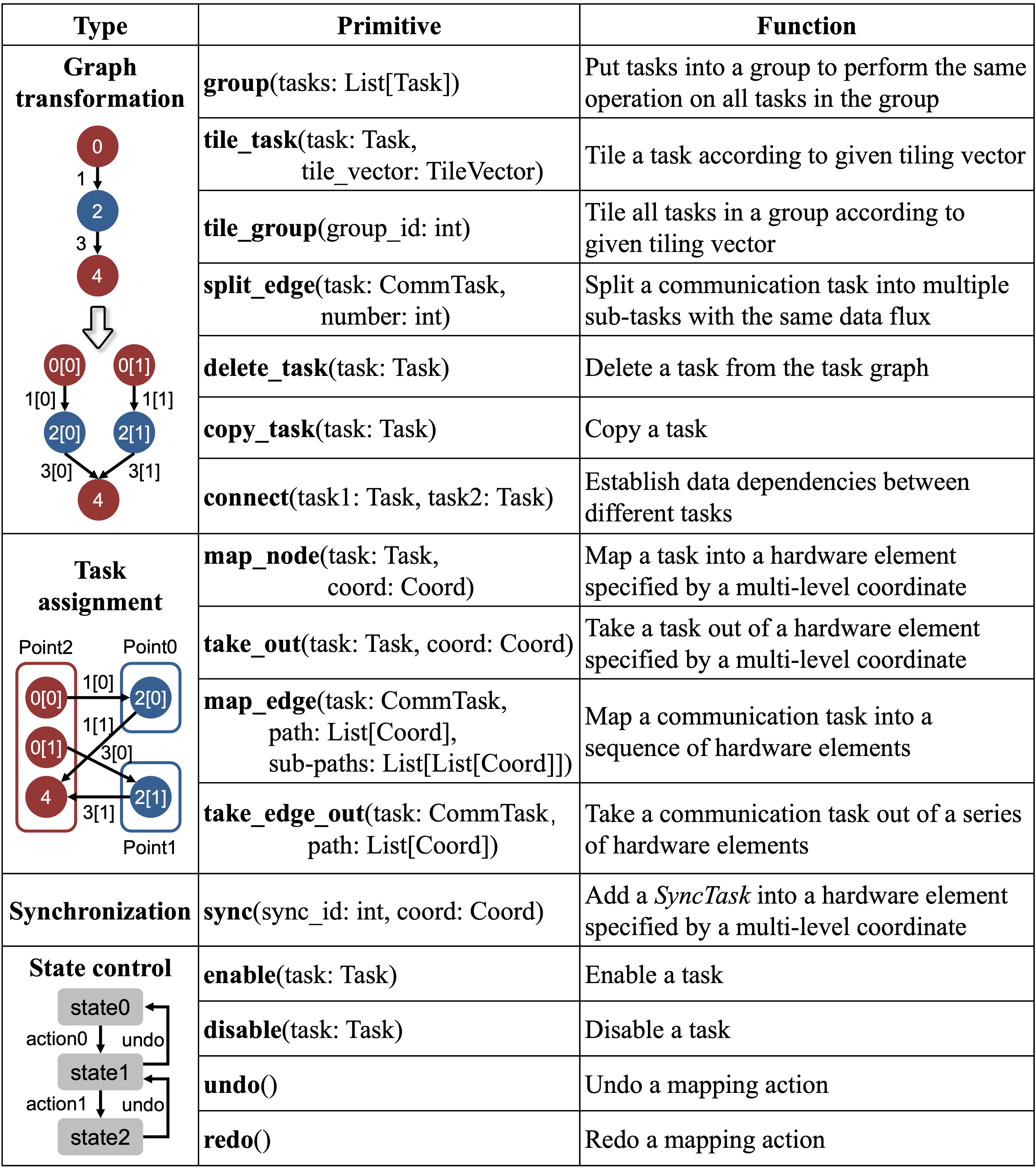}
  \label{fig:mapping_primitive}
\end{figure}

To address mapping optimization complexity\cite{singh2013mapping}, heuristic search algorithms are typically used. Instead of developing architecture-specific search algorithms, MLDSE introduces a versatile set of mapping primitives, enabling flexible construction of mapping search procedures for diverse multi-level hardware. We define four types of mapping primitives: \textbf{graph transformation}, transforms original task graphs into new ones (e.g., tiling); \textbf{task assignment}, allocates/rearranges tasks on \textit{SpacePoints}; \textbf{synchronization}, enforces timing dependencies between tasks; \textbf{state control}, manages the searching procedure (e.g., Monte Carlo tree search). Table \ref{fig:mapping_primitive} details representative primitives. Some primitives are commonly used in NN software toolchains, such as parallelization strategies\cite{zheng2022alpa, lin2024nnscaler} and computation/storage assignment\cite{tan2024cocco, zhang2024llmcompass, shi2023welder}. MLDSE extends these primitives for task graphs on multi-level hardware with multi-level space-time coordinates.

Notable additions are \texttt{map\_edge} and synchronization primitives. The \texttt{map\_edge} primitive facilitates fine-grained communication allocation (Figure \ref{fig:edge_mapping}(a,d)) for cross-level communication. Its arguments are \texttt{edge} (the communication task), \texttt{path} (critical cross-level coordinates decomposing \texttt{edge} into intra-level sub-tasks), and \texttt{sub-paths} (paths for each intra-level sub-task). For each intra-level sub-task created by the decomposition in \texttt{path}, a corresponding path must be provided in \texttt{sub-paths}; $N$ critical cross-level coordinates in \texttt{path} require $N-1$ paths in \texttt{sub-paths}. For instance, in Figure \ref{fig:edge_mapping}, $[(0, 0), (2, 0), (2, 2)]$ (relative coordinates within the current level) represents an intra-level sub-path for communication from \textbf{V0} to \textbf{V1}. The synchronization primitive \texttt{sync(sync\_id, space\_coord)} specifies sync/async mechanisms. To establish synchronization barriers across multiple tasks, \textit{SyncTasks} with the same \texttt{sync\_id} are inserted into task queues or memory pools of \textit{SpacePoints}, forming synchronization relationships. The barrier completes when all associated \textit{SyncTasks} are \texttt{Ready}. Multi-level space-time coordinates (Section \ref{sec:mapping_ir}) can also represent structured sync/async mechanisms derived from hardware execution models.

To leverage MLDSE's mapping primitives for mapping optimization, users need to create a mapping search algorithm that iteratively explores the mapping space by applying sequences of primitives. For each primitive, the algorithm dynamically determines its parameters (optionally using user-provided initial values), e.g., \texttt{tile\_vector} for \texttt{tile\_task} or \texttt{path} and \texttt{sub-paths} for \texttt{map\_edge}. Parameter exploration can employ heuristics, random sampling, or advanced search algorithms (using our state control primitives). After each primitive's parameter selection and application, the MLDSE simulator evaluates the resulting mapping, providing feedback to guide subsequent mapping search iterations. The search algorithm's implementation is user-defined, enabling architecture/workload-aware optimization. Developing specialized search algorithms is beyond this paper's scope and left for future work.

\section{Universal Multi-Level Simulator Generation}
\label{sec:simulator}

\subsection{Task-Level Event-Driven Simulation}
\label{sec:event_driven_simulation}

As a meta-DSE tool, MLDSE is designed to support diverse user-defined multi-level hardware, including hypothetical architectures. This precludes using pre-built, architecture-specific evaluators, especially analytical models that rely on real-world data fitting. Existing evaluators/simulators are often architecture-specific, limiting their applicability to arbitrary multi-level hardware. Modular simulators\cite{herbst2024switchboard, zhi2021methodology, matthews2020mosaicsim}, which can integrate existing tools, are also unsuitable. Even if individual \textit{SpacePoints} could be evaluated using specialized evaluators, their interactions introduce complex dependencies, resource contention, and dynamic synchronization overheads. Aggregating these individual evaluations to obtain an overall performance is \textbf{non-trivial}, as the composition process is highly dependent on specific architectures and workloads. This \textbf{composition} challenge is a core problem MLDSE addresses.

To tackle this, MLDSE leverages the inherent characteristics of NNs, where computation, memory, and communication tasks are at tensor granularity. We propose a tensor-task-level event-driven simulator that performs trace-trigger-schedule-simulation of tasks on-the-fly. This approach balances simulation speed and accuracy while maintaining versatility across diverse hardware. At the core, it incorporates a universal simulation generation framework (Section \ref{sec:event_driven_simulation}) and a formally analyzed task scheduling algorithm (Section \ref{sec:scheduler}) for handling general task-level resource contention. The runtime event-driven design of our simulator also naturally supports dynamic workloads\cite{krause2023fast, delcorro2023skipdecode} via the trace-and-simulate process.

The simulation's fundamental unit is the \textit{task}, and an \textit{event} signifies a task's completion. This completion triggers subsequent tasks through a simulator event termed a \textit{tick}. We represent the mapped task graph as a dependency graph, $G=(V, D)$. $V = \{v_1, v_2, \dots\}$ are computation, storage, and communication tasks. $D = \{d_1, d_2, \dots\}$ are data dependencies between tasks. $P = \{p_1, p_2, \dots\}$ are \textit{SpacePoints} (hardware resources). We define a mapping function $p = M(v)$ to link $G$ with $P$; $M^{-1}(p)$ is the set of tasks on \textit{SpacePoint} $p$.

\begin{figure}[!t]
  \centering
  \includegraphics[scale=0.44, keepaspectratio]{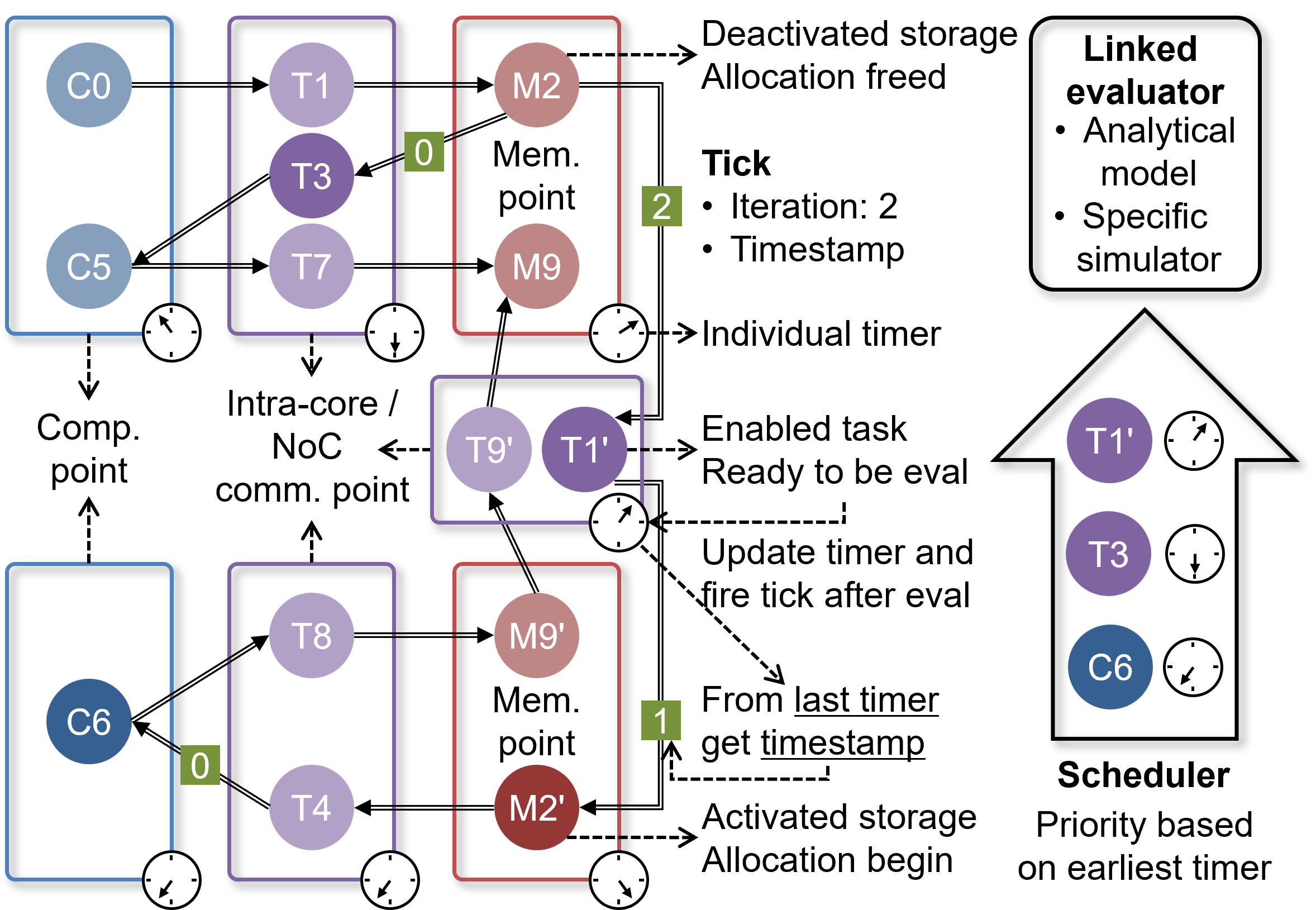}
  \caption{Task-level event-driven simulation.}
  \label{fig:simulation}
\end{figure}

Simulation starts with activating initial tasks through external inputs. A task activates when all its input data dependencies are met, making it ready for evaluation. During simulation, activated tasks can be scheduled to evaluation. One round of a task's evaluation consumes one tick from each of its input edge and fires one tick on each output edge, as shown in Figure \ref{fig:simulation}. A task deactivates if any input edge runs out of ticks. For storage tasks, the activation period determines its life cycle. When a storage task activates, it occupies the corresponding memory space, and each of its output edge always has a tick. A storage task deactivates (freed from memory) once all dependent computation tasks are completed. The generation and consumption of ticks drive overall simulation.

In NN scenarios, batches of input data can be streamed continuously into hardware. Therefore, each tick includes an iteration number to distinguish between batches and a timestamp to track when it fires. Each \textit{SpacePoint} $p$ maintains its own timer that tracks current simulation time $t_{current}$. When an activated task $v\in M^{-1}(p)$ is scheduled, its evaluation start time $\mathrm{Start}(v)$ is the maximum of its input tick timestamps and $t_{current}$. Upon completion, $t_{current}$ is updated to the task $v$’s end time $\mathrm{End}(v)$, which are also the timestamp for fired ticks of $v$. Formally:

\begin{equation}
\left\{\begin{matrix}
\mathrm{Start}(v) = \max\{t_1, t_2, \dots, t_k, t_{current} \} \\
\mathrm{End}(v) = \max\{t_1, t_2, \dots, t_k, t_{current} \} + E_{p}(v)
\end{matrix}\right.
\end{equation}

where $t_1$ to $t_k$ are the timestamps of ticks on $k$ input edges of $v$, and $E_{p}$ is the evaluation model on $p=M(v)$. Since the evaluation of each task in \textit{SpacePoints} is driven by ticks with various timestamps, timers across different \textit{SpacePoints} are at different time during simulation. \textit{SpacePoints} with later timers might still contain deactivated tasks that could be activated by the firing of tasks on \textit{SpacePoints} with earlier timers. These \textit{SpacePoints} should wait for earlier \textit{SpacePoints} with activated tasks to be simulated first to maintain the incremental update of the timer (to avoid latter tasks be evaluated before earlier tasks). Therefore, our simulator includes a scheduler that queues activated tasks, prioritizing those with the earliest timer of the corresponding \textit{SpacePoint} (Figure \ref{fig:simulation}).

Compared to computation and communication tasks, the evaluation of storage tasks focuses on memory occupancy size and duration. If a storage task $v$ receives $k$ ticks with timestamp $t_1, t_2, \dots, t_k$ and is accessed by $n$ tasks $v_1, v_2, \dots, v_n$ at time $t^{out}_{1}, t^{out}_{2}, \dots, t^{out}_{n}$, the survival duration of $v$ is:

\begin{equation}
\left\{\begin{matrix}
\mathrm{Start}(v)=\min_{i \in \{1, 2, \dots, k\}}(t_i)\\
\mathrm{End}(v)=\max_{j \in \{1, 2, \dots, n\}}(t^{out}_{j})
\end{matrix}\right.
\end{equation}

Although MLDSE uses a statically defined task graph, the graph itself can incorporate dynamic tasks, e.g., conditional branches. When evaluating dynamic workloads\cite{krause2023fast, delcorro2023skipdecode}, MLDSE's simulator collaborates with a task graph executor. The executor runs the graph step-by-step, and MLDSE uses its results to identify triggered tasks and simulates those tasks. This can be done in two modes: 1) online mode, the simulator dynamically identifies triggered tasks according to the executor's runtime results and immediately schedules their simulation, and 2) offline mode, the executor records a complete execution trace for subsequent simulation.

\subsection{Hardware-Consistent Scheduler}
\label{sec:scheduler}

\begin{figure}
  \centering
  \includegraphics[scale=0.28, keepaspectratio]{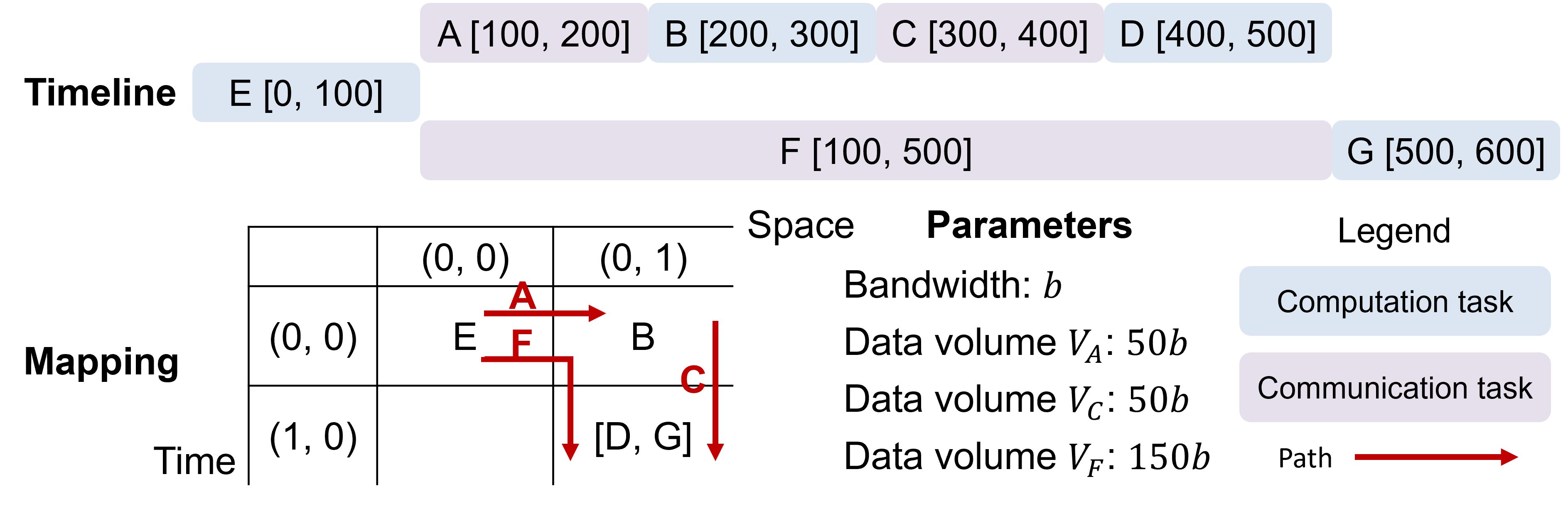}
  \caption{Example of inconsistency between task-level event-driven simulation (without hardware-consistent scheduler) and actual hardware execution due to resource competition.}
  \label{fig:competition}
\end{figure}

In task-level event-driven simulation, different \textit{SpacePoints} maintain \textbf{distributed and asynchronous timers}. While multiple simulation traversal orders on the task graph are valid as long as they do not violate data dependencies, resource competition among tasks can lead to inconsistencies between an dependency-respecting traversal and actual hardware behavior. Figure \ref{fig:competition} illustrates this: a traversal ($[\{E\}, \{A, F\}, \{B, G\}, \{C\}, \{D\}]$) triggers tasks $A$ and $F$ upon $E$'s completion ($t_E = 100$). Since $A$ and $F$'s first hop share a link and start simultaneously, they compete for resources (effective bandwidth for $A$ and $F$ becomes $0.5b$). Task $A$ completes at $t_A = t_E + V_A / 0.5b = 200$, but $F$ only transfers $1/3$ of its data. $F$ then completes at $t_F = t_A + (V_F - V_A) / b + V_F / b = 450$ with full bandwidth. However, task $C$, triggered at $t_B = 300$, actually contends with the already-evaluated task $F$. Since the evaluation of $F$ has already completed, if we schedule tasks in this order, evaluation results will be inconsistent with actual hardware. To address this, we formulate constraints for hardware-consistent simulation and develop a corresponding scheduling algorithm. We use $\le_d$ to denote the partial order on the dependency graph $G$, where $v_i <_d v_j$ means $v_j$ depends on $v_i$. Our simulator ensures the following constraint for any node $v$ in $G$:

\textbf{Constraint 1:} 

\begin{equation}
\mathrm{Start}(v)\ge \max_{w<_d v} \mathrm{End}(w)
\end{equation}

\begin{figure}
  \centering
  \includegraphics[scale=0.27, keepaspectratio]{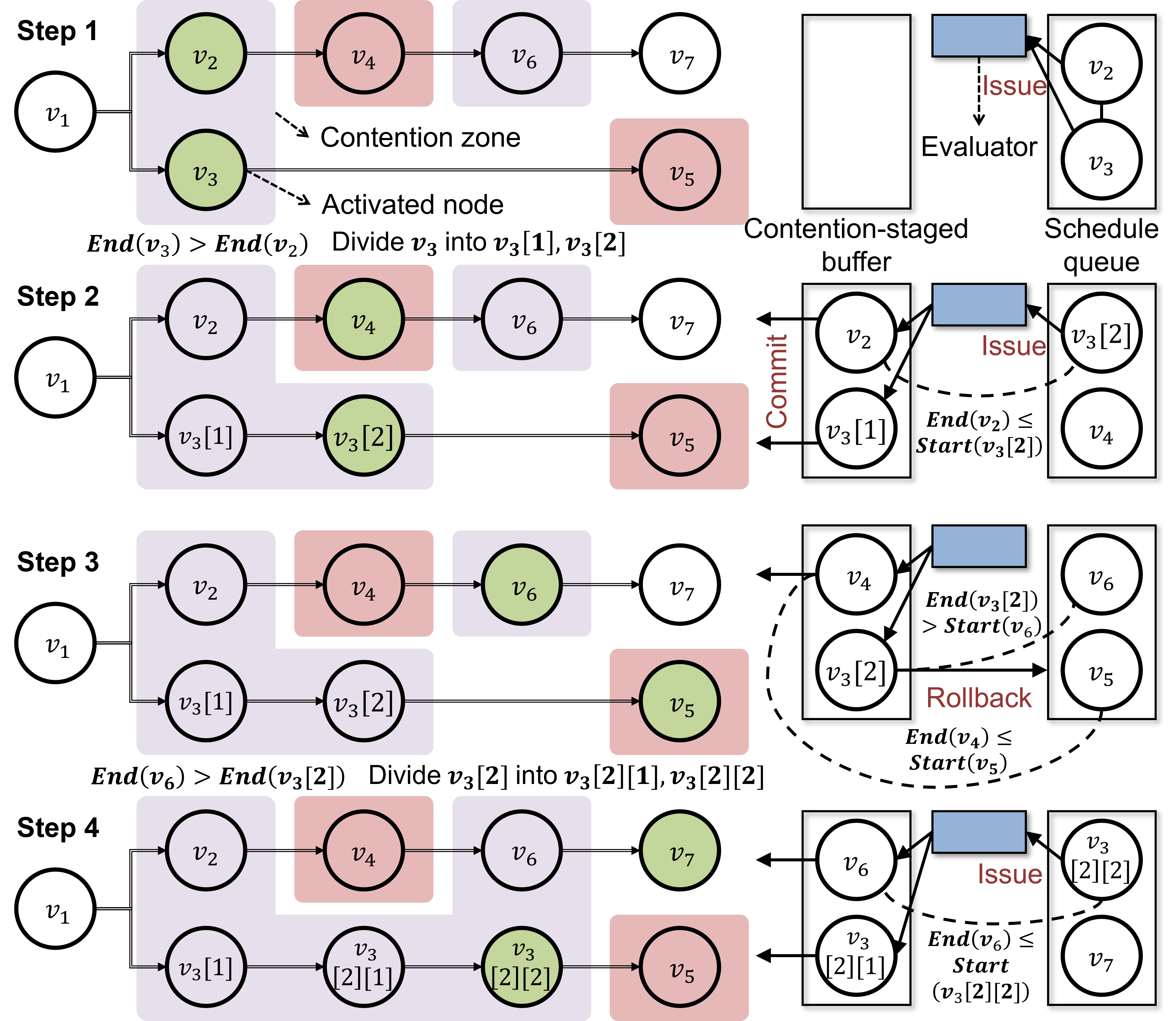}
  \caption{Illustrative example of the scheduling algorithm.}
  \label{fig:schedule}
\end{figure}

When resource contention occurs, multiple independent tasks may share the same hardware component, causing their execution time to overlap. In this case, tasks are no longer atomic evaluation units (e.g., $A$ and $F$ in Figure \ref{fig:comparison}). Assuming $v_2$ and $v_3$ in Figure \ref{fig:schedule} contend for resources, a straightforward solution is to bind $v_2$ and $v_3$ together for simultaneous scheduling and evaluation. The scheduling queue is then modified as $S=\left [ \left \{ v_1 \right \}, \left \{ v_2, v_3 \right \} ,\dots \right ]$. $\left \{ v_2, v_3 \right \}$ indicates $v_2$ and $v_3$ are mapped to the same \textit{SpacePoint} and will be scheduled together for simulation, denoted as $v_2 =_s v_3$. The schedule queue $S$ defines a partial order $\le_s$ to represent the simulation order (e.g., $v_1 <_s v_2$). Thus, solving the consistency problem is transformed into finding an $S$ such that for any $p$ and $v\in M^{-1}(p)$, the following constraints are satisfied:

\textbf{Constraint 2:} 

\begin{equation}
\mathrm{Start}(v)\ge \max_{w <_s v, w \in M^{-1}(p)}\mathrm{End}(w)
\end{equation}

Consider the worst case of constraint 1, constraint 2 turn into:

\begin{equation}
\max_{w <_d v} \mathrm{End}(w)\ge \max_{w <_s v, v \in M^{-1}(p)} \mathrm{End}(w)
\label{eq:condition2}
\end{equation}

Last, only activated tasks can be scheduled and it is natural for:

\textbf{Constraint 3:} 

\begin{equation}
w <_s v, \; \mathrm{if} \, w <_d v
\end{equation}

To satisfy Inequation \ref{eq:condition2}, we only need to consider tasks independent of $v$. We assume $u$ is such a task. Since $E(u)$ is irrelevant to $w<_dv$ and could be very large, $\max_{w <_d v} \mathrm{End}(w)\ge \mathrm{End}(u)$ can't be guaranteed for $u <_s v$ (vice versa for $v <_s u$). This suggests $u =_s v$. However, this could conflict with constraint 3. For example, in Figure \ref{fig:schedule}, $v_2$ and $v_6$ are irrelevant to $v_3$. We need to schedule them as $v_2=_s v_3=_s v_6$ if they are in one contention zone (a set of tasks that potentially share and compete for the same hardware resource). However, $v_6<_d v_2$ implies $v_6<_s v_2$, a contradiction.

We propose Algorithm \ref{alg:schedule} to address this problem. Consider Figure \ref{fig:schedule} as an example. In step 1, $v_2$ and $v_3$ are activated and mapped on the same \textit{SpacePoint}, making them share hardware resources simultaneously (in the same contention zone). Consequently, the current schedule queue is $S=[\{v_2, v_3\}]$, meaning $v_2$ and $v_3$ are issued together for evaluation. We assume $\mathrm{End}(v_3)>\mathrm{End}(v_2)$, meaning $v_3$ continues to occupy the resource after $v_2$ finishes. At the point of $v_2$ finishes, the simulation truncates the remaining portion of $v_3$ into a new task, $v_3[2]$, for future evaluation.

\begin{algorithm}
	\renewcommand{\algorithmicrequire}{\textbf{Input:}}
	\renewcommand{\algorithmicensure}{\textbf{Output:}}
	\caption{Hardware-Consistent Dynamic Task Scheduling}
	\label{alg:schedule}
	\begin{algorithmic}[1]
        \REQUIRE Dependency graph $G(V, D)$
		\STATE Initial empty queue $S$, and Contention-staged buffer $CSB$
		\WHILE{simulation not done}
        \STATE $A = $ Find all new activated tasks in $G$
        \FOR{all contention zone $\{v_1, \dots\}, \dots$ found in $A$}
            \STATE $S.push(\{v_1, \dots\})$
        \ENDFOR
        \STATE $issued_{tasks} = S.pop()$
        \STATE $simulated_{tasks}, truncated_{tasks}=simulate(issued_{tasks})$
        \IF{$truncated_{tasks}$ is not empty}
            \STATE $S.push(truncated_{tasks})$
        \ENDIF
        \STATE $CSB.add(simulated_{tasks})$ 
        \FOR{for $v$ in $CSB$}
            \IF{$can\_be\_committed(v)$}
                \STATE $CSB.commit(v)$
            \ENDIF
            \IF{$should\_be\_rollback(v)$}
                \STATE $CSB.remove(v)$
                \STATE $\{u, \dots\} =$ pop contending tasks related to $v$ from $S$
                \STATE $S.push(\{v, u, \dots\})$
            \ENDIF
        \ENDFOR
		\ENDWHILE
	\end{algorithmic}  
\end{algorithm}

In step 2, the newly activated tasks $v_3[2]$ and $v_4$ are added to $S$. Evaluated tasks $v_2$ and $v_3[1]$ enter the contention-staged buffer (CSB) for potential contention detection that might require reevaluation. Tasks in the CSB can either have their evaluation results committed or be rolled back to the schedule queue for reevaluation with newly activated tasks in the same contention zone. A task $v$ can be committed (the $can\_be\_committed$ condition in Algorithm \ref{alg:schedule}) if the earliest possible start time of all unissued tasks that might contend with $v$ is later than $\mathrm{End}(v)$. In step 2, the specific commit condition for $v_2$ is $\mathrm{End}(v_2) \leq \mathrm{Start}(v_3[2])$, and for $v_3[1]$, it is $\mathrm{End}(v_3[1]) = \mathrm{End}(v_2) \leq \mathrm{Start}(v_6)$. Both conditions are satisfied, so $v_2$ and $v_3[1]$ are committed.

In step 3, $v_4$ and $v_3[2]$ are issued successively. We assume $v_4$ and $v_5$ form another contention zone, and the evaluated $\mathrm{End}(v_3[2])$ is later than that of $v_4$. $v_4$ satisfies the commit condition after evaluation since $\mathrm{End}(v_4) \leq \mathrm{Start}(v_5)$. However, the shorter execution time of $v_4$ causes $v_6$ (in the same contention zone as $v_3[2]$) to activate before $\mathrm{End}(v_3[2])$. This means part of $v_3[2]$ needs to be evaluated together with $v_6$. Therefore, $v_3[2]$ retreats its evaluation result and rolls back to the schedule queue to be rescheduled with $v_6$. The rollback condition for a task $v$ ($should\_be\_rollback$ in Algorithm \ref{alg:schedule}) is when the start time of any potentially contending tasks in the schedule queue is earlier than the current evaluation of $\mathrm{End}(v)$.

In step 4, we assume $v_3[2]$ still takes longer than $v_6$. Consequently, $v_3[2]$ is further divided into $v_3[2][1]$ and $v_3[2][2]$. $v_6$ and $v_3[2][1]$ meet the commit conditions after evaluation, and no tasks need to roll back, allowing the simulation to proceed. Algorithm \ref{alg:schedule} ensures no unevaluated tasks share resources and potentially overlap with any committed task, satisfying constraints 1-3.

\section{Experiments}
\label{sec:exp}

\subsection{Experimental Setup}
\label{sec:setup}

Existing DSE tools showcase their capabilities on specific architectures, like single-core NN accelerators\cite{tan2024cocco, mei2021zigzag, juracy2021high, ahmad2020superslash, murali20243dnn, he2022design, krishnan2021siam, samajdar2018scale, kao2022digamma, jeong2021union, yang2020co, wang2023nicepim, samajdar2023airchitect, kwon2020maestro}, many-core accelerators \cite{fang2024palm, jia2021tensorlib, qi2023moela}, multi-chiplet accelerators\cite{tan2021nn, cai2024gemini, iff2023rapidchiplet, pal2020design, zhang2023indm, kim2019architecture}, wafer-scale chips\cite{zhu2024theseus, wang2024tmac}, and multi-device systems (GPUs, TPUs, NPUs)\cite{zhang2024llmcompass, rashidi2020astra, won2023astra, agrawal2023deap}, co-optimizing mapping and hardware parameters for optimal designs. To highlight MLDSE's unique capabilities in three-tier DSE (architecture-hardware parameter-mapping), we conducted a novel two-stage DSE: \textbf{cross-architecture DSE} and \textbf{spatial-level DSE}. Our DSE experiments (Section \ref{sec:cross_arch} \& \ref{sec:space_level}) used GPT3-6.7B\cite{brown2020language} (hidden dimension 4096) as the workload. Section \ref{sec:cross_arch} focuses on prefill (single layer, batch = 1, sequence length = 2048), while Section \ref{sec:space_level} examines decoding (generate the 2048th token, 8 layers). Accuracy evaluation (Section \ref{sec:accuracy}) used Llama2-70B\cite{touvron2023llama2}, Llama3-70B\cite{grattafiori2024llama3}, and Qwen-72B\cite{bai2023qwen} for workload diversity. The main differences between GPT-3 and these models are normalization, position encoding, and activation function. Section \ref{sec:accuracy} shows these have minimal performance impact. Attention, matmul, MLP, and communication collectives remain key performance drivers. Thus, we believe Section \ref{sec:cross_arch} \& \ref{sec:space_level}'s conclusions extend to other LLMs.

\textbf{Cross-architecture DSE} evaluated different architectures on a unified platform. We compared GPU-like shared memory (GSM) architecture\footnote{In this paper, "shared memory" refers to the L2 cache in GPUs or the global buffer in TPUs and "local memory" to both L1 cache and shared memory in real GPUs.} (Figure \ref{fig:exp}(a)) with distributed many-core (DMC) architecture (Figure \ref{fig:exp}(b)). For each architecture, we explored two parameter sets: 1) Compute-memory resources: local memory capacity, vector unit/systolic array size, and shared memory capacity (GSM), examining trade-offs between computation and memory resources under a fixed area budget. 2) Memory access and communication: local memory latency, local memory/NoC/shared memory bandwidth, investigating how memory access and communication parameters affect performance and comparing GSM and DMC memory access and communication patterns. \textbf{Spatial-level DSE} aims to study the impact of spatial hierarchy variations on performance. Starting with a DMC-based multi-package board, we incorporated chiplet technology to create new spatial levels (Figure \ref{fig:exp}(c)), analyzing how spatial hierarchies affect performance/cost trade-offs.

We used existing tools for latency, area, and cost analyses of \textit{SpacePoints}. Area evaluation used LLMCompass and CACTI\cite{balasubramonian2017cacti, zhang2024llmcompass}. Control logic and on-chip interconnect overhead was estimated using real hardware data\cite{zhang2024llmcompass, ma2022neuromorphic}. Chiplet cost evaluation used Chiplet Actuary\cite{feng2022chiplet}. GPU parameters were sourced from existing literature and NVIDIA's whitepapers\cite{nvidia2022hopper, luo2024benchmarking, choquette2021nvidia, choquette2023nvidia, nvidia2020ampere, nvidia2017volta, nvidia2018turing}. 

\begin{figure}[!t]
  \centering
  \includegraphics[scale=0.265, keepaspectratio]{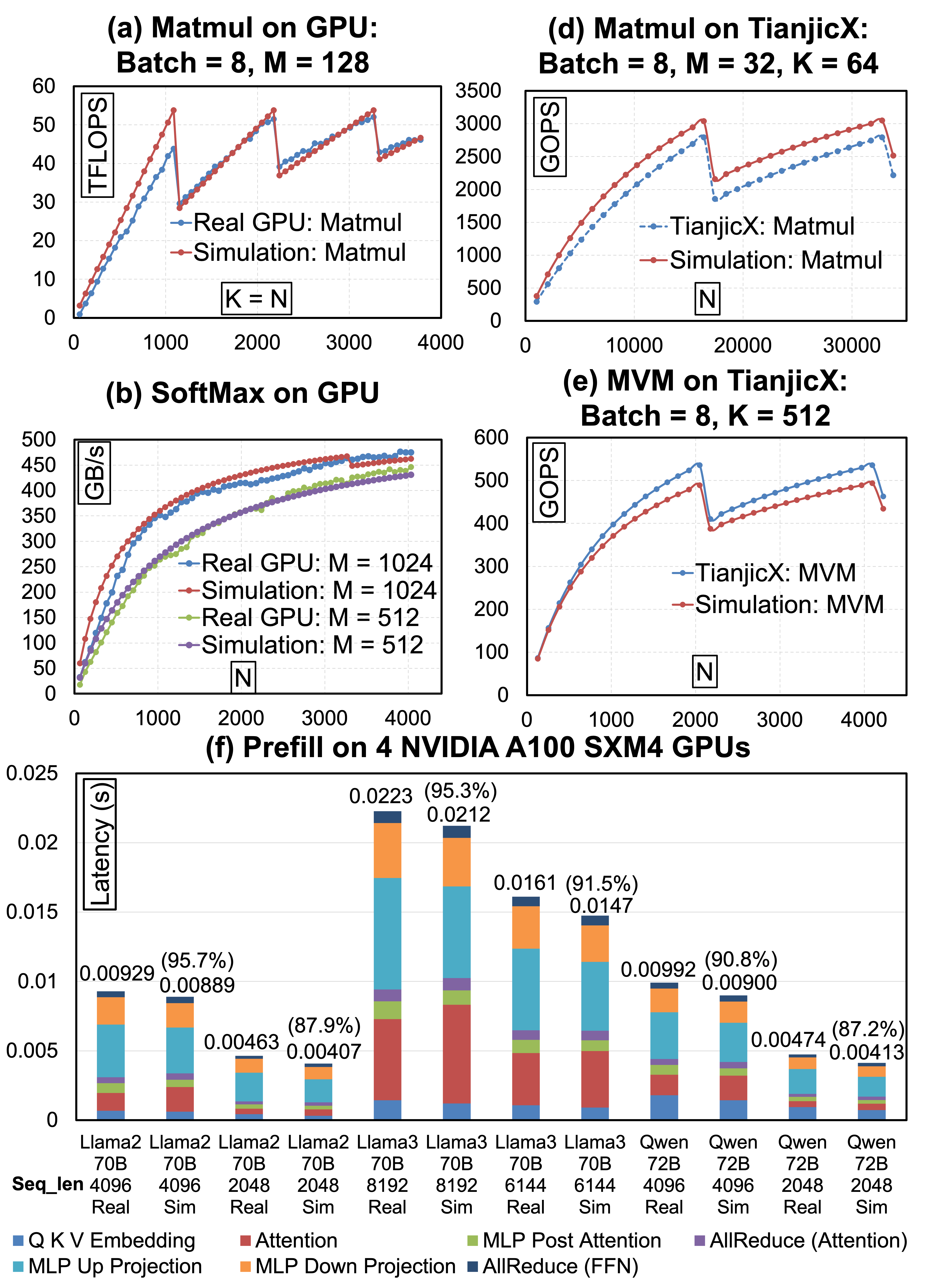}
  \caption{Kernel-level simulation accuracy of MLDSE on a GSM instance (2080Ti\cite{nvidia2018turing}) and a DMC instance (TianjicX\cite{ma2022neuromorphic}) and LLM simulation accuracy on a 4$\times$A100 system}.
  \label{fig:comparison}
\end{figure}

\subsection{Evaluation Accuracy and Speed}
\label{sec:accuracy}

MLDSE is a meta-DSE tool, requiring architecture-specific instantiation by modeling architectures and linking evaluators to \textit{SpacePoints}. The evaluators can be of any type, e.g., analytical models, functional/cycle-accurate simulators. Overall simulation accuracy depends on chosen evaluators (beyond the main scope of this work).

Figure \ref{fig:comparison}(a-f) shows kernel-level simulation accuracy for three LLM operators (Matmul, SoftMax, matrix-vector multiplication (MVM)) across GSM and DMC architectures\footnote{TianjicX is a neuromorphic chip that aligns with the DMC model. Subsequent evaluations use parameters resembling a Graphcore IPU, without directly modeling it.}. Using a roofline model\cite{williams2009roofline} with mapping, MLDSE can capture non-linear performance variations and align key transition points with real hardware measurements, with occasional discrepancies ($\sim$20\% error) at specific points. For LLM evaluation, we used LLMCompass's GPU simulator for device evaluation and a latency-bandwidth model\cite{nvidia2023nccl, bang2024vtrain} for communication analysis, e.g., the All-Reduce collective uses: 
\begin{equation*}
T = \underset{\text{Bidirectional ring reduce}}{(n-1)L + \frac{(n-1)S}{nB}} + \underset{\text{Fully-connected all-gather}}{L + \frac{2S}{B}}
\end{equation*}
where \(n\) = GPU count, \(L\) = link latency, \(S\) = data size, and \(B\) = bandwidth. This formula achieves <3\% error on a 4$\times$A100 system with full NVLink connectivity for Llama3-70B. Figure \ref{fig:comparison}(g) omits normalization, SiLU, and RoPE latencies (not supported in LLMCompass, contribute minimally to total latency, 5\%-6\%). MLDSE achieved 87\%--96\% accuracy for prefill (single layer) of Llama2-70B, Llama3-70B, and Qwen-72B across varied sequence lengths. This demonstrates MLDSE's ability to improve precision through strategic evaluator selection for different \textit{SpacePoints}.

\begin{table}[]
\centering
\caption{Different hardware configurations for computing and memory resources on GSM and DMC architectures.}
\label{tab:compute-memory}
\resizebox{\columnwidth}{!}{%
\begin{tabular}{|c|cccccccc|}
\hline
\multirow{2}{*}{Num.} & \multicolumn{8}{c|}{\textbf{Distributed Many-Core Architecture (DMC)}} \\ \cline{2-9} 
 & \multicolumn{3}{c|}{Local Memory} & \multicolumn{1}{c|}{\begin{tabular}[c]{@{}c@{}}Systolic\\ Array\end{tabular}} & \multicolumn{1}{c|}{\begin{tabular}[c]{@{}c@{}}Vector\\ Unit\end{tabular}} & \multicolumn{1}{c|}{\begin{tabular}[c]{@{}c@{}}Control\\ Logic Area\end{tabular}} & \multicolumn{1}{c|}{\begin{tabular}[c]{@{}c@{}}On-chip\\ Interconnect Area\end{tabular}} & \begin{tabular}[c]{@{}c@{}}Total\\ Area\end{tabular} \\ \hline
1 & \multicolumn{3}{c|}{1MB} & \multicolumn{1}{c|}{128×128} & \multicolumn{1}{c|}{512} & \multicolumn{1}{c|}{8.7} & \multicolumn{1}{c|}{43.7} & 926 \\ \hline
2 & \multicolumn{3}{c|}{2MB} & \multicolumn{1}{c|}{64×64} & \multicolumn{1}{c|}{512} & \multicolumn{1}{c|}{7.6} & \multicolumn{1}{c|}{38.1} & 808 \\ \hline
3 & \multicolumn{3}{c|}{2.5MB} & \multicolumn{1}{c|}{32×32} & \multicolumn{1}{c|}{128} & \multicolumn{1}{c|}{8.0} & \multicolumn{1}{c|}{39.9} & 845 \\ \hline
4 & \multicolumn{3}{c|}{3MB} & \multicolumn{1}{c|}{16×16} & \multicolumn{1}{c|}{128} & \multicolumn{1}{c|}{9.2} & \multicolumn{1}{c|}{46.1} & 978 \\ \hline
\multirow{2}{*}{Num.} & \multicolumn{8}{c|}{\textbf{GPU-Like Shared Memory Architecture (GSM)}} \\ \cline{2-9} 
 & \multicolumn{1}{c|}{\begin{tabular}[c]{@{}c@{}}L2\\ Cache\end{tabular}} & \multicolumn{1}{c|}{\begin{tabular}[c]{@{}c@{}}L1\\ Cache\end{tabular}} & \multicolumn{1}{c|}{\begin{tabular}[c]{@{}c@{}}Register\\ File\end{tabular}} & \multicolumn{1}{c|}{\begin{tabular}[c]{@{}c@{}}Systolic\\ Array\end{tabular}} & \multicolumn{1}{c|}{\begin{tabular}[c]{@{}c@{}}Vector\\ Unit\end{tabular}} & \multicolumn{1}{c|}{\begin{tabular}[c]{@{}c@{}}Control\\ Logic Area\end{tabular}} & \multicolumn{1}{c|}{\begin{tabular}[c]{@{}c@{}}On-chip \\ Interconnect Area\end{tabular}} & \begin{tabular}[c]{@{}c@{}}Total\\ Area\end{tabular} \\ \hline
1 & \multicolumn{1}{c|}{256MB} & \multicolumn{1}{c|}{128KB} & \multicolumn{1}{c|}{64KB} & \multicolumn{1}{c|}{16×16} & \multicolumn{1}{c|}{128} & \multicolumn{1}{c|}{134.7} & \multicolumn{1}{c|}{67.7} & 915 \\ \hline
2 & \multicolumn{1}{c|}{192MB} & \multicolumn{1}{c|}{256KB} & \multicolumn{1}{c|}{64KB} & \multicolumn{1}{c|}{32×32} & \multicolumn{1}{c|}{512} & \multicolumn{1}{c|}{119.5} & \multicolumn{1}{c|}{60.1} & 826 \\ \hline
3 & \multicolumn{1}{c|}{128MB} & \multicolumn{1}{c|}{512KB} & \multicolumn{1}{c|}{64KB} & \multicolumn{1}{c|}{64×64} & \multicolumn{1}{c|}{256} & \multicolumn{1}{c|}{123.2} & \multicolumn{1}{c|}{61.9} & 851 \\ \hline
4 & \multicolumn{1}{c|}{32MB} & \multicolumn{1}{c|}{128KB} & \multicolumn{1}{c|}{64KB} & \multicolumn{1}{c|}{128×128} & \multicolumn{1}{c|}{128} & \multicolumn{1}{c|}{132.5} & \multicolumn{1}{c|}{66.6} & 930 \\ \hline
\end{tabular}%
}
\end{table}

For NN workloads, tiling often produce identical task tiles. To enhance simulation efficiency, MLDSE can simulate only representative tasks, duplicating results for redundant ones\cite{fang2024palm}. We simulated 240 hardware configurations in 76 seconds, significantly outperforming existing packet-level or cycle-accurate simulations.

\subsection{Cross-Architecture DSE: From GPU-Like Shared Memory to Distributed Many-Core}
\label{sec:cross_arch}

This section presents DSE results of the DMC architecture (with limited GSM results for brevity) and cross-architecture DSE insights between GSM and DMC architectures.

\subsubsection{Computation-Memory Ratio}
\label{sec:compute-memory}

\begin{figure*}
  \centering
  \includegraphics[scale=0.27, keepaspectratio]{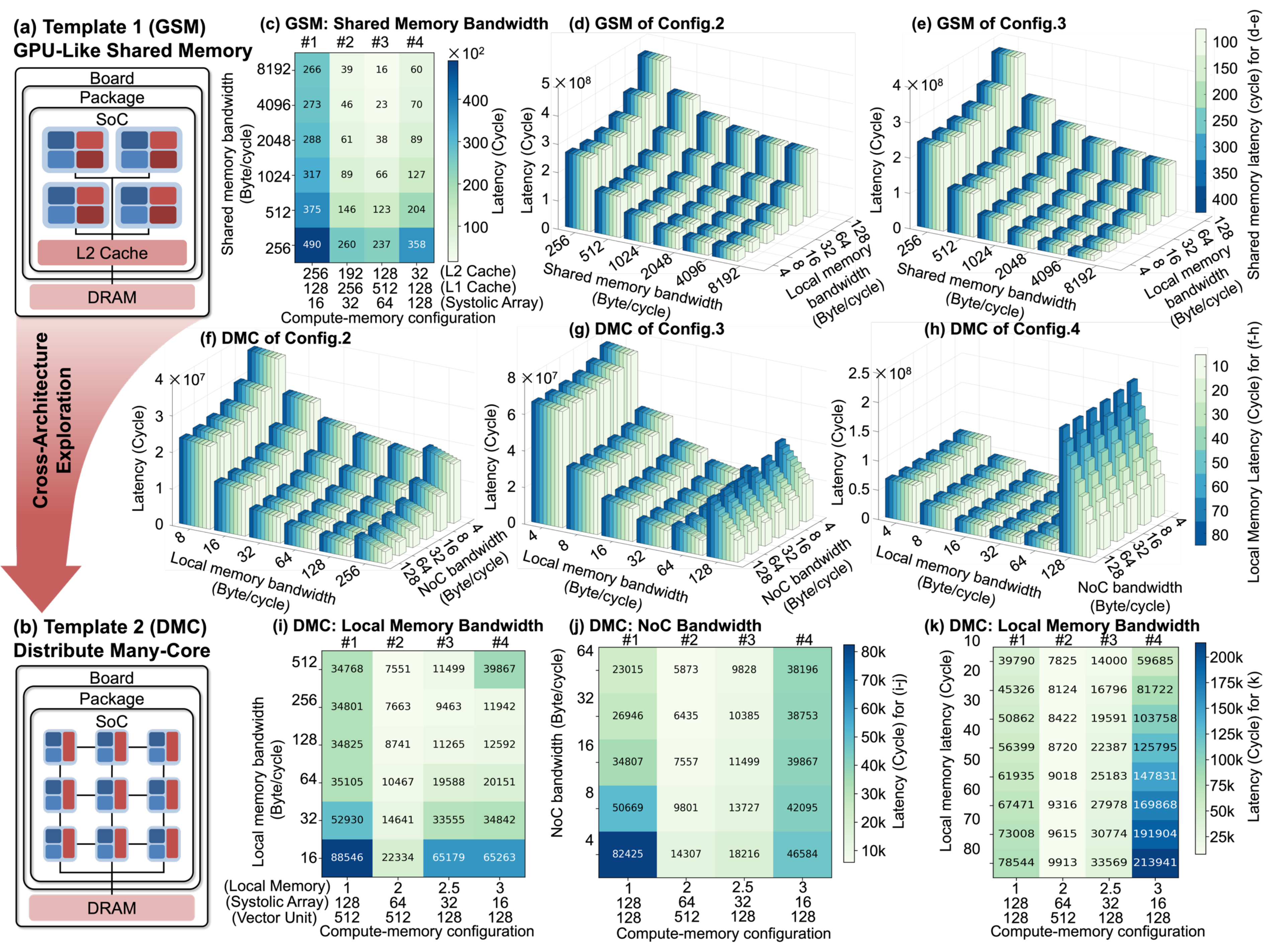}
  \caption{Cross-architecture DSE via MLDSE: GSM vs. DMC. (a-b) Illustration of GSM and DMC architectures. (c) Impact of shared memory bandwidth on GSM performance under 4 compute-memory configurations (Table \ref{tab:compute-memory}). (d-e) Impact of shared/local memory bandwidth and shared memory latency on GSM performance under config. 2 \& 3 in Table \ref{tab:compute-memory}. (f-h) Impact of local memory/NoC bandwidth and local memory latency on DMC performance under config. 2, 3, 4 in Table \ref{tab:compute-memory}. (i-k) Impact of local memory/NoC bandwidth and local memory latency on DMC performance under 4 compute-memory configurations (Table \ref{tab:compute-memory}).}
  \label{fig:cross_arch}
\end{figure*} 

We first examined how the computation-to-memory resource ratio affects prefill stage performance under a fixed total area budget ($\sim 858mm^2$). Table \ref{tab:compute-memory} lists hardware parameters for GSM and DMC, with some configurations exceeding $858mm^2$ for large on-chip memory or systolic arrays. Figure \ref{fig:cross_arch}(c, i-k) show that configurations 2 and 3 outperform configurations 1 and 4 for both GSM and DMC architectures by balancing computation and memory resources. For GSM (Figure \ref{fig:cross_arch}(c)), configuration 4's undersized shared memory tends to increase DRAM accesses, leading to performance degradation. Conversely, configuration 1 overallocates on-chip memory, limiting compute array area and degrading performance during compute-bound prefill stages\cite{zhang2024llmcompass}. For DMC (Figure \ref{fig:cross_arch}(i-k)), excessive local memory (3MB/core in configuration 4) limits compute area, resulting a reduction of throughput despite minimizing off-chip data movement. Conversely, oversized systolic arrays incur frequent DRAM accesses due to insufficient local memory, also leading to performance decrease. Notably, setting local memory to 2.5MB/core (320MB total) would achieve similar memory capacity similar to Graphcore IPU\cite{knowles2021graphcore} (304MB). Our results suggest reducing local memory to 2MB/core (256MB total) and increasing systolic array size (from 32×32 to 64×64) most likely achieves better prefill performance.

\subsubsection{Memory Access and Communication}

We evaluated how memory access and communication parameters affect performance of GSM and DMC, modeling memory bandwidth's effect on memory area. For given memory capacity, increased memory bandwidth increases memory area, resulting a reduction of available systolic array area within a fixed area budget. Figure \ref{fig:cross_arch}(d,e) shows how shared/local memory bandwidth and shared memory latency affect GSM performance under configurations 2 and 3. The impact of these parameters ranks as shared memory bandwidth > local memory bandwidth > shared memory latency. Compared to DMC, GSM’s small local memory could lead to frequent shared memory accesses, making shared memory bandwidth a dominant performance factor. For DMC (Figure \ref{fig:cross_arch}(f-h), configurations 2-4), local memory bandwidth has the most significant performance impact, followed by NoC bandwidth and local memory latency. We find: 1) Due to prefill stage's high memory access and computation demands, the performance impact of memory access and communication latencies tend to be minor. 2) In Figure \ref{fig:exp}(f-h), as local memory capacity increases, the performance dependency on NoC bandwidth would diminish, because larger local memory capacities can decrease inter-core data transfers. 3) Local memory bandwidth could directly influence computing throughput of systolic arrays, making it a critical performance factor. Interestingly, performance exhibits \textbf{non-linear} changes as local memory bandwidth increases due to area trade-offs: higher local memory bandwidth would reduce systolic array size to meet area constraints, creating an optimal performance point where local memory bandwidth and systolic array size are balanced. Larger local memory tends to amplify bandwidth-induced performance fluctuations. With a larger local memory, high bandwidth would impose greater area costs, causing significant systolic array size and overall performance variations.

\subsubsection{Insights: GPU-like Shared Memory vs Distributed Many-Core}

\begin{figure*}[!t]
  \centering
  \includegraphics[scale=0.27, keepaspectratio]{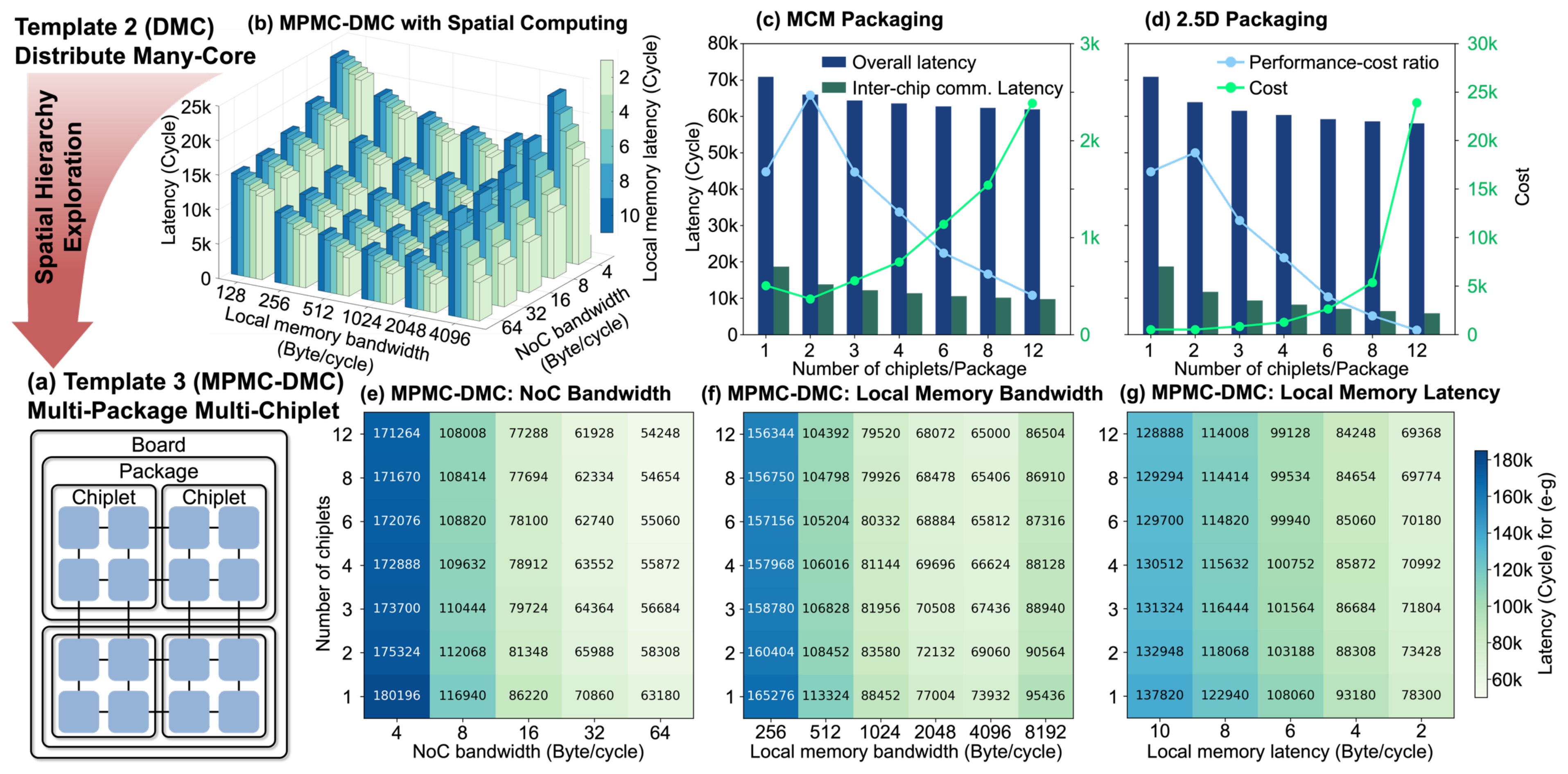}
  \caption{Spatial-level DSE: from DMC to MPMC-DMC. (a) Illustration of multi-package multi-chiplet DMC architecture (MPMC-DMC). (b) Impact of local memory/NoC bandwidth and local memory latency on MPMC-DMC performance when utilizing spatial computing. (d) Performance and cost variations of MPMC-DMC w.r.t. the number of chiplets/package under MCM and 2.5D packaging. (e-g) Impact of NoC/local memory bandwidth and local memory latency on MPMC-DMC performance across different numbers of chiplets/package.}
  \label{fig:space_level}
\end{figure*} 

With MLDSE, we compared GSM and DMC on a unified platform, revealing key insights: (1) DMC and GSM exhibit distinct memory access and communication patterns. GSM depends more on shared memory bandwidth than local memory bandwidth, while DMC relies more on local memory bandwidth than NoC bandwidth. (2) Both architectures require balanced compute-to-memory resource allocation to avoid under-utilization, making architecture-specific DSE essential for optimal performance. (3) DMC outperforms GSM under the same area budget due to: 1) On-chip memory: GSM's total on-chip memory capacity is smaller than DMC (Table \ref{tab:compute-memory}). This is because GSM allocates significant area to register files, which have better latency but lower bandwidth than DMC's local memory (64B/cycle for A100\cite{zhang2024llmcompass} vs. 152B/cycle\footnote{IPU has 1216 tiles and each tile can access 16B/cycle local memory. Here, we consider a equivalent hardware with only 128 tiles.} for Graphcore IPU\cite{knowles2021graphcore}). For workloads like LLMs with large, contiguous data access, the register file's low area efficiency leads to more frequent DRAM accesses in GSM. 2) Effective bandwidth utilization: While GSM has substantial shared memory bandwidth (e.g., 5120B/cycle for A100), DMC’s aggregate local memory bandwidth (e.g., 19456B/cycle for IPU) far exceeds it. Given the high spatial locality of LLM workloads, DMC can efficiently exploit the bandwidth of its individual tiles, thereby enhancing overall performance.

\subsection{Spatial-Level DSE: From Multi-Package to Multi-Package Multi-Chiplet}
\label{sec:space_level}

For the decode stage, we first evaluated the DMC architecture under temporal mapping, where inference of the 2048th token required 614272 cycles, with DRAM access being the primary bottleneck due to frequent weights and key-value (KV) cache transfers from DRAM to on-chip memory, significantly reducing computation utilization. To mitigate this, we employed spatial computing to minimize offchip data transfers. This approach used 24 DMC accelerators (128 cores and 128MB on-chip memory each), each housed in a separate package and interconnected via a board-level network (spatial hierarchy: PCB board $\to$ chip $\to$ core). The whole multi-package board was used to process 8 GPT3-6.7B layers. Each layer's attention and FFNs were mapped across three chips, keeping weights and KV cache on-chip. While spatial pipelining across 24 packages will reduce DRAM access, board-level communication introduces additional performance overhead. We further optimized this by introducing an \textbf{additional spatial level} using chiplet technology, converting DMC accelerators into chiplets integrated into a single package via advanced packaging (Figure \ref{fig:space_level}(a), spatial hierarchy: board $\to$ package $\to$ chiplet $\to$ core). Figure \ref{fig:space_level}(c,d) show results for MCM and 2.5D packaging. Increasing the number of chiplets per package reduces slower board-level communication by replacing it with NoP communication, which offers lower latency and higher bandwidth. However, additional chiplets increase costs, requiring a balance between performance and cost. Results showed that integrating two chiplets per package achieved the optimal cost-performance ratio, consistent with prior studies\cite{tan2021nn}. By introducing this new spatial level, board-level interconnects were complemented by local NoP interconnects, improving the performance-cost trade-off.

\begin{table*}[!t]
\centering
\caption{Comparison between different types of DSE tools}
\label{tab:comparison}
\resizebox{\textwidth}{!}{%
\begin{tabular}{|c|ccc|ccc|cccc|}
\hline
\textbf{Comparison Item} & \multicolumn{3}{c|}{\textbf{Modeling}} & \multicolumn{3}{c|}{\textbf{Mapping}} & \multicolumn{4}{c|}{\textbf{Evaluation}} \\ \hline
\textbf{\begin{tabular}[c]{@{}c@{}}Different\\ Types of Tools\end{tabular}} & \multicolumn{1}{c|}{\textbf{Param.}} & \multicolumn{1}{c|}{\textbf{\begin{tabular}[c]{@{}c@{}}Flexible\\ Organization\end{tabular}}} & \textbf{\begin{tabular}[c]{@{}c@{}}Flexible Spatial\\ Level\end{tabular}} & \multicolumn{1}{c|}{\textbf{\begin{tabular}[c]{@{}c@{}}Spatio-\\ temporal\end{tabular}}} & \multicolumn{1}{c|}{\textbf{\begin{tabular}[c]{@{}c@{}}Sync/\\ Async\end{tabular}}} & \textbf{Communication} & \multicolumn{1}{c|}{\textbf{Type}} & \multicolumn{1}{c|}{\textbf{\begin{tabular}[c]{@{}c@{}}Flexibility\\ (Hardware Scope)\end{tabular}}} & \multicolumn{1}{c|}{\textbf{\begin{tabular}[c]{@{}c@{}}Contention-\\ aware\end{tabular}}} & \textbf{\begin{tabular}[c]{@{}c@{}}Hardware-\\ consistent\end{tabular}} \\ \hline
MLDSE & \multicolumn{1}{c|}{\ding{52}} & \multicolumn{1}{c|}{\ding{52}} & \ding{52} & \multicolumn{1}{c|}{\ding{52}} & \multicolumn{1}{c|}{\ding{52}} & \ding{72}\ding{72} Cross-level & \multicolumn{1}{c|}{Hybrid} & \multicolumn{1}{c|}{\begin{tabular}[c]{@{}c@{}}Diverse multi-level\\ hardware\end{tabular}} & \multicolumn{1}{c|}{\ding{52}} & \begin{tabular}[c]{@{}c@{}}\ding{52}\\ Task-level\end{tabular} \\ \hline
\begin{tabular}[c]{@{}c@{}}Timeloop\cite{parashar2019timeloop}+\\ Accelergy\cite{wu2019accelergy}\end{tabular} & \multicolumn{1}{c|}{\ding{52}} & \multicolumn{1}{c|}{\begin{tabular}[c]{@{}c@{}}\ding{52}\rotatebox[origin=c]{-9.2}{\kern-0.7em\ding{55}} Memory \\ hierarchy\end{tabular}} & \begin{tabular}[c]{@{}c@{}}\ding{55} Lack support\\ for multi-level\\ networks\end{tabular} & \multicolumn{1}{c|}{\ding{55} Only temporal} & \multicolumn{1}{c|}{\ding{55}} & \begin{tabular}[c]{@{}c@{}}\ding{55} Single \\ accelerator\end{tabular} & \multicolumn{1}{c|}{Analytical} & \multicolumn{1}{c|}{\begin{tabular}[c]{@{}c@{}}Single-core accelerator\\ with multiple memory\\ hierarchies\end{tabular}} & \multicolumn{1}{c|}{\ding{55}} & \ding{55} \\ \hline
ASTRA-sim\cite{rashidi2020astra, won2023astra} & \multicolumn{1}{c|}{\ding{52}} & \multicolumn{1}{c|}{\begin{tabular}[c]{@{}c@{}}\ding{52}\rotatebox[origin=c]{-9.2}{\kern-0.7em\ding{55}} Topology\\ building blocks\end{tabular}} & \begin{tabular}[c]{@{}c@{}}\ding{52}\rotatebox[origin=c]{-9.2}{\kern-0.7em\ding{55}} Finest\\ granularity: NPU\end{tabular} & \multicolumn{1}{c|}{N/A} & \multicolumn{1}{c|}{N/A} & N/A & \multicolumn{1}{c|}{Hybrid} & \multicolumn{1}{c|}{\begin{tabular}[c]{@{}c@{}}NPU $\to$ package $\to$\\ server $\to$ cluster\end{tabular}} & \multicolumn{1}{c|}{\begin{tabular}[c]{@{}c@{}}ASTRA-sim1.0 \ding{52}\\ ASTRA-sim2.0 \ding{55}\end{tabular}} & \ding{55} \\ \hline
\begin{tabular}[c]{@{}c@{}}DSE Tool for \\ Specific Architecture\\ \cite{tan2021nn, hao2023monad, cai2024gemini, ardalani2024deepflow, fang2024palm, zhang2024llmcompass, wang2024tmac, zhu2024theseus}\end{tabular} & \multicolumn{1}{c|}{\ding{52}} & \multicolumn{1}{c|}{\ding{55}} & \ding{55} & \multicolumn{1}{c|}{\begin{tabular}[c]{@{}c@{}}\ding{55} Spatial\cite{tan2021nn, cai2024gemini, hao2023monad}\\ \ding{55} Temporal\cite{zhang2024llmcompass} \\ \ding{52} Spatiotemporal\cite{ardalani2024deepflow, wang2024tmac, zhu2024theseus}\end{tabular}} & \multicolumn{1}{c|}{\ding{55}} & \ding{72} Fixed level & \multicolumn{1}{c|}{\begin{tabular}[c]{@{}c@{}}Analytical\\ \cite{tan2021nn, cai2024gemini, hao2023monad, zhang2024llmcompass, wang2024tmac}\\ Hybrid\\ \cite{fang2024palm, zhu2024theseus}\end{tabular}} & \multicolumn{1}{c|}{\begin{tabular}[c]{@{}c@{}}Tiled accelerator\cite{fang2024palm}\\ Multi-chiplet accelerator\cite{tan2021nn, cai2024gemini, hao2023monad}\\ GPU/TPU systems\cite{zhang2024llmcompass}\\ Wafer-scale chips\cite{wang2024tmac,zhu2024theseus}\end{tabular}} & \multicolumn{1}{c|}{\begin{tabular}[c]{@{}c@{}}\ding{55}\cite{tan2021nn, cai2024gemini, zhang2024llmcompass, wang2024tmac}\\ \ding{52}\cite{fang2024palm, zhu2024theseus, hao2023monad}\end{tabular}} & \ding{55} \\ \hline
\begin{tabular}[c]{@{}c@{}}Modular\\ Simulator\cite{herbst2024switchboard, zhi2021methodology, matthews2020mosaicsim}\end{tabular} & \multicolumn{1}{c|}{\ding{52}} & \multicolumn{1}{c|}{\ding{52}} & \begin{tabular}[c]{@{}c@{}}\ding{52}\rotatebox[origin=c]{-9.2}{\kern-0.7em\ding{55}} Inflexible \\ communication\end{tabular} & \multicolumn{1}{c|}{N/A} & \multicolumn{1}{c|}{N/A} & N/A & \multicolumn{1}{c|}{Simulation} & \multicolumn{1}{c|}{Any composable hardware} & \multicolumn{1}{c|}{\ding{52}} & \ding{55} \\ \hline
\begin{tabular}[c]{@{}c@{}}Simulator for Specific\\ Multi-Level Hardware\cite{orenes2024muchisim, lin2024hex}\end{tabular} & \multicolumn{1}{c|}{\ding{52}} & \multicolumn{1}{c|}{\ding{55}} & \begin{tabular}[c]{@{}c@{}}\ding{55} Complex but fixed spatial\\ levels\end{tabular} & \multicolumn{1}{c|}{N/A} & \multicolumn{1}{c|}{N/A} & N/A & \multicolumn{1}{c|}{Simulation} & \multicolumn{1}{c|}{\begin{tabular}[c]{@{}c@{}}Specific multi-level hardware:\\ From tile to cluster\cite{orenes2024muchisim}\\ Multi-chiplet many-core NPU\cite{lin2024hex}\end{tabular}} & \multicolumn{1}{c|}{\ding{52}} & \begin{tabular}[c]{@{}c@{}}\ding{52}\cite{orenes2024muchisim, lin2024hex}\\ Cycle-level\end{tabular} \\ \hline
Cycle-Accurate / RTL & \multicolumn{1}{c|}{\ding{55}} & \multicolumn{1}{c|}{\ding{55}} & \ding{55} & \multicolumn{1}{c|}{N/A} & \multicolumn{1}{c|}{N/A} & N/A & \multicolumn{1}{c|}{Simulation} & \multicolumn{1}{c|}{Specific hardware} & \multicolumn{1}{c|}{\ding{52}} & \ding{52} \\ \hline
\end{tabular}%
}
\end{table*}

We also performed hardware parameter DSE for the decode stage. Figure \ref{fig:space_level}(b, e-g) shows spatial computing eliminates frequent offchip data transfers, shifting the performance impact to other memory access and communication parameters. Figure \ref{fig:space_level}(b) shows that higher local memory bandwidth makes local memory latency more impactful. With spatial computing, the decode stage involves relatively small data transfers, making memory latency a key factor when bandwidth becomes high. Figure \ref{fig:space_level}(e-g) rank parameter impact as NoC bandwidth > local memory bandwidth > local memory latency. This highlights NoC communication as a potential bottleneck during the compute-light decode stage.

\section{Related Works}
\label{sec:related_works}

We distinguish MLDSE from existing DSE tools across four aspects:

\textbf{Methodology:} The classical methodology employed by many existing DSE tools involves designing fixed hardware templates and extending them to accommodate a broader range of cases \cite{fang2024palm, zhang2024llmcompass, lin2024hex, orenes2024muchisim, rashidi2020astra, won2023astra}. In contrast, MLDSE is a \textbf{meta-DSE framework} that establishes composable rules for generalized scenarios. While it can be instantiated into specialized DSE tools (demonstrated in Section \ref{sec:cross_arch} and \ref{sec:space_level}), MLDSE is \textbf{not} a wrapper around existing tools. Instead, it functions as a dynamic composer, addressing the challenges of complete modeling and mapping languages and polymorphic real-time DSE process generation.

\textbf{Hardware/Mapping Representations:} MLDSE creates a hardware/mapping IR that can be instantiated into user-specified dialects. While these IRs support architectures and mapping strategies discussed in existing works\cite{orenes2024muchisim, rashidi2020astra, won2023astra, parashar2019timeloop, cai2024gemini, fang2024palm, zhang2024llmcompass}, the key innovation lies in the recursive and composable rules that enable our abstraction to represent and \textbf{operate on} flexible organizations of hardware components and arbitrary expansions of vertical levels. This capability is \textbf{not} offered by existing tools that are confined to specific, predefined architectures. For instance, Timeloop (relatively flexible compared to other tools) is limited to a tree hierarchy within a single vertical level, unable to model: 1) Mesh-connected many-core architectures, e.g., Timeloop is only used to simulate \href{https://github.com/Accelergy-Project/timeloop-accelergy-exercises/tree/master/workspace/example\_designs/example\_designs/simba\_like}{a single chip of a MCM hardware like Simba}\cite{shao2019simba}. 2) Two-level spatial hierarchies, e.g., a multi-chiplet many-core package. Existing tools for specific multi-level hardware\cite{orenes2024muchisim, rashidi2020astra, fang2024palm, zhang2024llmcompass, cai2024gemini} fail to adapt to changes in the number of levels (e.g., adding a new level in Section \ref{sec:space_level}), adjust organization within levels, or explore complex inter-level relationships (e.g., hardware elements at the same level have different spatial sub-structures). While ASTRA-Sim2.0\cite{won2023astra} introduces complete and powerful topology building rules for hierarchical network construction, it focuses on cluster-level simulation and cannot model hardware hierarchies below the NPU level. MLDSE extends its rules for more general computation, memory and communication blocks with \textit{SpaceMatrix} and \textit{SpacePoint} data structures and enables modeling from processing elements to clusters (or any other spatial levels) and supports mixed-granularity elements within different levels. Certain specific representations also remain unexplored in existing works, including fine-grained cross-level communication mapping and hierarchical synchronization. While Gemini\cite{cai2024gemini} claims to be the first work to address layer-pipelined spatial mapping, MLDSE extends this to spatiotemporal mapping in cross-level scenarios.

\textbf{Evaluation:} MLDSE has the ability to conduct accurate and efficient evaluations \textbf{without} relying on pre-defined architectures. It dynamically generates a customized simulator tailored to user-specified architecture. This contrasts with traditional analytical models\cite{isaev2023calculon, tan2021nn, cai2024gemini, hao2023monad, zhang2024llmcompass}, which often target at specific, fixed architectures and are thus unsuitable for broader scope. Some modular simulators\cite{matthews2020mosaicsim, herbst2024switchboard, zhi2021methodology} developed frameworks that link existing simulators (e.g., cycle-accurate simulators) for overall evaluation. However, even if each individual \textit{SpacePoint} could be evaluated with a specific tool, their combinations introduce complex inter-dependencies, resource contention, and dynamic synchronization overheads. How to tackle these complex orchestration in a general and polymorphic approach remains a challenge for these works.

\textbf{DSE Scope:} Existing works typically focus on single level hardware \cite{parashar2019timeloop} or assume a specific scope of multi-level hardware and then explore hardware parameters and mapping within that scope \cite{orenes2024muchisim, lin2024hex}. As demonstrated by our experiments, MLDSE is capable of performing \textbf{three-tier DSE}: architecture-level, parameter-level, and mapping DSE.

Due to the large number of existing DSE tools, we provide a more comprehensive comparison in Table \ref{tab:comparison}.

\section{Conclusion}

We introduce MLDSE, a \textbf{meta-DSE tool} for domain-specific design space exploration (DSE) of multi-level hardware, enabling the creation of specialized DSE tools for diverse multi-level architectures. MLDSE aims to support academia and industry in rapidly prototyping architecture-specific DSE tools, minimizing redundant efforts and accelerating hardware development. Its extensible design makes it a potential platform for architectural exploration and fair comparisons. By combining recursive hardware modeling, spatiotemporal mapping optimization, and task-level event simulation, MLDSE achieves a balance of efficiency, accuracy, and flexibility in multi-level hardware exploration. Future improvements could involve integrating more precise hardware-specific simulators and practicing across a wider range of hardware architectures.



\begin{thebibliography}{108}


\ifx \showCODEN    \undefined \def \showCODEN     #1{\unskip}     \fi
\ifx \showDOI      \undefined \def \showDOI       #1{#1}\fi
\ifx \showISBNx    \undefined \def \showISBNx     #1{\unskip}     \fi
\ifx \showISBNxiii \undefined \def \showISBNxiii  #1{\unskip}     \fi
\ifx \showISSN     \undefined \def \showISSN      #1{\unskip}     \fi
\ifx \showLCCN     \undefined \def \showLCCN      #1{\unskip}     \fi
\ifx \shownote     \undefined \def \shownote      #1{#1}          \fi
\ifx \showarticletitle \undefined \def \showarticletitle #1{#1}   \fi
\ifx \showURL      \undefined \def \showURL       {\relax}        \fi
\providecommand\bibfield[2]{#2}
\providecommand\bibinfo[2]{#2}
\providecommand\natexlab[1]{#1}
\providecommand\showeprint[2][]{arXiv:#2}

\bibitem[Abts et~al\mbox{.}(2022a)]%
        {abts2022groq}
\bibfield{author}{\bibinfo{person}{Dennis Abts}, \bibinfo{person}{John Kim}, \bibinfo{person}{Garrin Kimmell}, \bibinfo{person}{Matthew Boyd}, \bibinfo{person}{Kris Kang}, \bibinfo{person}{Sahil Parmar}, \bibinfo{person}{Andrew Ling}, \bibinfo{person}{Andrew Bitar}, \bibinfo{person}{Ibrahim Ahmed}, {and} \bibinfo{person}{Jonathan Ross}.} \bibinfo{year}{2022}\natexlab{a}.
\newblock \showarticletitle{The groq software-defined scale-out tensor streaming multiprocessor: From chips-to-systems architectural overview}. In \bibinfo{booktitle}{\emph{2022 IEEE Hot Chips 34 Symposium (HCS)}}. IEEE Computer Society, \bibinfo{pages}{1--69}.
\newblock


\bibitem[Abts et~al\mbox{.}(2022b)]%
        {abts2022software}
\bibfield{author}{\bibinfo{person}{Dennis Abts}, \bibinfo{person}{Garrin Kimmell}, \bibinfo{person}{Andrew Ling}, \bibinfo{person}{John Kim}, \bibinfo{person}{Matt Boyd}, \bibinfo{person}{Andrew Bitar}, \bibinfo{person}{Sahil Parmar}, \bibinfo{person}{Ibrahim Ahmed}, \bibinfo{person}{Roberto DiCecco}, \bibinfo{person}{David Han}, \bibinfo{person}{John Thompson}, \bibinfo{person}{Michael Bye}, \bibinfo{person}{Jennifer Hwang}, \bibinfo{person}{Jeremy Fowers}, \bibinfo{person}{Peter Lillian}, \bibinfo{person}{Ashwin Murthy}, \bibinfo{person}{Elyas Mehtabuddin}, \bibinfo{person}{Chetan Tekur}, \bibinfo{person}{Thomas Sohmers}, \bibinfo{person}{Kris Kang}, \bibinfo{person}{Stephen Maresh}, {and} \bibinfo{person}{Jonathan Ross}.} \bibinfo{year}{2022}\natexlab{b}.
\newblock \showarticletitle{A software-defined tensor streaming multiprocessor for large-scale machine learning}. In \bibinfo{booktitle}{\emph{Proceedings of the 49th Annual International Symposium on Computer Architecture}}. \bibinfo{pages}{567--580}.
\newblock


\bibitem[Abts et~al\mbox{.}(2020)]%
        {dennis2020think}
\bibfield{author}{\bibinfo{person}{Dennis Abts}, \bibinfo{person}{Jonathan Ross}, \bibinfo{person}{Jonathan Sparling}, \bibinfo{person}{Mark Wong-VanHaren}, \bibinfo{person}{Max Baker}, \bibinfo{person}{Tom Hawkins}, \bibinfo{person}{Andrew Bell}, \bibinfo{person}{John Thompson}, \bibinfo{person}{Temesghen Kahsai}, \bibinfo{person}{Garrin Kimmell}, \bibinfo{person}{Jennifer Hwang}, \bibinfo{person}{Rebekah Leslie-Hurd}, \bibinfo{person}{Michael Bye}, \bibinfo{person}{E.R. Creswick}, \bibinfo{person}{Matthew Boyd}, \bibinfo{person}{Mahitha Venigalla}, \bibinfo{person}{Evan Laforge}, \bibinfo{person}{Jon Purdy}, \bibinfo{person}{Purushotham Kamath}, \bibinfo{person}{Dinesh Maheshwari}, \bibinfo{person}{Michael Beidler}, \bibinfo{person}{Geert Rosseel}, \bibinfo{person}{Omar Ahmad}, \bibinfo{person}{Gleb Gagarin}, \bibinfo{person}{Richard Czekalski}, \bibinfo{person}{Ashay Rane}, \bibinfo{person}{Sahil Parmar}, \bibinfo{person}{Jeff Werner}, \bibinfo{person}{Jim Sproch}, \bibinfo{person}{Adrian Macias}, {and}
  \bibinfo{person}{Brian Kurtz}.} \bibinfo{year}{2020}\natexlab{}.
\newblock \showarticletitle{Think Fast: A Tensor Streaming Processor (TSP) for Accelerating Deep Learning Workloads}. In \bibinfo{booktitle}{\emph{2020 ACM/IEEE 47th Annual International Symposium on Computer Architecture (ISCA)}}. \bibinfo{pages}{145--158}.
\newblock
\urldef\tempurl%
\url{https://doi.org/10.1109/ISCA45697.2020.00023}
\showDOI{\tempurl}


\bibitem[Agrawal et~al\mbox{.}(2024a)]%
        {agrawal2024etalon}
\bibfield{author}{\bibinfo{person}{Amey Agrawal}, \bibinfo{person}{Anmol Agarwal}, \bibinfo{person}{Nitin Kedia}, \bibinfo{person}{Jayashree Mohan}, \bibinfo{person}{Souvik Kundu}, \bibinfo{person}{Nipun Kwatra}, \bibinfo{person}{Ramachandran Ramjee}, {and} \bibinfo{person}{Alexey Tumanov}.} \bibinfo{year}{2024}\natexlab{a}.
\newblock \showarticletitle{Etalon: Holistic Performance Evaluation Framework for LLM Inference Systems}.
\newblock \bibinfo{journal}{\emph{arXiv preprint arXiv:2407.07000}} (\bibinfo{year}{2024}).
\newblock


\bibitem[Agrawal et~al\mbox{.}(2024b)]%
        {agrawal2024vidur}
\bibfield{author}{\bibinfo{person}{Amey Agrawal}, \bibinfo{person}{Nitin Kedia}, \bibinfo{person}{Jayashree Mohan}, \bibinfo{person}{Ashish Panwar}, \bibinfo{person}{Nipun Kwatra}, \bibinfo{person}{Bhargav Gulavani}, \bibinfo{person}{Ramachandran Ramjee}, {and} \bibinfo{person}{Alexey Tumanov}.} \bibinfo{year}{2024}\natexlab{b}.
\newblock \showarticletitle{Vidur: A Large-Scale Simulation Framework For LLM Inference}.
\newblock \bibinfo{journal}{\emph{Proceedings of Machine Learning and Systems}}  \bibinfo{volume}{6} (\bibinfo{year}{2024}), \bibinfo{pages}{351--366}.
\newblock


\bibitem[Agrawal et~al\mbox{.}(2023)]%
        {agrawal2023sarathi}
\bibfield{author}{\bibinfo{person}{Amey Agrawal}, \bibinfo{person}{Ashish Panwar}, \bibinfo{person}{Jayashree Mohan}, \bibinfo{person}{Nipun Kwatra}, \bibinfo{person}{Bhargav~S Gulavani}, {and} \bibinfo{person}{Ramachandran Ramjee}.} \bibinfo{year}{2023}\natexlab{}.
\newblock \showarticletitle{Sarathi: Efficient llm inference by piggybacking decodes with chunked prefills}.
\newblock \bibinfo{journal}{\emph{arXiv preprint arXiv:2308.16369}} (\bibinfo{year}{2023}).
\newblock


\bibitem[Agrawal and Xu(2023)]%
        {agrawal2023deap}
\bibfield{author}{\bibinfo{person}{Ekansh Agrawal} {and} \bibinfo{person}{Xiangyu~Sam Xu}.} \bibinfo{year}{2023}\natexlab{}.
\newblock \showarticletitle{DEAP: Design Space Exploration for DNN Accelerator Parallelism}.
\newblock \bibinfo{journal}{\emph{arXiv preprint arXiv:2312.15388}} (\bibinfo{year}{2023}).
\newblock


\bibitem[Ahmad et~al\mbox{.}(2020)]%
        {ahmad2020superslash}
\bibfield{author}{\bibinfo{person}{Hazoor Ahmad}, \bibinfo{person}{Tabasher Arif}, \bibinfo{person}{Muhammad~Abdullah Hanif}, \bibinfo{person}{Rehan Hafiz}, {and} \bibinfo{person}{Muhammad Shafique}.} \bibinfo{year}{2020}\natexlab{}.
\newblock \showarticletitle{SuperSlash: A unified design space exploration and model compression methodology for design of deep learning accelerators with reduced off-chip memory access volume}.
\newblock \bibinfo{journal}{\emph{IEEE Transactions on Computer-Aided Design of Integrated Circuits and Systems}} \bibinfo{volume}{39}, \bibinfo{number}{11} (\bibinfo{year}{2020}), \bibinfo{pages}{4191--4204}.
\newblock


\bibitem[Ardalani et~al\mbox{.}(2024)]%
        {ardalani2024deepflow}
\bibfield{author}{\bibinfo{person}{Newsha Ardalani}, \bibinfo{person}{Saptadeep Pal}, {and} \bibinfo{person}{Puneet Gupta}.} \bibinfo{year}{2024}\natexlab{}.
\newblock \showarticletitle{DeepFlow: A cross-stack pathfinding framework for distributed ai systems}.
\newblock \bibinfo{journal}{\emph{ACM Transactions on Design Automation of Electronic Systems}} \bibinfo{volume}{29}, \bibinfo{number}{2} (\bibinfo{year}{2024}), \bibinfo{pages}{1--20}.
\newblock


\bibitem[Arunkumar et~al\mbox{.}(2017)]%
        {arunkumar2017mcm}
\bibfield{author}{\bibinfo{person}{Akhil Arunkumar}, \bibinfo{person}{Evgeny Bolotin}, \bibinfo{person}{Benjamin Cho}, \bibinfo{person}{Ugljesa Milic}, \bibinfo{person}{Eiman Ebrahimi}, \bibinfo{person}{Oreste Villa}, \bibinfo{person}{Aamer Jaleel}, \bibinfo{person}{Carole-Jean Wu}, {and} \bibinfo{person}{David Nellans}.} \bibinfo{year}{2017}\natexlab{}.
\newblock \showarticletitle{MCM-GPU: Multi-chip-module GPUs for continued performance scalability}.
\newblock \bibinfo{journal}{\emph{ACM SIGARCH Computer Architecture News}} \bibinfo{volume}{45}, \bibinfo{number}{2} (\bibinfo{year}{2017}), \bibinfo{pages}{320--332}.
\newblock


\bibitem[Bai et~al\mbox{.}(2023)]%
        {bai2023qwen}
\bibfield{author}{\bibinfo{person}{Jinze Bai}, \bibinfo{person}{Shuai Bai}, \bibinfo{person}{Yunfei Chu}, \bibinfo{person}{Zeyu Cui}, \bibinfo{person}{Kai Dang}, \bibinfo{person}{Xiaodong Deng}, \bibinfo{person}{Yang Fan}, \bibinfo{person}{Wenbin Ge}, \bibinfo{person}{Yu Han}, \bibinfo{person}{Fei Huang}, \bibinfo{person}{Binyuan Hui}, \bibinfo{person}{Luo Ji}, \bibinfo{person}{Mei Li}, \bibinfo{person}{Junyang Lin}, \bibinfo{person}{Runji Lin}, \bibinfo{person}{Dayiheng Liu}, \bibinfo{person}{Gao Liu}, \bibinfo{person}{Chengqiang Lu}, \bibinfo{person}{Keming Lu}, \bibinfo{person}{Jianxin Ma}, \bibinfo{person}{Rui Men}, \bibinfo{person}{Xingzhang Ren}, \bibinfo{person}{Xuancheng Ren}, \bibinfo{person}{Chuanqi Tan}, \bibinfo{person}{Sinan Tan}, \bibinfo{person}{Jianhong Tu}, \bibinfo{person}{Peng Wang}, \bibinfo{person}{Shijie Wang}, \bibinfo{person}{Wei Wang}, \bibinfo{person}{Shengguang Wu}, \bibinfo{person}{Benfeng Xu}, \bibinfo{person}{Jin Xu}, \bibinfo{person}{An Yang}, \bibinfo{person}{Hao Yang},
  \bibinfo{person}{Jian Yang}, \bibinfo{person}{Shusheng Yang}, \bibinfo{person}{Yang Yao}, \bibinfo{person}{Bowen Yu}, \bibinfo{person}{Hongyi Yuan}, \bibinfo{person}{Zheng Yuan}, \bibinfo{person}{Jianwei Zhang}, \bibinfo{person}{Xingxuan Zhang}, \bibinfo{person}{Yichang Zhang}, \bibinfo{person}{Zhenru Zhang}, \bibinfo{person}{Chang Zhou}, \bibinfo{person}{Jingren Zhou}, \bibinfo{person}{Xiaohuan Zhou}, {and} \bibinfo{person}{Tianhang Zhu}.} \bibinfo{year}{2023}\natexlab{}.
\newblock \showarticletitle{Qwen Technical Report}.
\newblock  (\bibinfo{year}{2023}).
\newblock
\showeprint[arxiv]{2309.16609}~[cs.CL]
\urldef\tempurl%
\url{https://arxiv.org/abs/2309.16609}
\showURL{%
\tempurl}


\bibitem[Balasubramonian et~al\mbox{.}(2017)]%
        {balasubramonian2017cacti}
\bibfield{author}{\bibinfo{person}{Rajeev Balasubramonian}, \bibinfo{person}{Andrew~B Kahng}, \bibinfo{person}{Naveen Muralimanohar}, \bibinfo{person}{Ali Shafiee}, {and} \bibinfo{person}{Vaishnav Srinivas}.} \bibinfo{year}{2017}\natexlab{}.
\newblock \showarticletitle{CACTI 7: New tools for interconnect exploration in innovative off-chip memories}.
\newblock \bibinfo{journal}{\emph{ACM Transactions on Architecture and Code Optimization (TACO)}} \bibinfo{volume}{14}, \bibinfo{number}{2} (\bibinfo{year}{2017}), \bibinfo{pages}{1--25}.
\newblock


\bibitem[Bang et~al\mbox{.}(2024)]%
        {bang2024vtrain}
\bibfield{author}{\bibinfo{person}{Jehyeon Bang}, \bibinfo{person}{Yujeong Choi}, \bibinfo{person}{Myeongwoo Kim}, \bibinfo{person}{Yongdeok Kim}, {and} \bibinfo{person}{Minsoo Rhu}.} \bibinfo{year}{2024}\natexlab{}.
\newblock \showarticletitle{vtrain: A simulation framework for evaluating cost-effective and compute-optimal large language model training}. In \bibinfo{booktitle}{\emph{2024 57th IEEE/ACM International Symposium on Microarchitecture (MICRO)}}. IEEE, \bibinfo{pages}{153--167}.
\newblock


\bibitem[Brown et~al\mbox{.}(2020)]%
        {brown2020language}
\bibfield{author}{\bibinfo{person}{Tom~B. Brown}, \bibinfo{person}{Benjamin Mann}, \bibinfo{person}{Nick Ryder}, \bibinfo{person}{Melanie Subbiah}, \bibinfo{person}{Jared Kaplan}, \bibinfo{person}{Prafulla Dhariwal}, \bibinfo{person}{Arvind Neelakantan}, \bibinfo{person}{Pranav Shyam}, \bibinfo{person}{Girish Sastry}, \bibinfo{person}{Amanda Askell}, \bibinfo{person}{Sandhini Agarwal}, \bibinfo{person}{Ariel Herbert{-}Voss}, \bibinfo{person}{Gretchen Krueger}, \bibinfo{person}{Tom Henighan}, \bibinfo{person}{Rewon Child}, \bibinfo{person}{Aditya Ramesh}, \bibinfo{person}{Daniel~M. Ziegler}, \bibinfo{person}{Jeffrey Wu}, \bibinfo{person}{Clemens Winter}, \bibinfo{person}{Christopher Hesse}, \bibinfo{person}{Mark Chen}, \bibinfo{person}{Eric Sigler}, \bibinfo{person}{Mateusz Litwin}, \bibinfo{person}{Scott Gray}, \bibinfo{person}{Benjamin Chess}, \bibinfo{person}{Jack Clark}, \bibinfo{person}{Christopher Berner}, \bibinfo{person}{Sam McCandlish}, \bibinfo{person}{Alec Radford}, \bibinfo{person}{Ilya Sutskever},
  {and} \bibinfo{person}{Dario Amodei}.} \bibinfo{year}{2020}\natexlab{}.
\newblock \showarticletitle{Language Models are Few-Shot Learners}.
\newblock \bibinfo{journal}{\emph{CoRR}}  \bibinfo{volume}{abs/2005.14165} (\bibinfo{year}{2020}).
\newblock
\showeprint[arXiv]{2005.14165}
\urldef\tempurl%
\url{https://arxiv.org/abs/2005.14165}
\showURL{%
\tempurl}


\bibitem[Cai et~al\mbox{.}(2024)]%
        {cai2024gemini}
\bibfield{author}{\bibinfo{person}{Jingwei Cai}, \bibinfo{person}{Zuotong Wu}, \bibinfo{person}{Sen Peng}, \bibinfo{person}{Yuchen Wei}, \bibinfo{person}{Zhanhong Tan}, \bibinfo{person}{Guiming Shi}, \bibinfo{person}{Mingyu Gao}, {and} \bibinfo{person}{Kaisheng Ma}.} \bibinfo{year}{2024}\natexlab{}.
\newblock \showarticletitle{Gemini: Mapping and Architecture Co-exploration for Large-scale DNN Chiplet Accelerators}. In \bibinfo{booktitle}{\emph{2024 IEEE International Symposium on High-Performance Computer Architecture (HPCA)}}. IEEE, \bibinfo{pages}{156--171}.
\newblock


\bibitem[Chen et~al\mbox{.}(2014a)]%
        {chen2014diannao}
\bibfield{author}{\bibinfo{person}{Tianshi Chen}, \bibinfo{person}{Zidong Du}, \bibinfo{person}{Ninghui Sun}, \bibinfo{person}{Jia Wang}, \bibinfo{person}{Chengyong Wu}, \bibinfo{person}{Yunji Chen}, {and} \bibinfo{person}{Olivier Temam}.} \bibinfo{year}{2014}\natexlab{a}.
\newblock \showarticletitle{Diannao: A small-footprint high-throughput accelerator for ubiquitous machine-learning}.
\newblock \bibinfo{journal}{\emph{ACM SIGARCH Computer Architecture News}} \bibinfo{volume}{42}, \bibinfo{number}{1} (\bibinfo{year}{2014}), \bibinfo{pages}{269--284}.
\newblock


\bibitem[Chen et~al\mbox{.}(2014b)]%
        {chen2014dadiannao}
\bibfield{author}{\bibinfo{person}{Yunji Chen}, \bibinfo{person}{Tao Luo}, \bibinfo{person}{Shaoli Liu}, \bibinfo{person}{Shijin Zhang}, \bibinfo{person}{Liqiang He}, \bibinfo{person}{Jia Wang}, \bibinfo{person}{Ling Li}, \bibinfo{person}{Tianshi Chen}, \bibinfo{person}{Zhiwei Xu}, \bibinfo{person}{Ninghui Sun}, {and} \bibinfo{person}{Olivier Temam}.} \bibinfo{year}{2014}\natexlab{b}.
\newblock \showarticletitle{Dadiannao: A machine-learning supercomputer}. In \bibinfo{booktitle}{\emph{2014 47th Annual IEEE/ACM International Symposium on Microarchitecture}}. IEEE, \bibinfo{pages}{609--622}.
\newblock


\bibitem[Choquette(2023)]%
        {choquette2023nvidia}
\bibfield{author}{\bibinfo{person}{Jack Choquette}.} \bibinfo{year}{2023}\natexlab{}.
\newblock \showarticletitle{Nvidia hopper h100 gpu: Scaling performance}.
\newblock \bibinfo{journal}{\emph{IEEE Micro}} (\bibinfo{year}{2023}).
\newblock


\bibitem[Choquette et~al\mbox{.}(2021)]%
        {choquette2021nvidia}
\bibfield{author}{\bibinfo{person}{Jack Choquette}, \bibinfo{person}{Wishwesh Gandhi}, \bibinfo{person}{Olivier Giroux}, \bibinfo{person}{Nick Stam}, {and} \bibinfo{person}{Ronny Krashinsky}.} \bibinfo{year}{2021}\natexlab{}.
\newblock \showarticletitle{Nvidia a100 tensor core gpu: Performance and innovation}.
\newblock \bibinfo{journal}{\emph{IEEE Micro}} \bibinfo{volume}{41}, \bibinfo{number}{2} (\bibinfo{year}{2021}), \bibinfo{pages}{29--35}.
\newblock


\bibitem[Cong et~al\mbox{.}(2024)]%
        {cong2024attentionlego}
\bibfield{author}{\bibinfo{person}{Rongqing Cong}, \bibinfo{person}{Wenyang He}, \bibinfo{person}{Mingxuan Li}, \bibinfo{person}{Bangning Luo}, \bibinfo{person}{Zebin Yang}, \bibinfo{person}{Yuchao Yang}, \bibinfo{person}{Ru Huang}, {and} \bibinfo{person}{Bonan Yan}.} \bibinfo{year}{2024}\natexlab{}.
\newblock \showarticletitle{AttentionLego: An Open-Source Building Block For Spatially-Scalable Large Language Model Accelerator With Processing-In-Memory Technology}.
\newblock \bibinfo{journal}{\emph{arXiv preprint arXiv:2401.11459}} (\bibinfo{year}{2024}).
\newblock


\bibitem[Corro et~al\mbox{.}(2023)]%
        {delcorro2023skipdecode}
\bibfield{author}{\bibinfo{person}{Luciano~Del Corro}, \bibinfo{person}{Allie~Del Giorno}, \bibinfo{person}{Sahaj Agarwal}, \bibinfo{person}{Bin Yu}, \bibinfo{person}{Ahmed Awadallah}, {and} \bibinfo{person}{Subhabrata Mukherjee}.} \bibinfo{year}{2023}\natexlab{}.
\newblock \showarticletitle{SkipDecode: Autoregressive Skip Decoding with Batching and Caching for Efficient LLM Inference}.
\newblock  (\bibinfo{year}{2023}).
\newblock
\showeprint[arxiv]{2307.02628}~[cs.CL]
\urldef\tempurl%
\url{https://arxiv.org/abs/2307.02628}
\showURL{%
\tempurl}


\bibitem[Das et~al\mbox{.}(2024)]%
        {das2024multi}
\bibfield{author}{\bibinfo{person}{Abhijit Das}, \bibinfo{person}{Enrico Russo}, {and} \bibinfo{person}{Maurizio Palesi}.} \bibinfo{year}{2024}\natexlab{}.
\newblock \showarticletitle{Multi-Objective Hardware-Mapping Co-Optimisation for Multi-DNN Workloads on Chiplet-based Accelerators}.
\newblock \bibinfo{journal}{\emph{IEEE Trans. Comput.}} (\bibinfo{year}{2024}).
\newblock


\bibitem[Fang et~al\mbox{.}(2024)]%
        {fang2024palm}
\bibfield{author}{\bibinfo{person}{Jiahao Fang}, \bibinfo{person}{Huizheng Wang}, \bibinfo{person}{Qize Yang}, \bibinfo{person}{Dehao Kong}, \bibinfo{person}{Xu Dai}, \bibinfo{person}{Jinyi Deng}, \bibinfo{person}{Yang Hu}, {and} \bibinfo{person}{Shouyi Yin}.} \bibinfo{year}{2024}\natexlab{}.
\newblock \showarticletitle{PALM: A Efficient Performance Simulator for Tiled Accelerators with Large-scale Model Training}.
\newblock \bibinfo{journal}{\emph{arXiv preprint arXiv:2406.03868}} (\bibinfo{year}{2024}).
\newblock


\bibitem[Feng and Ma(2022)]%
        {feng2022chiplet}
\bibfield{author}{\bibinfo{person}{Yinxiao Feng} {and} \bibinfo{person}{Kaisheng Ma}.} \bibinfo{year}{2022}\natexlab{}.
\newblock \showarticletitle{Chiplet actuary: A quantitative cost model and multi-chiplet architecture exploration}. In \bibinfo{booktitle}{\emph{Proceedings of the 59th ACM/IEEE Design Automation Conference}}. \bibinfo{pages}{121--126}.
\newblock


\bibitem[Grattafiori et~al\mbox{.}(2024)]%
        {grattafiori2024llama3}
\bibfield{author}{\bibinfo{person}{Aaron Grattafiori}, \bibinfo{person}{Abhimanyu Dubey}, \bibinfo{person}{Abhinav Jauhri}, \bibinfo{person}{Abhinav Pandey}, \bibinfo{person}{Abhishek Kadian}, \bibinfo{person}{Ahmad Al-Dahle}, \bibinfo{person}{Aiesha Letman}, \bibinfo{person}{Akhil Mathur}, \bibinfo{person}{Alan Schelten}, \bibinfo{person}{Alex Vaughan}, \bibinfo{person}{Amy Yang}, \bibinfo{person}{Angela Fan}, \bibinfo{person}{Anirudh Goyal}, \bibinfo{person}{Anthony Hartshorn}, \bibinfo{person}{Aobo Yang}, \bibinfo{person}{Archi Mitra}, \bibinfo{person}{Archie Sravankumar}, \bibinfo{person}{Artem Korenev}, \bibinfo{person}{Arthur Hinsvark}, \bibinfo{person}{Arun Rao}, \bibinfo{person}{Aston Zhang}, \bibinfo{person}{Aurelien Rodriguez}, \bibinfo{person}{Austen Gregerson}, \bibinfo{person}{Ava Spataru}, \bibinfo{person}{Baptiste Roziere}, \bibinfo{person}{Bethany Biron}, \bibinfo{person}{Binh Tang}, \bibinfo{person}{Bobbie Chern}, \bibinfo{person}{Charlotte Caucheteux}, \bibinfo{person}{Chaya Nayak},
  \bibinfo{person}{Chloe Bi}, \bibinfo{person}{Chris Marra}, \bibinfo{person}{Chris McConnell}, \bibinfo{person}{Christian Keller}, \bibinfo{person}{Christophe Touret}, \bibinfo{person}{Chunyang Wu}, \bibinfo{person}{Corinne Wong}, \bibinfo{person}{Cristian~Canton Ferrer}, \bibinfo{person}{Cyrus Nikolaidis}, \bibinfo{person}{Damien Allonsius}, \bibinfo{person}{Daniel Song}, \bibinfo{person}{Danielle Pintz}, \bibinfo{person}{Danny Livshits}, \bibinfo{person}{Danny Wyatt}, \bibinfo{person}{David Esiobu}, \bibinfo{person}{Dhruv Choudhary}, \bibinfo{person}{Dhruv Mahajan}, \bibinfo{person}{Diego Garcia-Olano}, \bibinfo{person}{Diego Perino}, \bibinfo{person}{Dieuwke Hupkes}, \bibinfo{person}{Egor Lakomkin}, \bibinfo{person}{Ehab AlBadawy}, \bibinfo{person}{Elina Lobanova}, \bibinfo{person}{Emily Dinan}, \bibinfo{person}{Eric~Michael Smith}, \bibinfo{person}{Filip Radenovic}, \bibinfo{person}{Francisco Guzmán}, \bibinfo{person}{Frank Zhang}, \bibinfo{person}{Gabriel Synnaeve}, \bibinfo{person}{Gabrielle Lee},
  \bibinfo{person}{Georgia~Lewis Anderson}, \bibinfo{person}{Govind Thattai}, \bibinfo{person}{Graeme Nail}, \bibinfo{person}{Gregoire Mialon}, \bibinfo{person}{Guan Pang}, \bibinfo{person}{Guillem Cucurell}, \bibinfo{person}{Hailey Nguyen}, \bibinfo{person}{Hannah Korevaar}, \bibinfo{person}{Hu Xu}, \bibinfo{person}{Hugo Touvron}, \bibinfo{person}{Iliyan Zarov}, \bibinfo{person}{Imanol~Arrieta Ibarra}, \bibinfo{person}{Isabel Kloumann}, \bibinfo{person}{Ishan Misra}, \bibinfo{person}{Ivan Evtimov}, \bibinfo{person}{Jack Zhang}, \bibinfo{person}{Jade Copet}, \bibinfo{person}{Jaewon Lee}, \bibinfo{person}{Jan Geffert}, \bibinfo{person}{Jana Vranes}, \bibinfo{person}{Jason Park}, \bibinfo{person}{Jay Mahadeokar}, \bibinfo{person}{Jeet Shah}, \bibinfo{person}{Jelmer van~der Linde}, \bibinfo{person}{Jennifer Billock}, \bibinfo{person}{Jenny Hong}, \bibinfo{person}{Jenya Lee}, \bibinfo{person}{Jeremy Fu}, \bibinfo{person}{Jianfeng Chi}, \bibinfo{person}{Jianyu Huang}, \bibinfo{person}{Jiawen Liu},
  \bibinfo{person}{Jie Wang}, \bibinfo{person}{Jiecao Yu}, \bibinfo{person}{Joanna Bitton}, \bibinfo{person}{Joe Spisak}, \bibinfo{person}{Jongsoo Park}, \bibinfo{person}{Joseph Rocca}, \bibinfo{person}{Joshua Johnstun}, \bibinfo{person}{Joshua Saxe}, \bibinfo{person}{Junteng Jia}, \bibinfo{person}{Kalyan~Vasuden Alwala}, \bibinfo{person}{Karthik Prasad}, \bibinfo{person}{Kartikeya Upasani}, \bibinfo{person}{Kate Plawiak}, \bibinfo{person}{Ke Li}, \bibinfo{person}{Kenneth Heafield}, \bibinfo{person}{Kevin Stone}, \bibinfo{person}{Khalid El-Arini}, \bibinfo{person}{Krithika Iyer}, \bibinfo{person}{Kshitiz Malik}, \bibinfo{person}{Kuenley Chiu}, \bibinfo{person}{Kunal Bhalla}, \bibinfo{person}{Kushal Lakhotia}, \bibinfo{person}{Lauren Rantala-Yeary}, \bibinfo{person}{Laurens van~der Maaten}, \bibinfo{person}{Lawrence Chen}, \bibinfo{person}{Liang Tan}, \bibinfo{person}{Liz Jenkins}, \bibinfo{person}{Louis Martin}, \bibinfo{person}{Lovish Madaan}, \bibinfo{person}{Lubo Malo}, \bibinfo{person}{Lukas Blecher},
  \bibinfo{person}{Lukas Landzaat}, \bibinfo{person}{Luke de Oliveira}, \bibinfo{person}{Madeline Muzzi}, \bibinfo{person}{Mahesh Pasupuleti}, \bibinfo{person}{Mannat Singh}, \bibinfo{person}{Manohar Paluri}, \bibinfo{person}{Marcin Kardas}, \bibinfo{person}{Maria Tsimpoukelli}, \bibinfo{person}{Mathew Oldham}, \bibinfo{person}{Mathieu Rita}, \bibinfo{person}{Maya Pavlova}, \bibinfo{person}{Melanie Kambadur}, \bibinfo{person}{Mike Lewis}, \bibinfo{person}{Min Si}, \bibinfo{person}{Mitesh~Kumar Singh}, \bibinfo{person}{Mona Hassan}, \bibinfo{person}{Naman Goyal}, \bibinfo{person}{Narjes Torabi}, \bibinfo{person}{Nikolay Bashlykov}, \bibinfo{person}{Nikolay Bogoychev}, \bibinfo{person}{Niladri Chatterji}, \bibinfo{person}{Ning Zhang}, \bibinfo{person}{Olivier Duchenne}, \bibinfo{person}{Onur Çelebi}, \bibinfo{person}{Patrick Alrassy}, \bibinfo{person}{Pengchuan Zhang}, \bibinfo{person}{Pengwei Li}, \bibinfo{person}{Petar Vasic}, \bibinfo{person}{Peter Weng}, \bibinfo{person}{Prajjwal Bhargava},
  \bibinfo{person}{Pratik Dubal}, \bibinfo{person}{Praveen Krishnan}, \bibinfo{person}{Punit~Singh Koura}, \bibinfo{person}{Puxin Xu}, \bibinfo{person}{Qing He}, \bibinfo{person}{Qingxiao Dong}, \bibinfo{person}{Ragavan Srinivasan}, \bibinfo{person}{Raj Ganapathy}, \bibinfo{person}{Ramon Calderer}, \bibinfo{person}{Ricardo~Silveira Cabral}, \bibinfo{person}{Robert Stojnic}, \bibinfo{person}{Roberta Raileanu}, \bibinfo{person}{Rohan Maheswari}, \bibinfo{person}{Rohit Girdhar}, \bibinfo{person}{Rohit Patel}, \bibinfo{person}{Romain Sauvestre}, \bibinfo{person}{Ronnie Polidoro}, \bibinfo{person}{Roshan Sumbaly}, \bibinfo{person}{Ross Taylor}, \bibinfo{person}{Ruan Silva}, \bibinfo{person}{Rui Hou}, \bibinfo{person}{Rui Wang}, \bibinfo{person}{Saghar Hosseini}, \bibinfo{person}{Sahana Chennabasappa}, \bibinfo{person}{Sanjay Singh}, \bibinfo{person}{Sean Bell}, \bibinfo{person}{Seohyun~Sonia Kim}, \bibinfo{person}{Sergey Edunov}, \bibinfo{person}{Shaoliang Nie}, \bibinfo{person}{Sharan Narang},
  \bibinfo{person}{Sharath Raparthy}, \bibinfo{person}{Sheng Shen}, \bibinfo{person}{Shengye Wan}, \bibinfo{person}{Shruti Bhosale}, \bibinfo{person}{Shun Zhang}, \bibinfo{person}{Simon Vandenhende}, \bibinfo{person}{Soumya Batra}, \bibinfo{person}{Spencer Whitman}, \bibinfo{person}{Sten Sootla}, \bibinfo{person}{Stephane Collot}, \bibinfo{person}{Suchin Gururangan}, \bibinfo{person}{Sydney Borodinsky}, \bibinfo{person}{Tamar Herman}, \bibinfo{person}{Tara Fowler}, \bibinfo{person}{Tarek Sheasha}, \bibinfo{person}{Thomas Georgiou}, \bibinfo{person}{Thomas Scialom}, \bibinfo{person}{Tobias Speckbacher}, \bibinfo{person}{Todor Mihaylov}, \bibinfo{person}{Tong Xiao}, \bibinfo{person}{Ujjwal Karn}, \bibinfo{person}{Vedanuj Goswami}, \bibinfo{person}{Vibhor Gupta}, \bibinfo{person}{Vignesh Ramanathan}, \bibinfo{person}{Viktor Kerkez}, \bibinfo{person}{Vincent Gonguet}, \bibinfo{person}{Virginie Do}, \bibinfo{person}{Vish Vogeti}, \bibinfo{person}{Vítor Albiero}, \bibinfo{person}{Vladan Petrovic},
  \bibinfo{person}{Weiwei Chu}, \bibinfo{person}{Wenhan Xiong}, \bibinfo{person}{Wenyin Fu}, \bibinfo{person}{Whitney Meers}, \bibinfo{person}{Xavier Martinet}, \bibinfo{person}{Xiaodong Wang}, \bibinfo{person}{Xiaofang Wang}, \bibinfo{person}{Xiaoqing~Ellen Tan}, \bibinfo{person}{Xide Xia}, \bibinfo{person}{Xinfeng Xie}, \bibinfo{person}{Xuchao Jia}, \bibinfo{person}{Xuewei Wang}, \bibinfo{person}{Yaelle Goldschlag}, \bibinfo{person}{Yashesh Gaur}, \bibinfo{person}{Yasmine Babaei}, \bibinfo{person}{Yi Wen}, \bibinfo{person}{Yiwen Song}, \bibinfo{person}{Yuchen Zhang}, \bibinfo{person}{Yue Li}, \bibinfo{person}{Yuning Mao}, \bibinfo{person}{Zacharie~Delpierre Coudert}, \bibinfo{person}{Zheng Yan}, \bibinfo{person}{Zhengxing Chen}, \bibinfo{person}{Zoe Papakipos}, \bibinfo{person}{Aaditya Singh}, \bibinfo{person}{Aayushi Srivastava}, \bibinfo{person}{Abha Jain}, \bibinfo{person}{Adam Kelsey}, \bibinfo{person}{Adam Shajnfeld}, \bibinfo{person}{Adithya Gangidi}, \bibinfo{person}{Adolfo Victoria},
  \bibinfo{person}{Ahuva Goldstand}, \bibinfo{person}{Ajay Menon}, \bibinfo{person}{Ajay Sharma}, \bibinfo{person}{Alex Boesenberg}, \bibinfo{person}{Alexei Baevski}, \bibinfo{person}{Allie Feinstein}, \bibinfo{person}{Amanda Kallet}, \bibinfo{person}{Amit Sangani}, \bibinfo{person}{Amos Teo}, \bibinfo{person}{Anam Yunus}, \bibinfo{person}{Andrei Lupu}, \bibinfo{person}{Andres Alvarado}, \bibinfo{person}{Andrew Caples}, \bibinfo{person}{Andrew Gu}, \bibinfo{person}{Andrew Ho}, \bibinfo{person}{Andrew Poulton}, \bibinfo{person}{Andrew Ryan}, \bibinfo{person}{Ankit Ramchandani}, \bibinfo{person}{Annie Dong}, \bibinfo{person}{Annie Franco}, \bibinfo{person}{Anuj Goyal}, \bibinfo{person}{Aparajita Saraf}, \bibinfo{person}{Arkabandhu Chowdhury}, \bibinfo{person}{Ashley Gabriel}, \bibinfo{person}{Ashwin Bharambe}, \bibinfo{person}{Assaf Eisenman}, \bibinfo{person}{Azadeh Yazdan}, \bibinfo{person}{Beau James}, \bibinfo{person}{Ben Maurer}, \bibinfo{person}{Benjamin Leonhardi}, \bibinfo{person}{Bernie Huang},
  \bibinfo{person}{Beth Loyd}, \bibinfo{person}{Beto~De Paola}, \bibinfo{person}{Bhargavi Paranjape}, \bibinfo{person}{Bing Liu}, \bibinfo{person}{Bo Wu}, \bibinfo{person}{Boyu Ni}, \bibinfo{person}{Braden Hancock}, \bibinfo{person}{Bram Wasti}, \bibinfo{person}{Brandon Spence}, \bibinfo{person}{Brani Stojkovic}, \bibinfo{person}{Brian Gamido}, \bibinfo{person}{Britt Montalvo}, \bibinfo{person}{Carl Parker}, \bibinfo{person}{Carly Burton}, \bibinfo{person}{Catalina Mejia}, \bibinfo{person}{Ce Liu}, \bibinfo{person}{Changhan Wang}, \bibinfo{person}{Changkyu Kim}, \bibinfo{person}{Chao Zhou}, \bibinfo{person}{Chester Hu}, \bibinfo{person}{Ching-Hsiang Chu}, \bibinfo{person}{Chris Cai}, \bibinfo{person}{Chris Tindal}, \bibinfo{person}{Christoph Feichtenhofer}, \bibinfo{person}{Cynthia Gao}, \bibinfo{person}{Damon Civin}, \bibinfo{person}{Dana Beaty}, \bibinfo{person}{Daniel Kreymer}, \bibinfo{person}{Daniel Li}, \bibinfo{person}{David Adkins}, \bibinfo{person}{David Xu}, \bibinfo{person}{Davide Testuggine},
  \bibinfo{person}{Delia David}, \bibinfo{person}{Devi Parikh}, \bibinfo{person}{Diana Liskovich}, \bibinfo{person}{Didem Foss}, \bibinfo{person}{Dingkang Wang}, \bibinfo{person}{Duc Le}, \bibinfo{person}{Dustin Holland}, \bibinfo{person}{Edward Dowling}, \bibinfo{person}{Eissa Jamil}, \bibinfo{person}{Elaine Montgomery}, \bibinfo{person}{Eleonora Presani}, \bibinfo{person}{Emily Hahn}, \bibinfo{person}{Emily Wood}, \bibinfo{person}{Eric-Tuan Le}, \bibinfo{person}{Erik Brinkman}, \bibinfo{person}{Esteban Arcaute}, \bibinfo{person}{Evan Dunbar}, \bibinfo{person}{Evan Smothers}, \bibinfo{person}{Fei Sun}, \bibinfo{person}{Felix Kreuk}, \bibinfo{person}{Feng Tian}, \bibinfo{person}{Filippos Kokkinos}, \bibinfo{person}{Firat Ozgenel}, \bibinfo{person}{Francesco Caggioni}, \bibinfo{person}{Frank Kanayet}, \bibinfo{person}{Frank Seide}, \bibinfo{person}{Gabriela~Medina Florez}, \bibinfo{person}{Gabriella Schwarz}, \bibinfo{person}{Gada Badeer}, \bibinfo{person}{Georgia Swee}, \bibinfo{person}{Gil Halpern},
  \bibinfo{person}{Grant Herman}, \bibinfo{person}{Grigory Sizov}, \bibinfo{person}{Guangyi}, \bibinfo{person}{Zhang}, \bibinfo{person}{Guna Lakshminarayanan}, \bibinfo{person}{Hakan Inan}, \bibinfo{person}{Hamid Shojanazeri}, \bibinfo{person}{Han Zou}, \bibinfo{person}{Hannah Wang}, \bibinfo{person}{Hanwen Zha}, \bibinfo{person}{Haroun Habeeb}, \bibinfo{person}{Harrison Rudolph}, \bibinfo{person}{Helen Suk}, \bibinfo{person}{Henry Aspegren}, \bibinfo{person}{Hunter Goldman}, \bibinfo{person}{Hongyuan Zhan}, \bibinfo{person}{Ibrahim Damlaj}, \bibinfo{person}{Igor Molybog}, \bibinfo{person}{Igor Tufanov}, \bibinfo{person}{Ilias Leontiadis}, \bibinfo{person}{Irina-Elena Veliche}, \bibinfo{person}{Itai Gat}, \bibinfo{person}{Jake Weissman}, \bibinfo{person}{James Geboski}, \bibinfo{person}{James Kohli}, \bibinfo{person}{Janice Lam}, \bibinfo{person}{Japhet Asher}, \bibinfo{person}{Jean-Baptiste Gaya}, \bibinfo{person}{Jeff Marcus}, \bibinfo{person}{Jeff Tang}, \bibinfo{person}{Jennifer Chan},
  \bibinfo{person}{Jenny Zhen}, \bibinfo{person}{Jeremy Reizenstein}, \bibinfo{person}{Jeremy Teboul}, \bibinfo{person}{Jessica Zhong}, \bibinfo{person}{Jian Jin}, \bibinfo{person}{Jingyi Yang}, \bibinfo{person}{Joe Cummings}, \bibinfo{person}{Jon Carvill}, \bibinfo{person}{Jon Shepard}, \bibinfo{person}{Jonathan McPhie}, \bibinfo{person}{Jonathan Torres}, \bibinfo{person}{Josh Ginsburg}, \bibinfo{person}{Junjie Wang}, \bibinfo{person}{Kai Wu}, \bibinfo{person}{Kam~Hou U}, \bibinfo{person}{Karan Saxena}, \bibinfo{person}{Kartikay Khandelwal}, \bibinfo{person}{Katayoun Zand}, \bibinfo{person}{Kathy Matosich}, \bibinfo{person}{Kaushik Veeraraghavan}, \bibinfo{person}{Kelly Michelena}, \bibinfo{person}{Keqian Li}, \bibinfo{person}{Kiran Jagadeesh}, \bibinfo{person}{Kun Huang}, \bibinfo{person}{Kunal Chawla}, \bibinfo{person}{Kyle Huang}, \bibinfo{person}{Lailin Chen}, \bibinfo{person}{Lakshya Garg}, \bibinfo{person}{Lavender A}, \bibinfo{person}{Leandro Silva}, \bibinfo{person}{Lee Bell}, \bibinfo{person}{Lei
  Zhang}, \bibinfo{person}{Liangpeng Guo}, \bibinfo{person}{Licheng Yu}, \bibinfo{person}{Liron Moshkovich}, \bibinfo{person}{Luca Wehrstedt}, \bibinfo{person}{Madian Khabsa}, \bibinfo{person}{Manav Avalani}, \bibinfo{person}{Manish Bhatt}, \bibinfo{person}{Martynas Mankus}, \bibinfo{person}{Matan Hasson}, \bibinfo{person}{Matthew Lennie}, \bibinfo{person}{Matthias Reso}, \bibinfo{person}{Maxim Groshev}, \bibinfo{person}{Maxim Naumov}, \bibinfo{person}{Maya Lathi}, \bibinfo{person}{Meghan Keneally}, \bibinfo{person}{Miao Liu}, \bibinfo{person}{Michael~L. Seltzer}, \bibinfo{person}{Michal Valko}, \bibinfo{person}{Michelle Restrepo}, \bibinfo{person}{Mihir Patel}, \bibinfo{person}{Mik Vyatskov}, \bibinfo{person}{Mikayel Samvelyan}, \bibinfo{person}{Mike Clark}, \bibinfo{person}{Mike Macey}, \bibinfo{person}{Mike Wang}, \bibinfo{person}{Miquel~Jubert Hermoso}, \bibinfo{person}{Mo Metanat}, \bibinfo{person}{Mohammad Rastegari}, \bibinfo{person}{Munish Bansal}, \bibinfo{person}{Nandhini Santhanam},
  \bibinfo{person}{Natascha Parks}, \bibinfo{person}{Natasha White}, \bibinfo{person}{Navyata Bawa}, \bibinfo{person}{Nayan Singhal}, \bibinfo{person}{Nick Egebo}, \bibinfo{person}{Nicolas Usunier}, \bibinfo{person}{Nikhil Mehta}, \bibinfo{person}{Nikolay~Pavlovich Laptev}, \bibinfo{person}{Ning Dong}, \bibinfo{person}{Norman Cheng}, \bibinfo{person}{Oleg Chernoguz}, \bibinfo{person}{Olivia Hart}, \bibinfo{person}{Omkar Salpekar}, \bibinfo{person}{Ozlem Kalinli}, \bibinfo{person}{Parkin Kent}, \bibinfo{person}{Parth Parekh}, \bibinfo{person}{Paul Saab}, \bibinfo{person}{Pavan Balaji}, \bibinfo{person}{Pedro Rittner}, \bibinfo{person}{Philip Bontrager}, \bibinfo{person}{Pierre Roux}, \bibinfo{person}{Piotr Dollar}, \bibinfo{person}{Polina Zvyagina}, \bibinfo{person}{Prashant Ratanchandani}, \bibinfo{person}{Pritish Yuvraj}, \bibinfo{person}{Qian Liang}, \bibinfo{person}{Rachad Alao}, \bibinfo{person}{Rachel Rodriguez}, \bibinfo{person}{Rafi Ayub}, \bibinfo{person}{Raghotham Murthy}, \bibinfo{person}{Raghu
  Nayani}, \bibinfo{person}{Rahul Mitra}, \bibinfo{person}{Rangaprabhu Parthasarathy}, \bibinfo{person}{Raymond Li}, \bibinfo{person}{Rebekkah Hogan}, \bibinfo{person}{Robin Battey}, \bibinfo{person}{Rocky Wang}, \bibinfo{person}{Russ Howes}, \bibinfo{person}{Ruty Rinott}, \bibinfo{person}{Sachin Mehta}, \bibinfo{person}{Sachin Siby}, \bibinfo{person}{Sai~Jayesh Bondu}, \bibinfo{person}{Samyak Datta}, \bibinfo{person}{Sara Chugh}, \bibinfo{person}{Sara Hunt}, \bibinfo{person}{Sargun Dhillon}, \bibinfo{person}{Sasha Sidorov}, \bibinfo{person}{Satadru Pan}, \bibinfo{person}{Saurabh Mahajan}, \bibinfo{person}{Saurabh Verma}, \bibinfo{person}{Seiji Yamamoto}, \bibinfo{person}{Sharadh Ramaswamy}, \bibinfo{person}{Shaun Lindsay}, \bibinfo{person}{Shaun Lindsay}, \bibinfo{person}{Sheng Feng}, \bibinfo{person}{Shenghao Lin}, \bibinfo{person}{Shengxin~Cindy Zha}, \bibinfo{person}{Shishir Patil}, \bibinfo{person}{Shiva Shankar}, \bibinfo{person}{Shuqiang Zhang}, \bibinfo{person}{Shuqiang Zhang}, \bibinfo{person}{Sinong
  Wang}, \bibinfo{person}{Sneha Agarwal}, \bibinfo{person}{Soji Sajuyigbe}, \bibinfo{person}{Soumith Chintala}, \bibinfo{person}{Stephanie Max}, \bibinfo{person}{Stephen Chen}, \bibinfo{person}{Steve Kehoe}, \bibinfo{person}{Steve Satterfield}, \bibinfo{person}{Sudarshan Govindaprasad}, \bibinfo{person}{Sumit Gupta}, \bibinfo{person}{Summer Deng}, \bibinfo{person}{Sungmin Cho}, \bibinfo{person}{Sunny Virk}, \bibinfo{person}{Suraj Subramanian}, \bibinfo{person}{Sy Choudhury}, \bibinfo{person}{Sydney Goldman}, \bibinfo{person}{Tal Remez}, \bibinfo{person}{Tamar Glaser}, \bibinfo{person}{Tamara Best}, \bibinfo{person}{Thilo Koehler}, \bibinfo{person}{Thomas Robinson}, \bibinfo{person}{Tianhe Li}, \bibinfo{person}{Tianjun Zhang}, \bibinfo{person}{Tim Matthews}, \bibinfo{person}{Timothy Chou}, \bibinfo{person}{Tzook Shaked}, \bibinfo{person}{Varun Vontimitta}, \bibinfo{person}{Victoria Ajayi}, \bibinfo{person}{Victoria Montanez}, \bibinfo{person}{Vijai Mohan}, \bibinfo{person}{Vinay~Satish Kumar},
  \bibinfo{person}{Vishal Mangla}, \bibinfo{person}{Vlad Ionescu}, \bibinfo{person}{Vlad Poenaru}, \bibinfo{person}{Vlad~Tiberiu Mihailescu}, \bibinfo{person}{Vladimir Ivanov}, \bibinfo{person}{Wei Li}, \bibinfo{person}{Wenchen Wang}, \bibinfo{person}{Wenwen Jiang}, \bibinfo{person}{Wes Bouaziz}, \bibinfo{person}{Will Constable}, \bibinfo{person}{Xiaocheng Tang}, \bibinfo{person}{Xiaojian Wu}, \bibinfo{person}{Xiaolan Wang}, \bibinfo{person}{Xilun Wu}, \bibinfo{person}{Xinbo Gao}, \bibinfo{person}{Yaniv Kleinman}, \bibinfo{person}{Yanjun Chen}, \bibinfo{person}{Ye Hu}, \bibinfo{person}{Ye Jia}, \bibinfo{person}{Ye Qi}, \bibinfo{person}{Yenda Li}, \bibinfo{person}{Yilin Zhang}, \bibinfo{person}{Ying Zhang}, \bibinfo{person}{Yossi Adi}, \bibinfo{person}{Youngjin Nam}, \bibinfo{person}{Yu}, \bibinfo{person}{Wang}, \bibinfo{person}{Yu Zhao}, \bibinfo{person}{Yuchen Hao}, \bibinfo{person}{Yundi Qian}, \bibinfo{person}{Yunlu Li}, \bibinfo{person}{Yuzi He}, \bibinfo{person}{Zach Rait}, \bibinfo{person}{Zachary
  DeVito}, \bibinfo{person}{Zef Rosnbrick}, \bibinfo{person}{Zhaoduo Wen}, \bibinfo{person}{Zhenyu Yang}, \bibinfo{person}{Zhiwei Zhao}, {and} \bibinfo{person}{Zhiyu Ma}.} \bibinfo{year}{2024}\natexlab{}.
\newblock \bibinfo{title}{The Llama 3 Herd of Models}.
\newblock
\newblock
\showeprint[arxiv]{2407.21783}~[cs.AI]
\urldef\tempurl%
\url{https://arxiv.org/abs/2407.21783}
\showURL{%
\tempurl}


\bibitem[Han et~al\mbox{.}(2023)]%
        {han2023big}
\bibfield{author}{\bibinfo{person}{Yinhe Han}, \bibinfo{person}{Haobo Xu}, \bibinfo{person}{Meixuan Lu}, \bibinfo{person}{Haoran Wang}, \bibinfo{person}{Junpei Huang}, \bibinfo{person}{Ying Wang}, \bibinfo{person}{Yujie Wang}, \bibinfo{person}{Feng Min}, \bibinfo{person}{Qi Liu}, \bibinfo{person}{Ming Liu}, {and} \bibinfo{person}{Ninghui Sun}.} \bibinfo{year}{2023}\natexlab{}.
\newblock \showarticletitle{The Big Chip: Challenge, Model and Architecture}.
\newblock \bibinfo{journal}{\emph{Fundamental Research}} (\bibinfo{year}{2023}).
\newblock


\bibitem[Hao et~al\mbox{.}(2023)]%
        {hao2023monad}
\bibfield{author}{\bibinfo{person}{Xiaochen Hao}, \bibinfo{person}{Zijian Ding}, \bibinfo{person}{Jieming Yin}, \bibinfo{person}{Yuan Wang}, {and} \bibinfo{person}{Yun Liang}.} \bibinfo{year}{2023}\natexlab{}.
\newblock \showarticletitle{Monad: Towards Cost-Effective Specialization for Chiplet-Based Spatial Accelerators}. In \bibinfo{booktitle}{\emph{2023 IEEE/ACM International Conference on Computer Aided Design (ICCAD)}}. IEEE, \bibinfo{pages}{1--9}.
\newblock


\bibitem[He et~al\mbox{.}(2022)]%
        {he2022design}
\bibfield{author}{\bibinfo{person}{Kang He}, \bibinfo{person}{Indranil Chakraborty}, \bibinfo{person}{Cheng Wang}, {and} \bibinfo{person}{Kaushik Roy}.} \bibinfo{year}{2022}\natexlab{}.
\newblock \showarticletitle{Design Space and Memory Technology Co-Exploration for In-Memory Computing Based Machine Learning Accelerators}. In \bibinfo{booktitle}{\emph{Proceedings of the 41st IEEE/ACM International Conference on Computer-Aided Design}} (San Diego, California) \emph{(\bibinfo{series}{ICCAD '22})}. \bibinfo{publisher}{Association for Computing Machinery}, \bibinfo{address}{New York, NY, USA}, Article \bibinfo{articleno}{91}, \bibinfo{numpages}{9}~pages.
\newblock
\showISBNx{9781450392174}
\urldef\tempurl%
\url{https://doi.org/10.1145/3508352.3549453}
\showDOI{\tempurl}


\bibitem[Heo et~al\mbox{.}(2024)]%
        {heo2024neupims}
\bibfield{author}{\bibinfo{person}{Guseul Heo}, \bibinfo{person}{Sangyeop Lee}, \bibinfo{person}{Jaehong Cho}, \bibinfo{person}{Hyunmin Choi}, \bibinfo{person}{Sanghyeon Lee}, \bibinfo{person}{Hyungkyu Ham}, \bibinfo{person}{Gwangsun Kim}, \bibinfo{person}{Divya Mahajan}, {and} \bibinfo{person}{Jongse Park}.} \bibinfo{year}{2024}\natexlab{}.
\newblock \showarticletitle{Neupims: Npu-pim heterogeneous acceleration for batched llm inferencing}. In \bibinfo{booktitle}{\emph{Proceedings of the 29th ACM International Conference on Architectural Support for Programming Languages and Operating Systems, Volume 3}}. \bibinfo{pages}{722--737}.
\newblock


\bibitem[Herbst et~al\mbox{.}(2024)]%
        {herbst2024switchboard}
\bibfield{author}{\bibinfo{person}{Steven Herbst}, \bibinfo{person}{Noah Moroze}, \bibinfo{person}{Edgar Iglesias}, {and} \bibinfo{person}{Andreas Olofsson}.} \bibinfo{year}{2024}\natexlab{}.
\newblock \showarticletitle{Switchboard: An Open-Source Framework for Modular Simulation of Large Hardware Systems}.
\newblock \bibinfo{journal}{\emph{arXiv preprint arXiv:2407.20537}} (\bibinfo{year}{2024}).
\newblock


\bibitem[Huang et~al\mbox{.}(2023)]%
        {huang2023hierarch}
\bibfield{author}{\bibinfo{person}{Jyun-Siou Huang}, \bibinfo{person}{Ting-Han Chou}, \bibinfo{person}{Juin-Ming Lu}, \bibinfo{person}{Chih-Tsun Huang}, {and} \bibinfo{person}{Jing-Jia Liou}.} \bibinfo{year}{2023}\natexlab{}.
\newblock \showarticletitle{HierArch: A Cluster-Based DNN Accelerator with Hierarchical Buses for Design Space Exploration}. In \bibinfo{booktitle}{\emph{2023 IEEE 36th International System-on-Chip Conference (SOCC)}}. IEEE, \bibinfo{pages}{1--6}.
\newblock


\bibitem[Huang et~al\mbox{.}(2024)]%
        {huang2024hecaton}
\bibfield{author}{\bibinfo{person}{Zongle Huang}, \bibinfo{person}{Shupei Fan}, \bibinfo{person}{Chen Tang}, \bibinfo{person}{Xinyuan Lin}, \bibinfo{person}{Shuwen Deng}, {and} \bibinfo{person}{Yongpan Liu}.} \bibinfo{year}{2024}\natexlab{}.
\newblock \showarticletitle{Hecaton: Training and Finetuning Large Language Models with Scalable Chiplet Systems}.
\newblock \bibinfo{journal}{\emph{arXiv preprint arXiv:2407.05784}} (\bibinfo{year}{2024}).
\newblock


\bibitem[Hwang et~al\mbox{.}(2020)]%
        {hwang2020centaur}
\bibfield{author}{\bibinfo{person}{Ranggi Hwang}, \bibinfo{person}{Taehun Kim}, \bibinfo{person}{Youngeun Kwon}, {and} \bibinfo{person}{Minsoo Rhu}.} \bibinfo{year}{2020}\natexlab{}.
\newblock \showarticletitle{Centaur: A chiplet-based, hybrid sparse-dense accelerator for personalized recommendations}. In \bibinfo{booktitle}{\emph{2020 ACM/IEEE 47th Annual International Symposium on Computer Architecture (ISCA)}}. IEEE, \bibinfo{pages}{968--981}.
\newblock


\bibitem[Iff et~al\mbox{.}(2023)]%
        {iff2023rapidchiplet}
\bibfield{author}{\bibinfo{person}{Patrick Iff}, \bibinfo{person}{Benigna Bruggmann}, \bibinfo{person}{Maciej Besta}, \bibinfo{person}{Luca Benini}, {and} \bibinfo{person}{Torsten Hoefler}.} \bibinfo{year}{2023}\natexlab{}.
\newblock \showarticletitle{RapidChiplet: A Toolchain for Rapid Design Space Exploration of Chiplet Architectures}.
\newblock \bibinfo{journal}{\emph{arXiv preprint arXiv:2311.06081}} (\bibinfo{year}{2023}).
\newblock


\bibitem[Isaev et~al\mbox{.}(2023)]%
        {isaev2023calculon}
\bibfield{author}{\bibinfo{person}{Mikhail Isaev}, \bibinfo{person}{Nic McDonald}, \bibinfo{person}{Larry Dennison}, {and} \bibinfo{person}{Richard Vuduc}.} \bibinfo{year}{2023}\natexlab{}.
\newblock \showarticletitle{Calculon: a methodology and tool for high-level co-design of systems and large language models}. In \bibinfo{booktitle}{\emph{Proceedings of the International Conference for High Performance Computing, Networking, Storage and Analysis}}. \bibinfo{pages}{1--14}.
\newblock


\bibitem[Jeong et~al\mbox{.}(2021)]%
        {jeong2021union}
\bibfield{author}{\bibinfo{person}{Geonhwa Jeong}, \bibinfo{person}{Gokcen Kestor}, \bibinfo{person}{Prasanth Chatarasi}, \bibinfo{person}{Angshuman Parashar}, \bibinfo{person}{Po-An Tsai}, \bibinfo{person}{Sivasankaran Rajamanickam}, \bibinfo{person}{Roberto Gioiosa}, {and} \bibinfo{person}{Tushar Krishna}.} \bibinfo{year}{2021}\natexlab{}.
\newblock \showarticletitle{Union: A unified HW-SW co-design ecosystem in MLIR for evaluating tensor operations on spatial accelerators}. In \bibinfo{booktitle}{\emph{2021 30th International Conference on Parallel Architectures and Compilation Techniques (PACT)}}. IEEE, \bibinfo{pages}{30--44}.
\newblock


\bibitem[Jia et~al\mbox{.}(2021)]%
        {jia2021tensorlib}
\bibfield{author}{\bibinfo{person}{Liancheng Jia}, \bibinfo{person}{Zizhang Luo}, \bibinfo{person}{Liqiang Lu}, {and} \bibinfo{person}{Yun Liang}.} \bibinfo{year}{2021}\natexlab{}.
\newblock \showarticletitle{TensorLib: A Spatial Accelerator Generation Framework for Tensor Algebra}. In \bibinfo{booktitle}{\emph{2021 58th ACM/IEEE Design Automation Conference (DAC)}}. \bibinfo{pages}{865--870}.
\newblock
\urldef\tempurl%
\url{https://doi.org/10.1109/DAC18074.2021.9586329}
\showDOI{\tempurl}


\bibitem[Jia et~al\mbox{.}(2019)]%
        {jia2019dissecting}
\bibfield{author}{\bibinfo{person}{Zhe Jia}, \bibinfo{person}{Blake Tillman}, \bibinfo{person}{Marco Maggioni}, {and} \bibinfo{person}{Daniele~Paolo Scarpazza}.} \bibinfo{year}{2019}\natexlab{}.
\newblock \showarticletitle{Dissecting the graphcore ipu architecture via microbenchmarking}.
\newblock \bibinfo{journal}{\emph{arXiv preprint arXiv:1912.03413}} (\bibinfo{year}{2019}).
\newblock


\bibitem[Juracy et~al\mbox{.}(2021)]%
        {juracy2021high}
\bibfield{author}{\bibinfo{person}{Leonardo~Rezende Juracy}, \bibinfo{person}{Matheus~Trevisan Moreira}, \bibinfo{person}{Alexandre de Morais~Amory}, \bibinfo{person}{Alexandre~F. Hampel}, {and} \bibinfo{person}{Fernando~Gehm Moraes}.} \bibinfo{year}{2021}\natexlab{}.
\newblock \showarticletitle{A High-Level Modeling Framework for Estimating Hardware Metrics of CNN Accelerators}.
\newblock \bibinfo{journal}{\emph{IEEE Transactions on Circuits and Systems I: Regular Papers}} \bibinfo{volume}{68}, \bibinfo{number}{11} (\bibinfo{year}{2021}), \bibinfo{pages}{4783--4795}.
\newblock
\urldef\tempurl%
\url{https://doi.org/10.1109/TCSI.2021.3104644}
\showDOI{\tempurl}


\bibitem[Kao et~al\mbox{.}(2022)]%
        {kao2022digamma}
\bibfield{author}{\bibinfo{person}{Sheng-Chun Kao}, \bibinfo{person}{Michael Pellauer}, \bibinfo{person}{Angshuman Parashar}, {and} \bibinfo{person}{Tushar Krishna}.} \bibinfo{year}{2022}\natexlab{}.
\newblock \showarticletitle{Digamma: Domain-aware genetic algorithm for hw-mapping co-optimization for dnn accelerators}. In \bibinfo{booktitle}{\emph{2022 Design, Automation \& Test in Europe Conference \& Exhibition (DATE)}}. IEEE, \bibinfo{pages}{232--237}.
\newblock


\bibitem[Kaplan et~al\mbox{.}(2020)]%
        {kaplan2020scaling}
\bibfield{author}{\bibinfo{person}{Jared Kaplan}, \bibinfo{person}{Sam McCandlish}, \bibinfo{person}{Tom Henighan}, \bibinfo{person}{Tom~B Brown}, \bibinfo{person}{Benjamin Chess}, \bibinfo{person}{Rewon Child}, \bibinfo{person}{Scott Gray}, \bibinfo{person}{Alec Radford}, \bibinfo{person}{Jeffrey Wu}, {and} \bibinfo{person}{Dario Amodei}.} \bibinfo{year}{2020}\natexlab{}.
\newblock \showarticletitle{Scaling laws for neural language models}.
\newblock \bibinfo{journal}{\emph{arXiv preprint arXiv:2001.08361}} (\bibinfo{year}{2020}).
\newblock


\bibitem[Kim et~al\mbox{.}(2019)]%
        {kim2019architecture}
\bibfield{author}{\bibinfo{person}{Jinwoo Kim}, \bibinfo{person}{Gauthaman Murali}, \bibinfo{person}{Heechun Park}, \bibinfo{person}{Eric Qin}, \bibinfo{person}{Hyoukjun Kwon}, \bibinfo{person}{Venkata Chaitanya}, \bibinfo{person}{Krishna Chekuri}, \bibinfo{person}{Nihar Dasari}, \bibinfo{person}{Arvind Singh}, \bibinfo{person}{Minah Lee}, \bibinfo{person}{Hakki~Mert Torun}, \bibinfo{person}{Kallol Roy}, \bibinfo{person}{Madhavan Swaminathan}, \bibinfo{person}{Saibal Mukhopadhyay}, \bibinfo{person}{Tushar Krishna}, {and} \bibinfo{person}{Sung~Kyu Lim}.} \bibinfo{year}{2019}\natexlab{}.
\newblock \showarticletitle{Architecture, Chip, and Package Co-design Flow for 2.5D IC Design Enabling Heterogeneous IP Reuse}. In \bibinfo{booktitle}{\emph{Proceedings of the 56th Annual Design Automation Conference 2019}} (Las Vegas, NV, USA) \emph{(\bibinfo{series}{DAC '19})}. \bibinfo{publisher}{Association for Computing Machinery}, \bibinfo{address}{New York, NY, USA}, Article \bibinfo{articleno}{178}, \bibinfo{numpages}{6}~pages.
\newblock
\showISBNx{9781450367257}
\urldef\tempurl%
\url{https://doi.org/10.1145/3316781.3317775}
\showDOI{\tempurl}


\bibitem[Knowles(2021)]%
        {knowles2021graphcore}
\bibfield{author}{\bibinfo{person}{Simon Knowles}.} \bibinfo{year}{2021}\natexlab{}.
\newblock \showarticletitle{Graphcore}. In \bibinfo{booktitle}{\emph{2021 IEEE Hot Chips 33 Symposium (HCS)}}. IEEE, \bibinfo{pages}{1--25}.
\newblock


\bibitem[Krishnan et~al\mbox{.}(2021)]%
        {krishnan2021siam}
\bibfield{author}{\bibinfo{person}{Gokul Krishnan}, \bibinfo{person}{Sumit~K. Mandal}, \bibinfo{person}{Manvitha Pannala}, \bibinfo{person}{Chaitali Chakrabarti}, \bibinfo{person}{Jae-Sun Seo}, \bibinfo{person}{Umit~Y. Ogras}, {and} \bibinfo{person}{Yu Cao}.} \bibinfo{year}{2021}\natexlab{}.
\newblock \showarticletitle{SIAM: Chiplet-based Scalable In-Memory Acceleration with Mesh for Deep Neural Networks}.
\newblock \bibinfo{journal}{\emph{ACM Trans. Embed. Comput. Syst.}} \bibinfo{volume}{20}, \bibinfo{number}{5s}, Article \bibinfo{articleno}{68} (\bibinfo{date}{Sept.} \bibinfo{year}{2021}), \bibinfo{numpages}{24}~pages.
\newblock
\showISSN{1539-9087}
\urldef\tempurl%
\url{https://doi.org/10.1145/3476999}
\showDOI{\tempurl}


\bibitem[Kundu et~al\mbox{.}(2024)]%
        {kundu2024performance}
\bibfield{author}{\bibinfo{person}{Joyjit Kundu}, \bibinfo{person}{Wenzhe Guo}, \bibinfo{person}{Ali BanaGozar}, \bibinfo{person}{Udari De~Alwis}, \bibinfo{person}{Sourav Sengupta}, \bibinfo{person}{Puneet Gupta}, {and} \bibinfo{person}{Arindam Mallik}.} \bibinfo{year}{2024}\natexlab{}.
\newblock \showarticletitle{Performance Modeling and Workload Analysis of Distributed Large Language Model Training and Inference}.
\newblock \bibinfo{journal}{\emph{arXiv preprint arXiv:2407.14645}} (\bibinfo{year}{2024}).
\newblock


\bibitem[Kwon et~al\mbox{.}(2020)]%
        {kwon2020maestro}
\bibfield{author}{\bibinfo{person}{Hyoukjun Kwon}, \bibinfo{person}{Prasanth Chatarasi}, \bibinfo{person}{Vivek Sarkar}, \bibinfo{person}{Tushar Krishna}, \bibinfo{person}{Michael Pellauer}, {and} \bibinfo{person}{Angshuman Parashar}.} \bibinfo{year}{2020}\natexlab{}.
\newblock \showarticletitle{MAESTRO: A Data-Centric Approach to Understand Reuse, Performance, and Hardware Cost of DNN Mappings}.
\newblock \bibinfo{journal}{\emph{IEEE Micro}} \bibinfo{volume}{40}, \bibinfo{number}{3} (\bibinfo{year}{2020}), \bibinfo{pages}{20--29}.
\newblock
\urldef\tempurl%
\url{https://doi.org/10.1109/MM.2020.2985963}
\showDOI{\tempurl}


\bibitem[Kwon et~al\mbox{.}(2023)]%
        {kwon2023efficient}
\bibfield{author}{\bibinfo{person}{Woosuk Kwon}, \bibinfo{person}{Zhuohan Li}, \bibinfo{person}{Siyuan Zhuang}, \bibinfo{person}{Ying Sheng}, \bibinfo{person}{Lianmin Zheng}, \bibinfo{person}{Cody~Hao Yu}, \bibinfo{person}{Joseph Gonzalez}, \bibinfo{person}{Hao Zhang}, {and} \bibinfo{person}{Ion Stoica}.} \bibinfo{year}{2023}\natexlab{}.
\newblock \showarticletitle{Efficient memory management for large language model serving with pagedattention}. In \bibinfo{booktitle}{\emph{Proceedings of the 29th Symposium on Operating Systems Principles}}. \bibinfo{pages}{611--626}.
\newblock


\bibitem[Lattner et~al\mbox{.}(2021)]%
        {lattner2021mlir}
\bibfield{author}{\bibinfo{person}{Chris Lattner}, \bibinfo{person}{Mehdi Amini}, \bibinfo{person}{Uday Bondhugula}, \bibinfo{person}{Albert Cohen}, \bibinfo{person}{Andy Davis}, \bibinfo{person}{Jacques Pienaar}, \bibinfo{person}{River Riddle}, \bibinfo{person}{Tatiana Shpeisman}, \bibinfo{person}{Nicolas Vasilache}, {and} \bibinfo{person}{Oleksandr Zinenko}.} \bibinfo{year}{2021}\natexlab{}.
\newblock \showarticletitle{MLIR: Scaling compiler infrastructure for domain specific computation}. In \bibinfo{booktitle}{\emph{2021 IEEE/ACM International Symposium on Code Generation and Optimization (CGO)}}. IEEE, \bibinfo{pages}{2--14}.
\newblock


\bibitem[Lee et~al\mbox{.}(2024a)]%
        {lee2024cost}
\bibfield{author}{\bibinfo{person}{Hyungdeok Lee}, \bibinfo{person}{Guhyun Kim}, \bibinfo{person}{Dayeon Yun}, \bibinfo{person}{Ilkon Kim}, \bibinfo{person}{Yongkee Kwon}, {and} \bibinfo{person}{Euicheol Lim}.} \bibinfo{year}{2024}\natexlab{a}.
\newblock \showarticletitle{Cost-effective llm accelerator using processing in memory technology}. In \bibinfo{booktitle}{\emph{2024 IEEE Symposium on VLSI Technology and Circuits (VLSI Technology and Circuits)}}. IEEE, \bibinfo{pages}{1--2}.
\newblock


\bibitem[Lee et~al\mbox{.}(2024b)]%
        {lee2024tender}
\bibfield{author}{\bibinfo{person}{Jungi Lee}, \bibinfo{person}{Wonbeom Lee}, {and} \bibinfo{person}{Jaewoong Sim}.} \bibinfo{year}{2024}\natexlab{b}.
\newblock \showarticletitle{Tender: Accelerating Large Language Models via Tensor Decomposition and Runtime Requantization}.
\newblock \bibinfo{journal}{\emph{arXiv preprint arXiv:2406.12930}} (\bibinfo{year}{2024}).
\newblock


\bibitem[Lee et~al\mbox{.}(2024c)]%
        {lee2024infinigen}
\bibfield{author}{\bibinfo{person}{Wonbeom Lee}, \bibinfo{person}{Jungi Lee}, \bibinfo{person}{Junghwan Seo}, {and} \bibinfo{person}{Jaewoong Sim}.} \bibinfo{year}{2024}\natexlab{c}.
\newblock \showarticletitle{InfiniGen: Efficient generative inference of large language models with dynamic $\{$KV$\}$ cache management}. In \bibinfo{booktitle}{\emph{18th USENIX Symposium on Operating Systems Design and Implementation (OSDI 24)}}. \bibinfo{pages}{155--172}.
\newblock


\bibitem[Leviathan et~al\mbox{.}(2023)]%
        {krause2023fast}
\bibfield{author}{\bibinfo{person}{Yaniv Leviathan}, \bibinfo{person}{Matan Kalman}, {and} \bibinfo{person}{Yossi Matias}.} \bibinfo{year}{2023}\natexlab{}.
\newblock \showarticletitle{Fast Inference from Transformers via Speculative Decoding}. In \bibinfo{booktitle}{\emph{Proceedings of the 40th International Conference on Machine Learning}} \emph{(\bibinfo{series}{Proceedings of Machine Learning Research}, Vol.~\bibinfo{volume}{202})}, \bibfield{editor}{\bibinfo{person}{Andreas Krause}, \bibinfo{person}{Emma Brunskill}, \bibinfo{person}{Kyunghyun Cho}, \bibinfo{person}{Barbara Engelhardt}, \bibinfo{person}{Sivan Sabato}, {and} \bibinfo{person}{Jonathan Scarlett}} (Eds.). \bibinfo{publisher}{PMLR}, \bibinfo{pages}{19274--19286}.
\newblock
\urldef\tempurl%
\url{https://proceedings.mlr.press/v202/leviathan23a.html}
\showURL{%
\tempurl}


\bibitem[Li et~al\mbox{.}(2022)]%
        {li2022spacx}
\bibfield{author}{\bibinfo{person}{Yuan Li}, \bibinfo{person}{Ahmed Louri}, {and} \bibinfo{person}{Avinash Karanth}.} \bibinfo{year}{2022}\natexlab{}.
\newblock \showarticletitle{SPACX: Silicon photonics-based scalable chiplet accelerator for DNN inference}. In \bibinfo{booktitle}{\emph{2022 IEEE International Symposium on High-Performance Computer Architecture (HPCA)}}. IEEE, \bibinfo{pages}{831--845}.
\newblock


\bibitem[Liao et~al\mbox{.}(2021)]%
        {liao2021ascend}
\bibfield{author}{\bibinfo{person}{Heng Liao}, \bibinfo{person}{Jiajin Tu}, \bibinfo{person}{Jing Xia}, \bibinfo{person}{Hu Liu}, \bibinfo{person}{Xiping Zhou}, \bibinfo{person}{Honghui Yuan}, {and} \bibinfo{person}{Yuxing Hu}.} \bibinfo{year}{2021}\natexlab{}.
\newblock \showarticletitle{Ascend: a scalable and unified architecture for ubiquitous deep neural network computing: Industry track paper}. In \bibinfo{booktitle}{\emph{2021 IEEE International Symposium on High-Performance Computer Architecture (HPCA)}}. IEEE, \bibinfo{pages}{789--801}.
\newblock


\bibitem[Lie(2021)]%
        {lie2021multi}
\bibfield{author}{\bibinfo{person}{Sean Lie}.} \bibinfo{year}{2021}\natexlab{}.
\newblock \showarticletitle{Multi-Million Core, Multi-Wafer AI Cluster.}. In \bibinfo{booktitle}{\emph{HCS}}. \bibinfo{pages}{1--41}.
\newblock


\bibitem[Lie(2022)]%
        {lie2022cerebras}
\bibfield{author}{\bibinfo{person}{Sean Lie}.} \bibinfo{year}{2022}\natexlab{}.
\newblock \showarticletitle{Cerebras Architecture Deep Dive: First Look Inside the HW/SW Co-Design for Deep Learning : Cerebras Systems}. In \bibinfo{booktitle}{\emph{2022 IEEE Hot Chips 34 Symposium (HCS)}}. \bibinfo{pages}{1--34}.
\newblock
\urldef\tempurl%
\url{https://doi.org/10.1109/HCS55958.2022.9895479}
\showDOI{\tempurl}


\bibitem[Lin et~al\mbox{.}(2023)]%
        {lin2023songc}
\bibfield{author}{\bibinfo{person}{Junfeng Lin}, \bibinfo{person}{Huanyu Qu}, \bibinfo{person}{Songchen Ma}, \bibinfo{person}{Xinglong Ji}, \bibinfo{person}{Hongyi Li}, \bibinfo{person}{Xiaochuan Li}, \bibinfo{person}{Chenhang Song}, {and} \bibinfo{person}{Weihao Zhang}.} \bibinfo{year}{2023}\natexlab{}.
\newblock \showarticletitle{SongC: A Compiler for Hybrid Near-Memory and In-Memory Many-Core Architecture}.
\newblock \bibinfo{journal}{\emph{IEEE Trans. Comput.}} (\bibinfo{year}{2023}).
\newblock


\bibitem[Lin et~al\mbox{.}(2024c)]%
        {lin2024hex}
\bibfield{author}{\bibinfo{person}{Xinquan Lin}, \bibinfo{person}{Haobo Xu}, \bibinfo{person}{Yinhe Han}, {and} \bibinfo{person}{Yiming Gan}.} \bibinfo{year}{2024}\natexlab{c}.
\newblock \showarticletitle{HEX-SIM: Evaluating Multi-modal Large Language Models on Multi-chiplet NPUs}. In \bibinfo{booktitle}{\emph{2024 IEEE International Symposium on Workload Characterization (IISWC)}}. IEEE, \bibinfo{pages}{108--120}.
\newblock


\bibitem[Lin et~al\mbox{.}(2024a)]%
        {lin2024tessel}
\bibfield{author}{\bibinfo{person}{Zhiqi Lin}, \bibinfo{person}{Youshan Miao}, \bibinfo{person}{Guanbin Xu}, \bibinfo{person}{Cheng Li}, \bibinfo{person}{Olli Saarikivi}, \bibinfo{person}{Saeed Maleki}, {and} \bibinfo{person}{Fan Yang}.} \bibinfo{year}{2024}\natexlab{a}.
\newblock \showarticletitle{Tessel: Boosting Distributed Execution of Large DNN Models via Flexible Schedule Search}. In \bibinfo{booktitle}{\emph{2024 IEEE International Symposium on High-Performance Computer Architecture (HPCA)}}. IEEE, \bibinfo{pages}{803--816}.
\newblock


\bibitem[Lin et~al\mbox{.}(2024b)]%
        {lin2024nnscaler}
\bibfield{author}{\bibinfo{person}{Zhiqi Lin}, \bibinfo{person}{Youshan Miao}, \bibinfo{person}{Quanlu Zhang}, \bibinfo{person}{Fan Yang}, \bibinfo{person}{Yi Zhu}, \bibinfo{person}{Cheng Li}, \bibinfo{person}{Saeed Maleki}, \bibinfo{person}{Xu Cao}, \bibinfo{person}{Ning Shang}, \bibinfo{person}{Yilei Yang}, \bibinfo{person}{Weijiang Xu}, \bibinfo{person}{Mao Yang}, \bibinfo{person}{Lintao Zhang}, {and} \bibinfo{person}{Lidong Zhou}.} \bibinfo{year}{2024}\natexlab{b}.
\newblock \showarticletitle{{nnScaler}: {Constraint-Guided} Parallelization Plan Generation for Deep Learning Training}. In \bibinfo{booktitle}{\emph{18th USENIX Symposium on Operating Systems Design and Implementation (OSDI 24)}}. \bibinfo{publisher}{USENIX Association}, \bibinfo{address}{Santa Clara, CA}, \bibinfo{pages}{347--363}.
\newblock
\showISBNx{978-1-939133-40-3}
\urldef\tempurl%
\url{https://www.usenix.org/conference/osdi24/presentation/lin-zhiqi}
\showURL{%
\tempurl}


\bibitem[Luo et~al\mbox{.}(2024)]%
        {luo2024benchmarking}
\bibfield{author}{\bibinfo{person}{Weile Luo}, \bibinfo{person}{Ruibo Fan}, \bibinfo{person}{Zeyu Li}, \bibinfo{person}{Dayou Du}, \bibinfo{person}{Qiang Wang}, {and} \bibinfo{person}{Xiaowen Chu}.} \bibinfo{year}{2024}\natexlab{}.
\newblock \showarticletitle{Benchmarking and dissecting the nvidia hopper gpu architecture}.
\newblock \bibinfo{journal}{\emph{arXiv preprint arXiv:2402.13499}} (\bibinfo{year}{2024}).
\newblock


\bibitem[Ma et~al\mbox{.}(2022)]%
        {ma2022neuromorphic}
\bibfield{author}{\bibinfo{person}{Songchen Ma}, \bibinfo{person}{Jing Pei}, \bibinfo{person}{Weihao Zhang}, \bibinfo{person}{Guanrui Wang}, \bibinfo{person}{Dahu Feng}, \bibinfo{person}{Fangwen Yu}, \bibinfo{person}{Chenhang Song}, \bibinfo{person}{Huanyu Qu}, \bibinfo{person}{Cheng Ma}, \bibinfo{person}{Mingsheng Lu}, \bibinfo{person}{Faqiang Liu}, \bibinfo{person}{Zhou Wenhao}, \bibinfo{person}{Wu Yujie}, \bibinfo{person}{Lin Yihan}, \bibinfo{person}{Li Hongyi}, \bibinfo{person}{Wang Taoyi}, \bibinfo{person}{Song Jiuru}, \bibinfo{person}{Liu Xue}, \bibinfo{person}{Li Guoqi}, \bibinfo{person}{Zhao Rong}, {and} \bibinfo{person}{Shi Luping}.} \bibinfo{year}{2022}\natexlab{}.
\newblock \showarticletitle{Neuromorphic computing chip with spatiotemporal elasticity for multi-intelligent-tasking robots}.
\newblock \bibinfo{journal}{\emph{Science Robotics}} \bibinfo{volume}{7}, \bibinfo{number}{67} (\bibinfo{year}{2022}), \bibinfo{pages}{eabk2948}.
\newblock


\bibitem[Matthews et~al\mbox{.}(2020)]%
        {matthews2020mosaicsim}
\bibfield{author}{\bibinfo{person}{Opeoluwa Matthews}, \bibinfo{person}{Aninda Manocha}, \bibinfo{person}{Davide Giri}, \bibinfo{person}{Marcelo Orenes-Vera}, \bibinfo{person}{Esin Tureci}, \bibinfo{person}{Tyler Sorensen}, \bibinfo{person}{Tae~Jun Ham}, \bibinfo{person}{Juan~L. Aragon}, \bibinfo{person}{Luca~P. Carloni}, {and} \bibinfo{person}{Margaret Martonosi}.} \bibinfo{year}{2020}\natexlab{}.
\newblock \showarticletitle{MosaicSim: A Lightweight, Modular Simulator for Heterogeneous Systems}. In \bibinfo{booktitle}{\emph{2020 IEEE International Symposium on Performance Analysis of Systems and Software (ISPASS)}}. \bibinfo{pages}{136--148}.
\newblock
\urldef\tempurl%
\url{https://doi.org/10.1109/ISPASS48437.2020.00029}
\showDOI{\tempurl}


\bibitem[Mei et~al\mbox{.}(2021)]%
        {mei2021zigzag}
\bibfield{author}{\bibinfo{person}{Linyan Mei}, \bibinfo{person}{Pouya Houshmand}, \bibinfo{person}{Vikram Jain}, \bibinfo{person}{Sebastian Giraldo}, {and} \bibinfo{person}{Marian Verhelst}.} \bibinfo{year}{2021}\natexlab{}.
\newblock \showarticletitle{ZigZag: Enlarging joint architecture-mapping design space exploration for DNN accelerators}.
\newblock \bibinfo{journal}{\emph{IEEE Trans. Comput.}} \bibinfo{volume}{70}, \bibinfo{number}{8} (\bibinfo{year}{2021}), \bibinfo{pages}{1160--1174}.
\newblock


\bibitem[Modha et~al\mbox{.}(2023)]%
        {modha2023neural}
\bibfield{author}{\bibinfo{person}{Dharmendra~S. Modha}, \bibinfo{person}{Filipp Akopyan}, \bibinfo{person}{Alexander Andreopoulos}, \bibinfo{person}{Rathinakumar Appuswamy}, \bibinfo{person}{John~V. Arthur}, \bibinfo{person}{Andrew~S. Cassidy}, \bibinfo{person}{Pallab Datta}, \bibinfo{person}{Michael~V. DeBole}, \bibinfo{person}{Steven~K. Esser}, \bibinfo{person}{Carlos~Ortega Otero}, \bibinfo{person}{Jun Sawada}, \bibinfo{person}{Brian Taba}, \bibinfo{person}{Arnon Amir}, \bibinfo{person}{Deepika Bablani}, \bibinfo{person}{Peter~J. Carlson}, \bibinfo{person}{Myron~D. Flickner}, \bibinfo{person}{Rajamohan Gandhasri}, \bibinfo{person}{Guillaume~J. Garreau}, \bibinfo{person}{Megumi Ito}, \bibinfo{person}{Jennifer~L. Klamo}, \bibinfo{person}{Jeffrey~A. Kusnitz}, \bibinfo{person}{Nathaniel~J. McClatchey}, \bibinfo{person}{Jeffrey~L. McKinstry}, \bibinfo{person}{Yutaka Nakamura}, \bibinfo{person}{Tapan~K. Nayak}, \bibinfo{person}{William~P. Risk}, \bibinfo{person}{Kai Schleupen}, \bibinfo{person}{Ben Shaw},
  \bibinfo{person}{Jay Sivagnaname}, \bibinfo{person}{Daniel~F. Smith}, \bibinfo{person}{Ignacio Terrizzano}, {and} \bibinfo{person}{Takanori Ueda}.} \bibinfo{year}{2023}\natexlab{}.
\newblock \showarticletitle{Neural inference at the frontier of energy, space, and time}.
\newblock \bibinfo{journal}{\emph{Science}} \bibinfo{volume}{382}, \bibinfo{number}{6668} (\bibinfo{year}{2023}), \bibinfo{pages}{329--335}.
\newblock


\bibitem[Murali et~al\mbox{.}(2024)]%
        {murali20243dnn}
\bibfield{author}{\bibinfo{person}{Gauthaman Murali}, \bibinfo{person}{Min~Gyu Park}, {and} \bibinfo{person}{Sung~Kyu Lim}.} \bibinfo{year}{2024}\natexlab{}.
\newblock \showarticletitle{3DNN-Xplorer: A Machine Learning Framework for Design Space Exploration of Heterogeneous 3-D DNN Accelerators}.
\newblock \bibinfo{journal}{\emph{IEEE Transactions on Very Large Scale Integration (VLSI) Systems}} (\bibinfo{year}{2024}).
\newblock


\bibitem[NVIDIA(2017a)]%
        {nvidia2017volta}
\bibfield{author}{\bibinfo{person}{NVIDIA}.} \bibinfo{year}{2017}\natexlab{a}.
\newblock \bibinfo{title}{NVIDIA TESLA V100 GPU Architecture}.
\newblock \bibinfo{howpublished}{\url{https://images.nvidia.com/content/volta-architecture/pdf/volta-architecture-whitepaper.pdf}}.
\newblock


\bibitem[NVIDIA(2017b)]%
        {nvidia2018turing}
\bibfield{author}{\bibinfo{person}{NVIDIA}.} \bibinfo{year}{2017}\natexlab{b}.
\newblock \bibinfo{title}{NVIDIA Turing GPU Architecture}.
\newblock \bibinfo{howpublished}{\url{https://images.nvidia.com/aem-dam/en-zz/Solutions/design-visualization/technologies/turing-architecture/NVIDIA-Turing-Architecture-Whitepaper.pdf}}.
\newblock


\bibitem[NVIDIA(2022a)]%
        {nvidia2020ampere}
\bibfield{author}{\bibinfo{person}{NVIDIA}.} \bibinfo{year}{2022}\natexlab{a}.
\newblock \bibinfo{title}{NVIDIA A100 Tensor Core GPU Architecture}.
\newblock \bibinfo{howpublished}{\url{https://images.nvidia.com/aem-dam/en-zz/Solutions/data-center/nvidia-ampere-architecture-whitepaper.pdf}}.
\newblock


\bibitem[NVIDIA(2022b)]%
        {nvidia2022hopper}
\bibfield{author}{\bibinfo{person}{NVIDIA}.} \bibinfo{year}{2022}\natexlab{b}.
\newblock \bibinfo{title}{NVIDIA Hopper Architecture In-Depth}.
\newblock \bibinfo{howpublished}{\url{https://developer.nvidia.com/blog/nvidia-hopper-architecture-in-depth}}.
\newblock


\bibitem[NVIDIA(2023)]%
        {nvidia2023nccl}
\bibfield{author}{\bibinfo{person}{NVIDIA}.} \bibinfo{year}{2023}\natexlab{}.
\newblock \bibinfo{title}{Performance reported by NCCL tests}.
\newblock \bibinfo{howpublished}{\url{https://github.com/ NVIDIA/nccl-tests/blob/master/doc/PERFORMANCE.md}}.
\newblock


\bibitem[OpenAI(2022)]%
        {openai2022chatgpt}
\bibfield{author}{\bibinfo{person}{OpenAI}.} \bibinfo{year}{2022}\natexlab{}.
\newblock \bibinfo{title}{Introducing ChatGPT}.
\newblock \bibinfo{howpublished}{\url{https://openai.com/blog/chatgpt}}.
\newblock


\bibitem[OpenAI(2023)(2023)]%
        {achiam2023gpt}
\bibfield{author}{\bibinfo{person}{OpenAI(2023)}.} \bibinfo{year}{2023}\natexlab{}.
\newblock \showarticletitle{Gpt-4 technical report}.
\newblock \bibinfo{journal}{\emph{arXiv preprint arXiv:2303.08774}} (\bibinfo{year}{2023}).
\newblock


\bibitem[Orenes-Vera et~al\mbox{.}(2024)]%
        {orenes2024muchisim}
\bibfield{author}{\bibinfo{person}{Marcelo Orenes-Vera}, \bibinfo{person}{Esin Tureci}, \bibinfo{person}{Margaret Martonosi}, {and} \bibinfo{person}{David Wentzlaff}.} \bibinfo{year}{2024}\natexlab{}.
\newblock \showarticletitle{Muchisim: A simulation framework for design exploration of multi-chip manycore systems}. In \bibinfo{booktitle}{\emph{2024 IEEE International Symposium on Performance Analysis of Systems and Software (ISPASS)}}. IEEE, \bibinfo{pages}{48--60}.
\newblock


\bibitem[Pal et~al\mbox{.}(2020)]%
        {pal2020design}
\bibfield{author}{\bibinfo{person}{Saptadeep Pal}, \bibinfo{person}{Daniel Petrisko}, \bibinfo{person}{Rakesh Kumar}, {and} \bibinfo{person}{Puneet Gupta}.} \bibinfo{year}{2020}\natexlab{}.
\newblock \showarticletitle{Design space exploration for chiplet-assembly-based processors}.
\newblock \bibinfo{journal}{\emph{IEEE Transactions on Very Large Scale Integration (VLSI) Systems}} \bibinfo{volume}{28}, \bibinfo{number}{4} (\bibinfo{year}{2020}), \bibinfo{pages}{1062--1073}.
\newblock


\bibitem[Parashar et~al\mbox{.}(2019)]%
        {parashar2019timeloop}
\bibfield{author}{\bibinfo{person}{Angshuman Parashar}, \bibinfo{person}{Priyanka Raina}, \bibinfo{person}{Yakun~Sophia Shao}, \bibinfo{person}{Yu-Hsin Chen}, \bibinfo{person}{Victor~A Ying}, \bibinfo{person}{Anurag Mukkara}, \bibinfo{person}{Rangharajan Venkatesan}, \bibinfo{person}{Brucek Khailany}, \bibinfo{person}{Stephen~W Keckler}, {and} \bibinfo{person}{Joel Emer}.} \bibinfo{year}{2019}\natexlab{}.
\newblock \showarticletitle{Timeloop: A systematic approach to dnn accelerator evaluation}. In \bibinfo{booktitle}{\emph{2019 IEEE international symposium on performance analysis of systems and software (ISPASS)}}. IEEE, \bibinfo{pages}{304--315}.
\newblock


\bibitem[Pei et~al\mbox{.}(2019)]%
        {pei2019towards}
\bibfield{author}{\bibinfo{person}{Jing Pei}, \bibinfo{person}{Lei Deng}, \bibinfo{person}{Sen Song}, \bibinfo{person}{Mingguo Zhao}, \bibinfo{person}{Youhui Zhang}, \bibinfo{person}{Shuang Wu}, \bibinfo{person}{Guanrui Wang}, \bibinfo{person}{Zhe Zou}, \bibinfo{person}{Zhenzhi Wu}, \bibinfo{person}{Wei He}, \bibinfo{person}{Feng Chen}, \bibinfo{person}{Ning Deng}, \bibinfo{person}{Si Wu}, \bibinfo{person}{Yu Wang}, \bibinfo{person}{Yujie Wu}, \bibinfo{person}{Zheyu Yang}, \bibinfo{person}{Cheng Ma}, \bibinfo{person}{Guoqi Li}, \bibinfo{person}{Wentao Han}, \bibinfo{person}{Huanglong Li}, \bibinfo{person}{Huaqiang Wu}, \bibinfo{person}{Rong Zhao}, \bibinfo{person}{Yuan Xie}, {and} \bibinfo{person}{Luping Shi}.} \bibinfo{year}{2019}\natexlab{}.
\newblock \showarticletitle{Towards artificial general intelligence with hybrid Tianjic chip architecture}.
\newblock \bibinfo{journal}{\emph{Nature}} \bibinfo{volume}{572}, \bibinfo{number}{7767} (\bibinfo{year}{2019}), \bibinfo{pages}{106--111}.
\newblock


\bibitem[Peng et~al\mbox{.}(2023)]%
        {peng2023chiplet}
\bibfield{author}{\bibinfo{person}{Huwan Peng}, \bibinfo{person}{Scott Davidson}, \bibinfo{person}{Richard Shi}, \bibinfo{person}{Shuaiwen~Leon Song}, {and} \bibinfo{person}{Michael Taylor}.} \bibinfo{year}{2023}\natexlab{}.
\newblock \showarticletitle{Chiplet cloud: Building ai supercomputers for serving large generative language models}.
\newblock \bibinfo{journal}{\emph{arXiv preprint arXiv:2307.02666}} (\bibinfo{year}{2023}).
\newblock


\bibitem[Qi et~al\mbox{.}(2023)]%
        {qi2023moela}
\bibfield{author}{\bibinfo{person}{Sirui Qi}, \bibinfo{person}{Yingheng Li}, \bibinfo{person}{Sudeep Pasricha}, {and} \bibinfo{person}{Ryan~Gary Kim}.} \bibinfo{year}{2023}\natexlab{}.
\newblock \showarticletitle{MOELA: A Multi-Objective Evolutionary/Learning Design Space Exploration Framework for 3D Heterogeneous Manycore Platforms}. In \bibinfo{booktitle}{\emph{2023 Design, Automation \& Test in Europe Conference \& Exhibition (DATE)}}. \bibinfo{pages}{1--6}.
\newblock
\urldef\tempurl%
\url{https://doi.org/10.23919/DATE56975.2023.10137276}
\showDOI{\tempurl}


\bibitem[Qin et~al\mbox{.}(2024)]%
        {qin2024mecla}
\bibfield{author}{\bibinfo{person}{Yubin Qin}, \bibinfo{person}{Yang Wang}, \bibinfo{person}{Zhiren Zhao}, \bibinfo{person}{Xiaolong Yang}, \bibinfo{person}{Yang Zhou}, \bibinfo{person}{Shaojun Wei}, \bibinfo{person}{Yang Hu}, {and} \bibinfo{person}{Shouyi Yin}.} \bibinfo{year}{2024}\natexlab{}.
\newblock \showarticletitle{MECLA: Memory-Compute-Efficient LLM Accelerator with Scaling Sub-matrix Partition}. In \bibinfo{booktitle}{\emph{2024 ACM/IEEE 51st Annual International Symposium on Computer Architecture (ISCA)}}. \bibinfo{pages}{1032--1047}.
\newblock
\urldef\tempurl%
\url{https://doi.org/10.1109/ISCA59077.2024.00079}
\showDOI{\tempurl}


\bibitem[Rashidi et~al\mbox{.}(2023)]%
        {rashidi2023unico}
\bibfield{author}{\bibinfo{person}{Bahador Rashidi}, \bibinfo{person}{Chao Gao}, \bibinfo{person}{Shan Lu}, \bibinfo{person}{Zhisheng Wang}, \bibinfo{person}{Chunhua Zhou}, \bibinfo{person}{Di Niu}, {and} \bibinfo{person}{Fengyu Sun}.} \bibinfo{year}{2023}\natexlab{}.
\newblock \showarticletitle{UNICO: Unified Hardware Software Co-Optimization for Robust Neural Network Acceleration}. In \bibinfo{booktitle}{\emph{Proceedings of the 56th Annual IEEE/ACM International Symposium on Microarchitecture}}. \bibinfo{pages}{77--90}.
\newblock


\bibitem[Rashidi et~al\mbox{.}(2020)]%
        {rashidi2020astra}
\bibfield{author}{\bibinfo{person}{Saeed Rashidi}, \bibinfo{person}{Srinivas Sridharan}, \bibinfo{person}{Sudarshan Srinivasan}, {and} \bibinfo{person}{Tushar Krishna}.} \bibinfo{year}{2020}\natexlab{}.
\newblock \showarticletitle{Astra-sim: Enabling sw/hw co-design exploration for distributed dl training platforms}. In \bibinfo{booktitle}{\emph{2020 IEEE International Symposium on Performance Analysis of Systems and Software (ISPASS)}}. IEEE, \bibinfo{pages}{81--92}.
\newblock


\bibitem[Samajdar et~al\mbox{.}(2023)]%
        {samajdar2023airchitect}
\bibfield{author}{\bibinfo{person}{Ananda Samajdar}, \bibinfo{person}{Jan~Moritz Joseph}, {and} \bibinfo{person}{Tushar Krishna}.} \bibinfo{year}{2023}\natexlab{}.
\newblock \showarticletitle{AIrchitect: Automating Hardware Architecture and Mapping Optimization}. In \bibinfo{booktitle}{\emph{2023 Design, Automation \& Test in Europe Conference \& Exhibition (DATE)}}. \bibinfo{pages}{1--6}.
\newblock
\urldef\tempurl%
\url{https://doi.org/10.23919/DATE56975.2023.10137333}
\showDOI{\tempurl}


\bibitem[Samajdar et~al\mbox{.}(2018)]%
        {samajdar2018scale}
\bibfield{author}{\bibinfo{person}{Ananda Samajdar}, \bibinfo{person}{Yuhao Zhu}, \bibinfo{person}{Paul Whatmough}, \bibinfo{person}{Matthew Mattina}, {and} \bibinfo{person}{Tushar Krishna}.} \bibinfo{year}{2018}\natexlab{}.
\newblock \showarticletitle{Scale-sim: Systolic cnn accelerator simulator}.
\newblock \bibinfo{journal}{\emph{arXiv preprint arXiv:1811.02883}} (\bibinfo{year}{2018}).
\newblock


\bibitem[Shao et~al\mbox{.}(2019)]%
        {shao2019simba}
\bibfield{author}{\bibinfo{person}{Yakun~Sophia Shao}, \bibinfo{person}{Jason Clemons}, \bibinfo{person}{Rangharajan Venkatesan}, \bibinfo{person}{Brian Zimmer}, \bibinfo{person}{Matthew Fojtik}, \bibinfo{person}{Nan Jiang}, \bibinfo{person}{Ben Keller}, \bibinfo{person}{Alicia Klinefelter}, \bibinfo{person}{Nathaniel Pinckney}, \bibinfo{person}{Priyanka Raina}, \bibinfo{person}{Stephen~G. Tell}, \bibinfo{person}{Yanqing Zhang}, \bibinfo{person}{William~J. Dally}, \bibinfo{person}{Joel Emer}, \bibinfo{person}{C.~Thomas Gray}, \bibinfo{person}{Brucek Khailany}, {and} \bibinfo{person}{Stephen~W. Keckler}.} \bibinfo{year}{2019}\natexlab{}.
\newblock \showarticletitle{Simba: Scaling Deep-Learning Inference with Multi-Chip-Module-Based Architecture}. In \bibinfo{booktitle}{\emph{Proceedings of the 52nd Annual IEEE/ACM International Symposium on Microarchitecture}} (Columbus, OH, USA) \emph{(\bibinfo{series}{MICRO '52})}. \bibinfo{publisher}{Association for Computing Machinery}, \bibinfo{address}{New York, NY, USA}, \bibinfo{pages}{14–27}.
\newblock
\showISBNx{9781450369381}
\urldef\tempurl%
\url{https://doi.org/10.1145/3352460.3358302}
\showDOI{\tempurl}


\bibitem[Sharma et~al\mbox{.}(2023)]%
        {sharma2023heterogeneous}
\bibfield{author}{\bibinfo{person}{Harsh Sharma}, \bibinfo{person}{Pratyush Dhingra}, \bibinfo{person}{Janardhan~Rao Doppa}, \bibinfo{person}{Umit Ogras}, {and} \bibinfo{person}{Partha~Pratim Pande}.} \bibinfo{year}{2023}\natexlab{}.
\newblock \showarticletitle{A Heterogeneous Chiplet Architecture for Accelerating End-to-End Transformer Models}.
\newblock \bibinfo{journal}{\emph{arXiv preprint arXiv:2312.11750}} (\bibinfo{year}{2023}).
\newblock


\bibitem[Shi et~al\mbox{.}(2023)]%
        {shi2023welder}
\bibfield{author}{\bibinfo{person}{Yining Shi}, \bibinfo{person}{Zhi Yang}, \bibinfo{person}{Jilong Xue}, \bibinfo{person}{Lingxiao Ma}, \bibinfo{person}{Yuqing Xia}, \bibinfo{person}{Ziming Miao}, \bibinfo{person}{Yuxiao Guo}, \bibinfo{person}{Fan Yang}, {and} \bibinfo{person}{Lidong Zhou}.} \bibinfo{year}{2023}\natexlab{}.
\newblock \showarticletitle{Welder: Scheduling deep learning memory access via tile-graph}. In \bibinfo{booktitle}{\emph{17th USENIX Symposium on Operating Systems Design and Implementation (OSDI 23)}}. \bibinfo{pages}{701--718}.
\newblock


\bibitem[Singh et~al\mbox{.}(2013)]%
        {singh2013mapping}
\bibfield{author}{\bibinfo{person}{Amit~Kumar Singh}, \bibinfo{person}{Muhammad Shafique}, \bibinfo{person}{Akash Kumar}, {and} \bibinfo{person}{J{\"o}rg Henkel}.} \bibinfo{year}{2013}\natexlab{}.
\newblock \showarticletitle{Mapping on multi/many-core systems: Survey of current and emerging trends}. In \bibinfo{booktitle}{\emph{2013 50th ACM/EDAC/IEEE Design Automation Conference (DAC)}}. IEEE, \bibinfo{pages}{1--10}.
\newblock


\bibitem[Smith et~al\mbox{.}(2024)]%
        {smith202411}
\bibfield{author}{\bibinfo{person}{Alan Smith}, \bibinfo{person}{Eric Chapman}, \bibinfo{person}{Chintan Patel}, \bibinfo{person}{Raja Swaminathan}, \bibinfo{person}{John Wuu}, \bibinfo{person}{Tyrone Huang}, \bibinfo{person}{Wonjun Jung}, \bibinfo{person}{Alexander Kaganov}, \bibinfo{person}{Hugh McIntyre}, {and} \bibinfo{person}{Ramon Mangaser}.} \bibinfo{year}{2024}\natexlab{}.
\newblock \showarticletitle{11.1 AMD InstinctTM MI300 Series Modular Chiplet Package--HPC and AI Accelerator for Exa-Class Systems}. In \bibinfo{booktitle}{\emph{2024 IEEE International Solid-State Circuits Conference (ISSCC)}}, Vol.~\bibinfo{volume}{67}. IEEE, \bibinfo{pages}{490--492}.
\newblock


\bibitem[Talpes et~al\mbox{.}(2023)]%
        {talpes2023microarchitecture}
\bibfield{author}{\bibinfo{person}{Emil Talpes}, \bibinfo{person}{Debjit~Das Sarma}, \bibinfo{person}{Doug Williams}, \bibinfo{person}{Sahil Arora}, \bibinfo{person}{Thomas Kunjan}, \bibinfo{person}{Benjamin Floering}, \bibinfo{person}{Ankit Jalote}, \bibinfo{person}{Christopher Hsiong}, \bibinfo{person}{Chandrasekhar Poorna}, \bibinfo{person}{Vaidehi Samant}, \bibinfo{person}{John Sicilia}, \bibinfo{person}{Anantha~Kumar Nivarti}, \bibinfo{person}{Raghuvir Ramachandran}, \bibinfo{person}{Tim Fischer}, \bibinfo{person}{Ben Herzberg}, \bibinfo{person}{Bill McGee}, \bibinfo{person}{Ganesh Venkataramanan}, {and} \bibinfo{person}{Pete Banon}.} \bibinfo{year}{2023}\natexlab{}.
\newblock \showarticletitle{The microarchitecture of dojo, tesla’s exa-scale computer}.
\newblock \bibinfo{journal}{\emph{IEEE Micro}} (\bibinfo{year}{2023}).
\newblock


\bibitem[Talpes et~al\mbox{.}(2022)]%
        {talpes2022dojo}
\bibfield{author}{\bibinfo{person}{Emil Talpes}, \bibinfo{person}{Douglas Williams}, {and} \bibinfo{person}{Debjit~Das Sarma}.} \bibinfo{year}{2022}\natexlab{}.
\newblock \showarticletitle{Dojo: The microarchitecture of tesla’s exa-scale computer}. In \bibinfo{booktitle}{\emph{2022 IEEE Hot Chips 34 Symposium (HCS)}}. IEEE Computer Society, \bibinfo{pages}{1--28}.
\newblock


\bibitem[Tan et~al\mbox{.}(2021)]%
        {tan2021nn}
\bibfield{author}{\bibinfo{person}{Zhanhong Tan}, \bibinfo{person}{Hongyu Cai}, \bibinfo{person}{Runpei Dong}, {and} \bibinfo{person}{Kaisheng Ma}.} \bibinfo{year}{2021}\natexlab{}.
\newblock \showarticletitle{NN-baton: DNN workload orchestration and chiplet granularity exploration for multichip accelerators}. In \bibinfo{booktitle}{\emph{2021 ACM/IEEE 48th Annual International Symposium on Computer Architecture (ISCA)}}. IEEE, \bibinfo{pages}{1013--1026}.
\newblock


\bibitem[Tan et~al\mbox{.}(2024)]%
        {tan2024cocco}
\bibfield{author}{\bibinfo{person}{Zhanhong Tan}, \bibinfo{person}{Zijian Zhu}, {and} \bibinfo{person}{Kaisheng Ma}.} \bibinfo{year}{2024}\natexlab{}.
\newblock \showarticletitle{Cocco: Hardware-Mapping Co-Exploration towards Memory Capacity-Communication Optimization}. In \bibinfo{booktitle}{\emph{Proceedings of the 29th ACM International Conference on Architectural Support for Programming Languages and Operating Systems, Volume 1}}. \bibinfo{pages}{69--84}.
\newblock


\bibitem[Tirumala and Wong(2024)]%
        {tirumala2024nvidia}
\bibfield{author}{\bibinfo{person}{Ajay Tirumala} {and} \bibinfo{person}{Raymond Wong}.} \bibinfo{year}{2024}\natexlab{}.
\newblock \showarticletitle{NVIDIA Blackwell Platform: Advancing Generative AI and Accelerated Computing}. In \bibinfo{booktitle}{\emph{2024 IEEE Hot Chips 36 Symposium (HCS)}}. IEEE Computer Society, \bibinfo{pages}{1--33}.
\newblock


\bibitem[Touvron et~al\mbox{.}(2023)]%
        {touvron2023llama2}
\bibfield{author}{\bibinfo{person}{Hugo Touvron}, \bibinfo{person}{Louis Martin}, \bibinfo{person}{Kevin Stone}, \bibinfo{person}{Peter Albert}, \bibinfo{person}{Amjad Almahairi}, \bibinfo{person}{Yasmine Babaei}, \bibinfo{person}{Nikolay Bashlykov}, \bibinfo{person}{Soumya Batra}, \bibinfo{person}{Prajjwal Bhargava}, \bibinfo{person}{Shruti Bhosale}, \bibinfo{person}{Dan Bikel}, \bibinfo{person}{Lukas Blecher}, \bibinfo{person}{Cristian~Canton Ferrer}, \bibinfo{person}{Moya Chen}, \bibinfo{person}{Guillem Cucurull}, \bibinfo{person}{David Esiobu}, \bibinfo{person}{Jude Fernandes}, \bibinfo{person}{Jeremy Fu}, \bibinfo{person}{Wenyin Fu}, \bibinfo{person}{Brian Fuller}, \bibinfo{person}{Cynthia Gao}, \bibinfo{person}{Vedanuj Goswami}, \bibinfo{person}{Naman Goyal}, \bibinfo{person}{Anthony Hartshorn}, \bibinfo{person}{Saghar Hosseini}, \bibinfo{person}{Rui Hou}, \bibinfo{person}{Hakan Inan}, \bibinfo{person}{Marcin Kardas}, \bibinfo{person}{Viktor Kerkez}, \bibinfo{person}{Madian Khabsa},
  \bibinfo{person}{Isabel Kloumann}, \bibinfo{person}{Artem Korenev}, \bibinfo{person}{Punit~Singh Koura}, \bibinfo{person}{Marie-Anne Lachaux}, \bibinfo{person}{Thibaut Lavril}, \bibinfo{person}{Jenya Lee}, \bibinfo{person}{Diana Liskovich}, \bibinfo{person}{Yinghai Lu}, \bibinfo{person}{Yuning Mao}, \bibinfo{person}{Xavier Martinet}, \bibinfo{person}{Todor Mihaylov}, \bibinfo{person}{Pushkar Mishra}, \bibinfo{person}{Igor Molybog}, \bibinfo{person}{Yixin Nie}, \bibinfo{person}{Andrew Poulton}, \bibinfo{person}{Jeremy Reizenstein}, \bibinfo{person}{Rashi Rungta}, \bibinfo{person}{Kalyan Saladi}, \bibinfo{person}{Alan Schelten}, \bibinfo{person}{Ruan Silva}, \bibinfo{person}{Eric~Michael Smith}, \bibinfo{person}{Ranjan Subramanian}, \bibinfo{person}{Xiaoqing~Ellen Tan}, \bibinfo{person}{Binh Tang}, \bibinfo{person}{Ross Taylor}, \bibinfo{person}{Adina Williams}, \bibinfo{person}{Jian~Xiang Kuan}, \bibinfo{person}{Puxin Xu}, \bibinfo{person}{Zheng Yan}, \bibinfo{person}{Iliyan Zarov}, \bibinfo{person}{Yuchen
  Zhang}, \bibinfo{person}{Angela Fan}, \bibinfo{person}{Melanie Kambadur}, \bibinfo{person}{Sharan Narang}, \bibinfo{person}{Aurelien Rodriguez}, \bibinfo{person}{Robert Stojnic}, \bibinfo{person}{Sergey Edunov}, {and} \bibinfo{person}{Thomas Scialom}.} \bibinfo{year}{2023}\natexlab{}.
\newblock \showarticletitle{Llama 2: Open Foundation and Fine-Tuned Chat Models}.
\newblock  (\bibinfo{year}{2023}).
\newblock
\showeprint[arxiv]{2307.09288}~[cs.CL]
\urldef\tempurl%
\url{https://arxiv.org/abs/2307.09288}
\showURL{%
\tempurl}


\bibitem[Vasiljevic et~al\mbox{.}(2021)]%
        {vasiljevic2021compute}
\bibfield{author}{\bibinfo{person}{Jasmina Vasiljevic}, \bibinfo{person}{Ljubisa Bajic}, \bibinfo{person}{Davor Capalija}, \bibinfo{person}{Stanislav Sokorac}, \bibinfo{person}{Dragoljub Ignjatovic}, \bibinfo{person}{Lejla Bajic}, \bibinfo{person}{Milos Trajkovic}, \bibinfo{person}{Ivan Hamer}, \bibinfo{person}{Ivan Matosevic}, \bibinfo{person}{Aleksandar Cejkov}, \bibinfo{person}{Utku Aydonat}, \bibinfo{person}{Tony Zhou}, \bibinfo{person}{Syed~Zohaib Gilani}, \bibinfo{person}{Armond Paiva}, \bibinfo{person}{Joseph Chu}, \bibinfo{person}{Djordje Maksimovic}, \bibinfo{person}{Stephen~Alexander Chin}, \bibinfo{person}{Zahi Moudallal}, \bibinfo{person}{Akhmed Rakhmati}, \bibinfo{person}{Sean Nijjar}, \bibinfo{person}{Almeet Bhullar}, \bibinfo{person}{Boris Drazic}, \bibinfo{person}{Charles Lee}, \bibinfo{person}{James Sun}, \bibinfo{person}{Kei-Ming Kwong}, \bibinfo{person}{James Connolly}, \bibinfo{person}{Miles Dooley}, \bibinfo{person}{Hassan Farooq}, \bibinfo{person}{Joy Yu~Ting Chen},
  \bibinfo{person}{Matthew Walker}, \bibinfo{person}{Keivan Dabiri}, \bibinfo{person}{Kyle Mabee}, \bibinfo{person}{Rakesh~Shaji Lal}, \bibinfo{person}{Namal Rajatheva}, \bibinfo{person}{Renjith Retnamma}, \bibinfo{person}{Shripad Karodi}, \bibinfo{person}{Daniel Rosen}, \bibinfo{person}{Emilio Munoz}, \bibinfo{person}{Andrew Lewycky}, \bibinfo{person}{Aleksandar Knezevic}, \bibinfo{person}{Raymond Kim}, \bibinfo{person}{Allan Rui}, \bibinfo{person}{Alexander Drouillard}, {and} \bibinfo{person}{David Thompson}.} \bibinfo{year}{2021}\natexlab{}.
\newblock \showarticletitle{Compute substrate for software 2.0}.
\newblock \bibinfo{journal}{\emph{IEEE micro}} \bibinfo{volume}{41}, \bibinfo{number}{2} (\bibinfo{year}{2021}), \bibinfo{pages}{50--55}.
\newblock


\bibitem[Wang et~al\mbox{.}(2024)]%
        {wang2024tmac}
\bibfield{author}{\bibinfo{person}{Huizheng Wang}, \bibinfo{person}{Qize Yang}, \bibinfo{person}{Taiquan Wei}, \bibinfo{person}{Xingmao Yu}, \bibinfo{person}{Chengran Li}, \bibinfo{person}{Jiahao Fang}, \bibinfo{person}{Guangyang Lu}, \bibinfo{person}{Xu Dai}, \bibinfo{person}{Liang Liu}, \bibinfo{person}{Shenfei Jiang}, \bibinfo{person}{Yang Hu}, \bibinfo{person}{Shouyi Yin}, {and} \bibinfo{person}{Shaojun Wei}.} \bibinfo{year}{2024}\natexlab{}.
\newblock \showarticletitle{TMAC: Training-Targeted Mapping and Architecture Co-Exploration for Wafer-Scale Chips}.
\newblock \bibinfo{journal}{\emph{Integrated Circuits and Systems}} \bibinfo{volume}{1}, \bibinfo{number}{4} (\bibinfo{year}{2024}), \bibinfo{pages}{178--195}.
\newblock
\urldef\tempurl%
\url{https://doi.org/10.23919/ICS.2024.3515003}
\showDOI{\tempurl}


\bibitem[Wang et~al\mbox{.}(2023)]%
        {wang2023nicepim}
\bibfield{author}{\bibinfo{person}{Junpeng Wang}, \bibinfo{person}{Mengke Ge}, \bibinfo{person}{Bo Ding}, \bibinfo{person}{Qi Xu}, \bibinfo{person}{Song Chen}, {and} \bibinfo{person}{Yi Kang}.} \bibinfo{year}{2023}\natexlab{}.
\newblock \showarticletitle{Nicepim: Design space exploration for processing-in-memory dnn accelerators with 3d-stacked-dram}.
\newblock \bibinfo{journal}{\emph{IEEE Transactions on Computer-Aided Design of Integrated Circuits and Systems}} (\bibinfo{year}{2023}).
\newblock


\bibitem[Williams et~al\mbox{.}(2009)]%
        {williams2009roofline}
\bibfield{author}{\bibinfo{person}{Samuel Williams}, \bibinfo{person}{Andrew Waterman}, {and} \bibinfo{person}{David Patterson}.} \bibinfo{year}{2009}\natexlab{}.
\newblock \showarticletitle{Roofline: an insightful visual performance model for multicore architectures}.
\newblock \bibinfo{journal}{\emph{Commun. ACM}} \bibinfo{volume}{52}, \bibinfo{number}{4} (\bibinfo{year}{2009}), \bibinfo{pages}{65--76}.
\newblock


\bibitem[Won et~al\mbox{.}(2023)]%
        {won2023astra}
\bibfield{author}{\bibinfo{person}{William Won}, \bibinfo{person}{Taekyung Heo}, \bibinfo{person}{Saeed Rashidi}, \bibinfo{person}{Srinivas Sridharan}, \bibinfo{person}{Sudarshan Srinivasan}, {and} \bibinfo{person}{Tushar Krishna}.} \bibinfo{year}{2023}\natexlab{}.
\newblock \showarticletitle{Astra-sim2. 0: Modeling hierarchical networks and disaggregated systems for large-model training at scale}. In \bibinfo{booktitle}{\emph{2023 IEEE International Symposium on Performance Analysis of Systems and Software (ISPASS)}}. IEEE, \bibinfo{pages}{283--294}.
\newblock


\bibitem[Wu et~al\mbox{.}(2019)]%
        {wu2019accelergy}
\bibfield{author}{\bibinfo{person}{Yannan~Nellie Wu}, \bibinfo{person}{Joel~S Emer}, {and} \bibinfo{person}{Vivienne Sze}.} \bibinfo{year}{2019}\natexlab{}.
\newblock \showarticletitle{Accelergy: An architecture-level energy estimation methodology for accelerator designs}. In \bibinfo{booktitle}{\emph{2019 IEEE/ACM International Conference on Computer-Aided Design (ICCAD)}}. IEEE, \bibinfo{pages}{1--8}.
\newblock


\bibitem[Yang et~al\mbox{.}(2020)]%
        {yang2020co}
\bibfield{author}{\bibinfo{person}{Lei Yang}, \bibinfo{person}{Zheyu Yan}, \bibinfo{person}{Meng Li}, \bibinfo{person}{Hyoukjun Kwon}, \bibinfo{person}{Liangzhen Lai}, \bibinfo{person}{Tushar Krishna}, \bibinfo{person}{Vikas Chandra}, \bibinfo{person}{Weiwen Jiang}, {and} \bibinfo{person}{Yiyu Shi}.} \bibinfo{year}{2020}\natexlab{}.
\newblock \showarticletitle{Co-exploration of neural architectures and heterogeneous asic accelerator designs targeting multiple tasks}. In \bibinfo{booktitle}{\emph{2020 57th ACM/IEEE Design Automation Conference (DAC)}}. IEEE, \bibinfo{pages}{1--6}.
\newblock


\bibitem[Zhang et~al\mbox{.}(2024)]%
        {zhang2024llmcompass}
\bibfield{author}{\bibinfo{person}{Hengrui Zhang}, \bibinfo{person}{August Ning}, \bibinfo{person}{Rohan~Baskar Prabhakar}, {and} \bibinfo{person}{David Wentzlaff}.} \bibinfo{year}{2024}\natexlab{}.
\newblock \showarticletitle{LLMCompass: Enabling Efficient Hardware Design for Large Language Model Inference}. In \bibinfo{booktitle}{\emph{2024 ACM/IEEE 51st Annual International Symposium on Computer Architecture (ISCA)}}. IEEE, \bibinfo{pages}{1080--1096}.
\newblock


\bibitem[Zhang et~al\mbox{.}(2023)]%
        {zhang2023indm}
\bibfield{author}{\bibinfo{person}{Jinming Zhang}, \bibinfo{person}{Xi Fan}, \bibinfo{person}{Yaoyao Ye}, \bibinfo{person}{Xuyan Wang}, \bibinfo{person}{Guojie Xiong}, \bibinfo{person}{Xianglun Leng}, \bibinfo{person}{Ningyi Xu}, \bibinfo{person}{Yong Lian}, {and} \bibinfo{person}{Guanghui He}.} \bibinfo{year}{2023}\natexlab{}.
\newblock \showarticletitle{INDM: Chiplet-Based Interconnect Network and Dataflow Mapping for DNN Accelerators}.
\newblock \bibinfo{journal}{\emph{IEEE Transactions on Computer-Aided Design of Integrated Circuits and Systems}} (\bibinfo{year}{2023}).
\newblock


\bibitem[Zheng et~al\mbox{.}(2022)]%
        {zheng2022alpa}
\bibfield{author}{\bibinfo{person}{Lianmin Zheng}, \bibinfo{person}{Zhuohan Li}, \bibinfo{person}{Hao Zhang}, \bibinfo{person}{Yonghao Zhuang}, \bibinfo{person}{Zhifeng Chen}, \bibinfo{person}{Yanping Huang}, \bibinfo{person}{Yida Wang}, \bibinfo{person}{Yuanzhong Xu}, \bibinfo{person}{Danyang Zhuo}, \bibinfo{person}{Eric~P. Xing}, \bibinfo{person}{Joseph~E. Gonzalez}, {and} \bibinfo{person}{Ion Stoica}.} \bibinfo{year}{2022}\natexlab{}.
\newblock \showarticletitle{Alpa: Automating Inter- and {Intra-Operator} Parallelism for Distributed Deep Learning}. In \bibinfo{booktitle}{\emph{16th USENIX Symposium on Operating Systems Design and Implementation (OSDI 22)}}. \bibinfo{publisher}{USENIX Association}, \bibinfo{address}{Carlsbad, CA}, \bibinfo{pages}{559--578}.
\newblock
\showISBNx{978-1-939133-28-1}
\urldef\tempurl%
\url{https://www.usenix.org/conference/osdi22/presentation/zheng-lianmin}
\showURL{%
\tempurl}


\bibitem[Zhi et~al\mbox{.}(2021)]%
        {zhi2021methodology}
\bibfield{author}{\bibinfo{person}{Haocong Zhi}, \bibinfo{person}{Xianuo Xu}, \bibinfo{person}{Weijian Han}, \bibinfo{person}{Zhilin Gao}, \bibinfo{person}{Xiaohang Wang}, \bibinfo{person}{Maurizio Palesi}, \bibinfo{person}{Amit~Kumar Singh}, {and} \bibinfo{person}{Letian Huang}.} \bibinfo{year}{2021}\natexlab{}.
\newblock \showarticletitle{A methodology for simulating multi-chiplet systems using open-source simulators}. In \bibinfo{booktitle}{\emph{Proceedings of the Eight Annual ACM International Conference on Nanoscale Computing and Communication}}. \bibinfo{pages}{1--6}.
\newblock


\bibitem[Zhong et~al\mbox{.}(2024)]%
        {zhong2024distserve}
\bibfield{author}{\bibinfo{person}{Yinmin Zhong}, \bibinfo{person}{Shengyu Liu}, \bibinfo{person}{Junda Chen}, \bibinfo{person}{Jianbo Hu}, \bibinfo{person}{Yibo Zhu}, \bibinfo{person}{Xuanzhe Liu}, \bibinfo{person}{Xin Jin}, {and} \bibinfo{person}{Hao Zhang}.} \bibinfo{year}{2024}\natexlab{}.
\newblock \showarticletitle{DistServe: Disaggregating Prefill and Decoding for Goodput-optimized Large Language Model Serving}. In \bibinfo{booktitle}{\emph{18th USENIX Symposium on Operating Systems Design and Implementation (OSDI 24)}}. \bibinfo{pages}{193--210}.
\newblock


\bibitem[Zhu et~al\mbox{.}(2024)]%
        {zhu2024theseus}
\bibfield{author}{\bibinfo{person}{Jingchen Zhu}, \bibinfo{person}{Chenhao Xue}, \bibinfo{person}{Yiqi Chen}, \bibinfo{person}{Zhao Wang}, {and} \bibinfo{person}{Guangyu Sun}.} \bibinfo{year}{2024}\natexlab{}.
\newblock \showarticletitle{Theseus: Towards High-Efficiency Wafer-Scale Chip Design Space Exploration for Large Language Models}.
\newblock \bibinfo{journal}{\emph{arXiv preprint arXiv:2407.02079}} (\bibinfo{year}{2024}).
\newblock


\end{thebibliography}

\end{document}